\documentclass[pdflatex,sn-basic]{sn-jnl}

\usepackage{graphicx}%
\usepackage{multirow}%
\usepackage{amsmath,amssymb,amsfonts}%
\usepackage{amsthm}%
\usepackage{mathrsfs}%
\usepackage[title]{appendix}%
\usepackage{xcolor}%
\usepackage{textcomp}%
\usepackage{manyfoot}%
\usepackage{booktabs}%
\usepackage{algorithm}%
\usepackage{algorithmicx}%
\usepackage{algpseudocode}%
\usepackage{listings}%

\usepackage{mathtools}%
\usepackage{orcidlink}
\usepackage{setspace}
\usepackage{lmodern}
\usepackage{anyfontsize}

\usepackage{xr}
\makeatletter
\newcommand*{\addFileDependency}[1]{
\typeout{(#1)}
\@addtofilelist{#1}
\IfFileExists{#1}{}{\typeout{No file #1.}}
}\makeatother

\newcommand*{\myexternaldocument}[1]{%
\externaldocument{#1}%
\addFileDependency{#1.tex}%
\addFileDependency{#1.aux}%
}

\myexternaldocument{supp}

\newcommand{\dd}{\text{d}}

\usepackage{xr-hyper}
\hypersetup{colorlinks,
    linkcolor={blue},
    filecolor={blue},
    urlcolor={blue},
    citecolor={blue}
}

\begin{document}

\title{Approximate solutions of a general stochastic velocity-jump \textcolor{black}{model} subject to discrete-time noisy observations}

\author*[1]{\fnm{Arianna} \sur{Ceccarelli}\orcidlink{0000-0002-9598-8845}}\email{arianna.ceccarelli@maths.ox.ac.uk}

\author[1]{\fnm{Alexander P.} \sur{Browning}\orcidlink{0000-0002-8753-1538}}

\author[1]{\fnm{Ruth E.} \sur{Baker}\orcidlink{0000-0002-6304-9333}}

\affil[1]{\centering
Mathematical Institute, University of Oxford,\\Woodstock Road, Oxford, OX2 6GG, UK}

\abstract{\textcolor{black}{Advances in experimental techniques allow the collection of high-resolution spatio-temporal data that track individual motile entities over time. These tracking data motivate the use of mathematical models to characterise the motion observed. In this paper, we aim to describe the solutions of velocity-jump models for single-agent motion in one spatial dimension, characterised by successive Markovian transitions within a finite network of \textit{n} states, each with a specified velocity and a fixed rate of switching to every other state. In particular, we focus on obtaining the solutions of the model subject to noisy, discrete-time, observations, with no direct access to the agent state. The lack of direct observation of the hidden state makes the problem of finding the exact distributions generally intractable. Therefore, we derive a series of approximations for the data distributions. We verify the accuracy of these approximations by comparing them to the empirical distributions generated through simulations of four example model structures. These comparisons confirm that the approximations are accurate given sufficiently infrequent state switching relative to the imaging frequency. The approximate distributions computed can be used to obtain fast forwards predictions, to give guidelines on experimental design, and as likelihoods for inference and model selection.}}


\maketitle

\section{Introduction}

Mathematical modelling of motility is an area of huge interest across numerous scientific domains, including ecology, biochemistry and cancer science. Modelling has been used, for example, to characterize bacterial chemotaxis~\citep{salek2019bacterial, rosser2014modelling, rosser2013novel, erban2004individual, berg1972chemotaxis}, the motion of molecular motors~\citep{han2024semi,hughes2012kinesins,hughes2011matrix,clancy2011universal,kutys2010monte}, axonal transport~\citep{cho2020fast,xue2017recent, bressloff2013stochastic, popovic2011stochastic, blum1988transport, blum1985model},
RNA motility~\citep{miles2024inferring, harrison2019testing, harrison2018impact, ciocanel2018modeling, ciocanel2017analysis}, cell migration~\citep{patel2018unique, jones2015inference} and animal movement~\citep{pike2023simulating, powalla2022numerical, taylor2015birds, preisler2004modeling, medvinsky2002spatiotemporal, bovet1988spatial, kareiva1983analyzing}.

\begin{figure*}[!ht]
    \centering
    \includegraphics[width=1\textwidth]{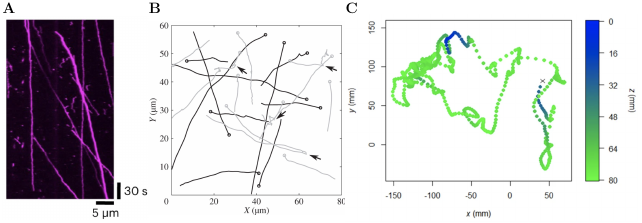}
    \caption{Examples of motion processes in biology. \textbf{A}. \textcolor{black}{One-dimensional movement of a kinesin molecular motor (Alexa Fluor 647 (A647)-labelled Khc FL) along a microtubule, taken from \cite{heber2024tropomyosin}. This kymograph is obtained projecting the position of the particles over time along an individual microtubule whose plus and minus ends are oriented towards the right and left, respectively.} \textbf{B}. Bacterial tracks with angular reorientations, taken from \cite{rosser2014modelling}. \textbf{C}. Tracks of fish movement in a tank, in which each dot represents a measured location in three dimensions at 0.1 s intervals, taken from \cite{pike2023simulating}. Panels \textbf{A} and \textbf{C} are reprinted under a     \href{http://creativecommons.org/licenses/by/4.0/}{Creative Commons Attribution 4.0 International License}.}
    \label{Fig:introduction}
\end{figure*}

\textcolor{black}{The spatial and temporal resolution of the data available to characterise motility is increasing thanks to the continued development of experimental imaging technologies~\citep{meijering2012methods}. In particular, experimental data that enables tracking of the location of individual motile entities over discrete time points is now readily available in many scenarios (see Figure~\ref{Fig:introduction}). These tracking data motivate the development and analysis of mathematical models to capture individual-level motion. However, mathematical models are often continuous in space and time, and the inherent constraints of experimental data, such as the collection of images at discrete time points and the introduction of errors in determining individual locations, are often neglected. In this work, we present a continuous-time stochastic model suitable to describe single-agent motion, and we approximate the model solutions subject to discrete-time noisy observations.}

\textcolor{black}{Single-agent movement is often modelled as a stochastic process, usually characterised by random jumps in space or in velocity. On the one hand, space-jump models, also referred to as random-walk models, have been widely used to simulate single-agent motion~\citep{jones2015inference, wu2000modelling, bergman2000caribou, bovet1988spatial, kareiva1983analyzing}, and analytical tools have been developed to characterise the agent motion or estimate model parameters~\citep{rosser2013effect, codling2005sampling, wu2000modelling, berg1993random, skellam1991random}. On the other hand, motility in many systems is well-described by velocity-jump models, in which the agent motion comprises a series of movements at constant velocity, separated by instantaneous reorientations during which the velocity is changed~\citep{treloar2011velocity,othmer1988models}. Velocity-jump models are fully characterised by the re-orientation kernels, velocities and waiting time distributions between reorientations. In this paper, we focus on stochastic models able to describe biological processes that exhibit velocity-jump-like motion and where data about the agent location can be collected at sufficiently high frequency. A number of frameworks have been provided to estimate asymptotic or mean quantities of interest~\citep{han2024semi,knoops2018motion,taylor2015birds} or to calibrate model parameters to data~\citep{harrison2018impact,rosser2013novel}. However, there is a lack of solutions for general stochastic velocity-jump models subject to discrete noisy observations, hence, in this work, we aim to contribute to the literature by obtaining analytical approximations for the model solutions.}

\textcolor{black}{We consider a general velocity-jump model in which an agent moves deterministically with a velocity prescribed by a discrete state. In particular, state $s$, associated with fixed velocity $v_s$ and fixed rates of switching to each other state, evolves according to a continuous-time Markov chain (CTMC). We focus on the case in which the data consists of a set of agent locations observed at discrete times and with measurement noise, such that the state is not directly observed, and the agent location does not, by itself, evolve according to a Markov process. In particular, the lack of direct observation of the hidden internal state impedes determination of the exact model solution. To capture the characteristics of imaging techniques, we propose a data collection model able to generate \textit{in silico} data tracks capturing one-dimensional motion, in which the locations are obtained at fixed time steps and with measurement noise. We then describe these data by approximating the probability distribution function (PDF) of the location increments, defined as difference between two subsequent measured locations.}

First, we compute an approximation for the marginal distribution of measuring a single location increment, and then we compute the joint distribution for a set of subsequent location increments. The challenges of computing these PDFs stem from the inherent noise and the experimental constraint that the agent location is measured with a discrete time frequency, preventing the determination of exact switching times which are necessary to directly compute velocities and switching rates from the data. Furthermore, we expect the approximate PDFs to be accurate for infrequent state transitions, thus to assess the validity of the approximations produced we compare them to the corresponding empirical PDFs for different model networks varying the switching rates.

The paper is organised as follows. In Section~\ref{Sec:2} we formulate the $n$-state velocity-jump model and propose a data collection model based on the intrinsic characteristics of typical experimental tracking data, obtained with a fixed time frequency and measurement noise. In Section~\ref{Sec:3} we provide approximations for the PDF of a single location increment, which are based on considering a finite number of state switches per measured interval. We plot these approximations for four network models and compare them with the empirical PDF for both infrequent and frequent state transitions. In Section~\ref{Sec:4} we produce an approximation for the joint PDF of a set of subsequent location increments, and we test its validity by comparing it with the empirical PDF for the models previously considered. In Section~\ref{Sec:5} we conclude that the approximate PDFs computed capture the distribution of the data when switching is sufficiently infrequent compared to the time resolution of the data. \textcolor{black}{The approximate solutions provided for the $n$-state velocity-jump model presented could be used in several applications, including obtaining model predictions without running simulations, informing details of experimental setup such as data collection frequency, and performing model calibration and selection using single-agent tracking data.}

\section{Formulation of the \textit{n}-state velocity-jump model}\label{Sec:2}

In this section, we present a one-dimensional $n$-state velocity-jump model suitable to be calibrated with typically measured data (see, for example, the tracking data in Figure~\ref{Fig:introduction}\textbf{A}). At any point in time $t\ge0$, the agent is in a state $S(t)=s$, with $s\in \{1,2,\ldots,n\}$, and in this state it moves with constant velocity $v_s\in\mathbb{R}$. We take velocities to be scalars for simplicity, however, the model as described holds for higher-dimensional velocity vectors, and the methods presented in this paper could be extended to higher spatial dimensions.

The sequence of states attained is a CTMC, a continuous stochastic process in which the time between changes of state is an exponential random variable, and the new state attained is sampled according to the probabilities specified in a stationary transition matrix~\citep{anderson2012continuous,liggett2010continuous,norris1998markov}. In particular, the amount of time spent in state $s$ before switching to a different state is an exponential random variable $\tau_s\sim\mathrm{Exp}{(\lambda_s)}$, for $\lambda_s\ne 0$ constant. We note that $\mathbb{E}[\tau_s] = 1/\lambda_s$ is the average time spent in state $s$. 

\textcolor{black}{The sequence of states is a discrete-time Markov chain (DTMC), often referred to as the embedded Markov chain of the continuous-time process.} We denote with $p_{su}$ the probability of switching from state $s$ to state $u$
\begin{equation}\label{Eq:p_su}
    p_{su}:= \mathbb{P}\left(S_{k}=u \;\middle|\; S_{k-1}=s\right),
\end{equation}
for $k\ge 2$, where $S_{k}$ denotes the $k$-th state attained. The probabilities $p_{su}$ are constant; thus they do not depend on $k$, and the Markovian property holds. By construction, $p_{ss}:=0$, and $\sum_{u=1}^n p_{su}:=1$. We write the transition matrix of the DTMC as
$$\boldsymbol{P}:=
\begin{bmatrix}
    0 & p_{12} & p_{13} & \hdots & p_{1n}
    \\
    p_{21} & 0 & p_{23} & \hdots & p_{2n}
    \\
    p_{31} & p_{32} & 0 & \hdots & p_{3n}
    \\
    \vdots & \vdots & \vdots & \ddots & \vdots
    \\
    p_{n1} & p_{n2} & p_{n3} & \hdots & 0
\end{bmatrix}.$$
For every state $u\ne s$, we define $q_{su}:=\lambda_s p_{su}$, and we define $q_{ss}:=-\lambda_s$ such that 
$$\sum_{u=1}^n q_{su}=0,$$
for all $s$.
These $q_{su}$ are then the entries of the transition-rate matrix of the CTMC
$$\boldsymbol{Q} :=
\begin{bmatrix}
    -\lambda_1 & \lambda_1 p_{12} & \lambda_1 p_{13} & \hdots & \lambda_1 p_{1n}
    \\
    \lambda_2 p_{21} & -\lambda_2 & \lambda_2 p_{23} & \hdots & \lambda_2 p_{2n}
    \\
    \lambda_3 p_{31} & \lambda_3 p_{32} & -\lambda_3 & \hdots & \lambda_3 p_{3n}
    \\
    \vdots & \vdots & \vdots & \ddots & \vdots
    \\
    \lambda_n p_{n1} & \lambda_n p_{n2} & \lambda_n p_{n3} & \hdots & -\lambda_n
\end{bmatrix}.$$
\textcolor{black}{Each entry $q_{su}$ for $u\ne s$ represents the probability that the chain moves from state $s$ to state $u$, divided by the expected time spent in state $s$.}

Without loss of generality, starting at a time $t_0=0$, we define the initial location of the agent as $x(0)=0$. Denoting $\hat{T}_s (t)$ for $s=1,2,\ldots,n$ as the total amount of time spent in state $s$ in the time interval $[0,t]$ we have
$$x(t) = \sum_{s=1}^n v_s \hat{T}_s (t).$$
We assume no knowledge of the agent initial state, which corresponds to assuming that the process is in equilibrium. The initial state $S_0:=S(0)$ is randomly sampled from the stationary  distribution of the CTMC, described by a vector $\boldsymbol{\pi} := [p_1, p_2, \ldots, p_n]$ defining the probability of being in each state $s=1,2,\ldots,n$, constructed such that $\boldsymbol{\pi}\boldsymbol{Q} = \boldsymbol{0}$. This choice of $\boldsymbol{\pi}$ guarantees the time independence of the process, which translates to having a constant marginal probability of being at each state over time 
\begin{equation}\label{Eq:P(S(t)=s)}
    \mathbb{P}(S(t)=s)=:\mathbb{P}(s)=p_s,
\end{equation} 
for all $t\ge 0$. The existence and uniqueness of the vector $\boldsymbol{\pi}$ are guaranteed by assuming that the chain is irreducible (see Supplementary Information Section~\ref{SI:Sec:Irreducibility} and \cite{liggett2010continuous} Proposition~2.59 and Corollary~2.67). In this assumption, the vector $\boldsymbol{\pi}$ can be explicitly computed by finding the kernel of $\boldsymbol{Q}^T$ (see Supplementary Information Section~\ref{SI:Sec:P(s)}).

In the model outlined here, the state evolution is a CTMC, the properties of which have been widely studied~\citep{ross2014introduction,anderson2012continuous,liggett2010continuous,norris1998markov}. However, existing mathematical models often overlook aspects of data collection, such as the fixed-time intervals and the experimental noise of the measured locations. In Section~\ref{Subsec:Data_collection} we introduce a measurement model for data collection, designed to capture these inherent traits of experimental setups.

Before proceeding, we first introduce four example networks that will be used throughout this work to illustrate the validity of the approximate solutions computed. These model networks are illustrated in Figure~\ref{Fig:in silico_track_examples}\textbf{A}-\textbf{D}, while the parameter sets are fully specified in Supplementary Information Figure~\ref{SI:Fig:networks}. In these figures, the nodes indicate the states and the black arrows represent a non-zero probability of switching between states, after a time sampled from an exponential distribution. The simplest model on the left is a two-state model (Figure~\ref{Fig:in silico_track_examples}\textbf{A}), in which the state switches between a forward state (\textit{F}), in which the agent has velocity $v_F>0$, and a backward state (\textit{B}), in which it has velocity $v_B<0$. The other networks presented are obtained adding stationary states to this two-state network. The three-state model (Figure~\ref{Fig:in silico_track_examples}\textbf{B}) is obtained by adding a stationary state with long average permanence (\textit{SL}), which corresponds to having a low switching rate compared to the other states. The four-state model (Figure~\ref{Fig:in silico_track_examples}\textbf{C}) is obtained by adding a stationary state with short average permanence (\textit{SS}) which corresponds to a higher switching rate compared to state \textit{SL}. In the four-state model (Figure~\ref{Fig:in silico_track_examples}\textbf{D}) the states \textit{F} and \textit{B} are not directly connected; rather, an agent needs to enter a stationary state before changing direction. Moreover, entering state \textit{SL} implies that the following state attained must be \textit{SS}, and therefore the total time spent in a stationary phase follows a hypoexponential distribution, as it is obtained as the sum of two exponential distributions. Finally, the six-state network is given a cyclic structure, and it incorporates stationary states followed by a very likely return to the previous velocity state (\textit{F} or \textit{B}), denoted as pause while moving forward (\textit{PF}) and pause while moving backward (\textit{PB}).

\begin{figure*}[!ht]
    \centering
    \includegraphics[width=1\textwidth]{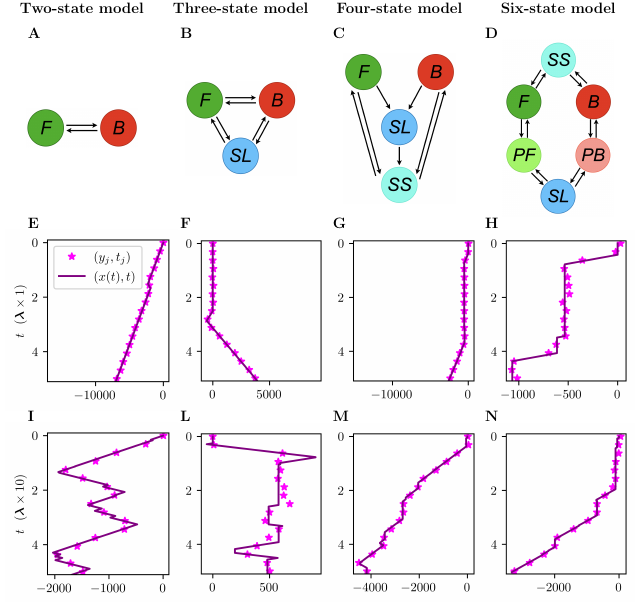}
    \caption{Panels \textbf{A}-\textbf{D} describe the model networks we use throughout the manuscript to demonstrate the results obtained, more details are included in Supplementary Information Figure~\ref{SI:Fig:networks}. Panels \textbf{E}-\textbf{N} show examples of \textit{in silico} data tracks from the model networks specified above, with two, three, four and six states. \textit{F} stands for forward state, \textit{B} for backward state, \textit{SL} for stationary state with long average permanence, and \textit{SS} for stationary state with short average permanence, \textit{PF} for pause while moving forward and \textit{PB} for pause while moving backward. The tracks are generated using the sets of parameters specified in Supplementary Information Figure~\ref{SI:Fig:networks}, except the switching rates $\boldsymbol{\lambda}=[\lambda_1, \lambda_2,\ldots,\lambda_n]$, multiplied by 10 to obtain the tracks in panels \textbf{I}-\textbf{N}. In an experimental setting increasing the switching rates is equivalent to increasing the time between collected images $\Delta t$.}
    \label{Fig:in silico_track_examples}
\end{figure*}

\subsection{Data collection model}\label{Subsec:Data_collection}

Here, we define a data collection model that mimics the data obtained from many experiments (see Figure~\ref{Fig:introduction}\textbf{A},\textbf{C}), in which measurements are noisy and can only be obtained at discrete times, and highlight the associated challenges. \textcolor{black}{In particular, we note that, together, $x(t)$ and $S(t)$ are a jointly Markov process, often referred to as the \textit{system model}. However, $x(t)$ alone is not a Markov process. Moreover, observed agent locations are collected at discrete time points, which does not allow us to directly determine transitions in $S(t)$. The data collection model is sometimes referred to as the observation model.}

\begin{figure*}
    \centering
    \includegraphics[width=0.7\textwidth]{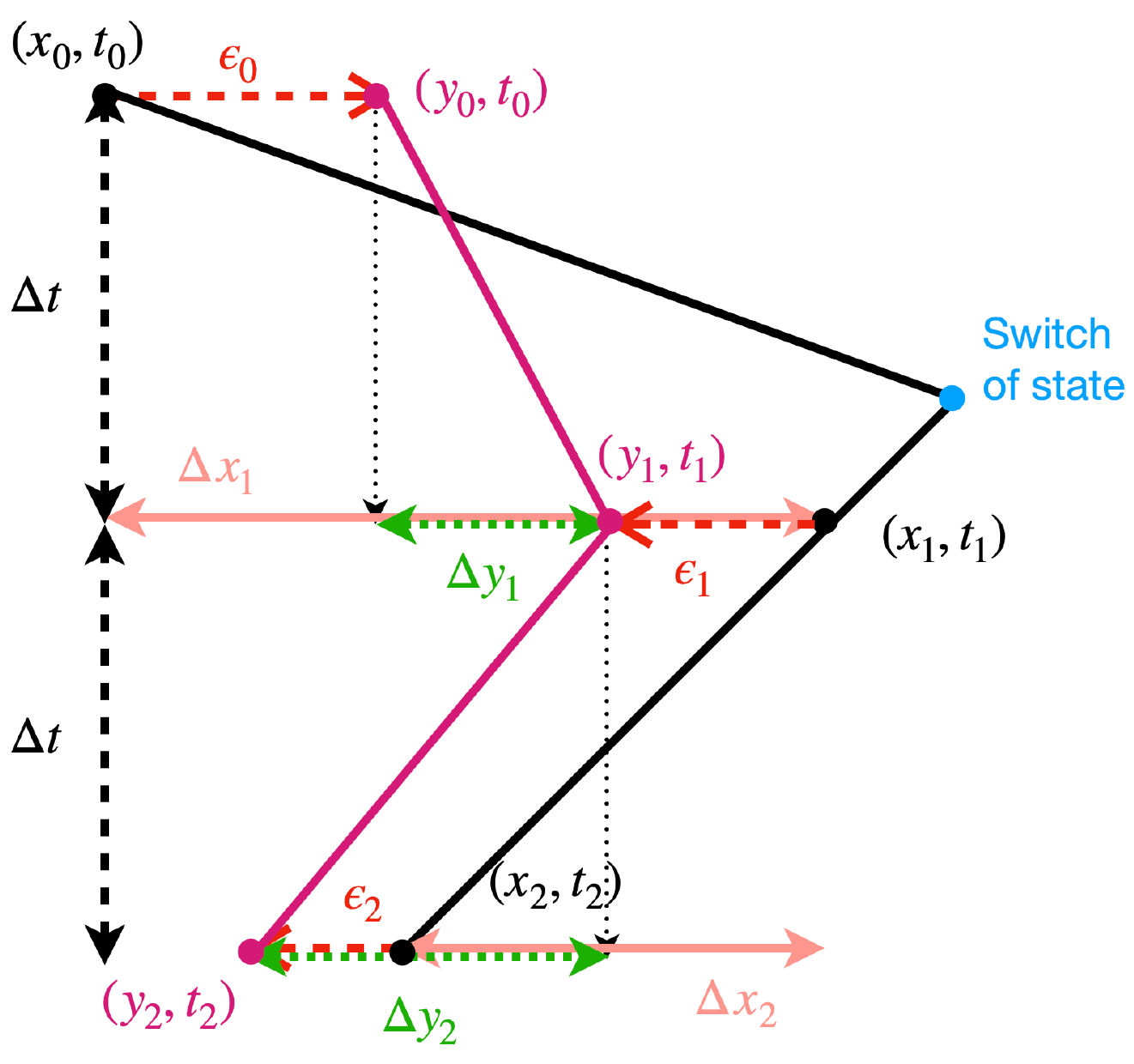}
    \caption{Visual description of the data collection model. At time $t_j=j\Delta t$ the agent is at location $x_j=x(j\Delta t)$. The $j$-th \textit{exact} location increment is defined as $\Delta x_j:=x_j - x_{j-1}$. The $j$-th \textit{noisy} location is defined as $y_j:= x_j+\epsilon_j$, where $\epsilon_j$ is the measurement error, assumed to be normally distributed. The $j$-th \textit{noisy} location increment is defined as $\Delta y_j:=y_j-y_{j-1}$.}\label{Fig:displacement_increments}
\end{figure*}

By simulating the CTMC we generate the agent location $x(t)$, for $t\in[0, N\Delta t]$ with $N\in \mathbb{N}$, and we obtain a set of $N+1$ data points which are the agent locations measured with fixed time frequency $\Delta t$ (see Figure~\ref{Fig:displacement_increments}). \textcolor{black}{For $j\in \{0,1,2,\ldots, N\}$, at time $t_j:=j\Delta t$, we denote the exact location of the agent as $x_j:=x(t_j)$, and the observed location as 
$$y_j:=y(t_j):=x_j + \epsilon_j,$$
where $\epsilon_j\sim \mathcal{N}(0,\sigma^2)$ denotes independent, normally distributed, measurement noise~\citep{pawitan2001all}. The set of $y_j$ for $j\in \{0,1,2,\ldots, N\}$ can also be referred to as the observation model.} The methods presented generalise to other noise model distributions for which the Markov property used still holds. Some examples of \textit{in silico} tracks are shown in Figure~\ref{Fig:in silico_track_examples}. Figure~\ref{Fig:displacement_increments} highlights that since the data is noisy and discrete we cannot determine the exact switching times.

We now present more notation that will be useful in the following sections. We define the total simulation time as $T:=N\Delta t$. For $j\in\{1,2,\ldots,N\}$, we define the \textit{exact} location increment as
$$\Delta x_j := x_j - x_{j-1},$$
and the \textit{noisy} location increment as
$$\Delta y_j := y_j - y_{j-1} = \Delta x_j + \Delta \epsilon_j \sim \mathcal{N}(\Delta x_j, 2\sigma^2),$$
since $\Delta\epsilon_j := \epsilon_j - \epsilon_{j-1} \sim \mathcal{N}(0, 2\sigma^2)$.

\textcolor{black}{In general, subsequent location increments will be correlated, as the previous increments provide information about the hidden state that encodes the velocity at the start of each interval (see Figure~\ref{Fig:displacement_increments}).} In particular, if the switching rates are small compared to the time interval between collected images then the state switches occur infrequently. Equivalently, $\mathbb{E}[\tau_s] = 1/\lambda_s \gg \Delta t$ for all $s$, and in this case the velocity at the beginning of an increment is likely to be the same as that at the beginning of the previous increment. We note that $1/\lambda_s \gg \Delta t$ can be obtained experimentally by appropriately choosing the time between captured images $\Delta t$.

The characteristics of the data collection model are considered in the following sections to compute approximations for the PDFs of both a single noisy location increment (Section~\ref{Sec:3}) and a set of subsequent noisy location increments obtained from a data track (Section~\ref{Sec:4}).

\section{Approximate solutions for the \textit{n}-state velocity-jump model}\label{Sec:3}

In this section, we aim to compute the solutions for the $n$-state velocity-jump model proposed in Section~\ref{Sec:2} to describe the distribution of measuring a single noisy location increment $\Delta y$, denoted as $\mathbb{P}(\Delta y)$. \textcolor{black}{We note that our methodology applies to the velocity-jump model with any network of $n$ states subject to discrete-time noisy observations.}

We can write the probability of measuring a single location increment, $\Delta y$,
\begin{equation}\label{Eq:P(Delta y)}
    \mathbb{P}(\Delta y) =\sum_{w=0}^{\infty} \mathbb{P}(\Delta y \,|\, W=w)\mathbb{P}(W=w),
\end{equation}
where $\mathbb{P}(\Delta y \,|\, W=w)$ is the density of $\Delta y$ conditioned on the number of switches $W$ during the given interval. The problem of finding an exact solution for the PDF $\mathbb{P}(\Delta y)$ is intractable. From Equation~\eqref{Eq:P(Delta y)} we can observe that there are an infinite number of terms to compute for which we cannot obtain a general formula. Hence, we define approximations that consider up-to-$m$ switches per measured interval limiting the amount of terms to compute. We rewrite Equation~\eqref{Eq:P(Delta y)} as
\begin{equation}\label{Eq:P(Delta y)_}
    \mathbb{P}(\Delta y) =\sum_{w=0}^{m} \mathbb{P}(\Delta y \,|\, W=w)\mathbb{P}(W=w) + \mathbb{P}(\Delta y \,|\, W>m)\mathbb{P}(W>m),
\end{equation}
and we define PDF approximations that assume
$$\mathbb{P}(\Delta y \,|\, W>m)\approx\mathbb{P}(\Delta y \,|\, W=m).$$
Using it, we obtain what we call an up-to-$m$-switch approximation $P_m(\Delta y)$, given by
\begin{equation}\label{Eq:P_m(Delta y)}
    P_m(\Delta y) :=\sum_{w=0}^{m-1} \mathbb{P}(\Delta y \,|\, W=w)\mathbb{P}(W=w) + \mathbb{P}(\Delta y \,|\, W=m)\mathbb{P}(W\ge m).
\end{equation}

We expect this approximation to work well either when $\mathbb{P}(\Delta y \,|\, W>m)\approx\mathbb{P}(\Delta y \,|\, W=m)$ or when $\mathbb{P}(W>m)\approx 0$. Alternatively, we could also define a truncated PDF approximation by considering only the terms with at most $m$ switches, using the approximation $\mathbb{P}(W>m)\approx 0$, but our investigations suggest that it provides a less accurate approximation compared to that in Equation~\eqref{Eq:P_m(Delta y)}. We now give some indication on the proportion of intervals with a set number of switches, $\mathbb{P}(W=w)$, by focusing on the two-state model and the three-state model previously introduced (see Supplementary Information Figure~\ref{SI:Fig:networks} for full details).

\begin{figure*}[!ht]
    \centering
    \includegraphics[width=1\textwidth]{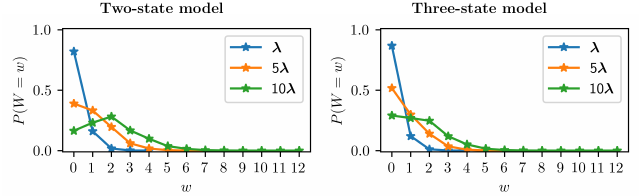}
    \caption{Empirical distributions of $\mathbb{P}(W=w)$, using the rates $\boldsymbol{\lambda}=[\lambda_1,\lambda_2,\ldots, \lambda_n]$ specified in Supplementary Information Figure~\ref{SI:Fig:networks}, for the two-state model (from  Figure~\ref{Fig:in silico_track_examples}\textbf{A}) and three-state model (from Figure~\ref{Fig:in silico_track_examples}\textbf{B}).}
    \label{Fig:P(K_w)}
\end{figure*}

The plots in Figure~\ref{Fig:P(K_w)} illustrate the probability of having a number of switches $w$, $\mathbb{P}(W=w)$, 
in an interval of fixed time length $\Delta t$, for the two-state model and three-state model with rates $\boldsymbol{\lambda}$ varied by the factors 1, 5 and 10. Parameters and networks are used as specified in Supplementary Information Figure~\ref{SI:Fig:networks}. 
Equivalently, these plots (Figure~\ref{Fig:P(K_w)}) indicate how the probability $\mathbb{P}(W=w)$ varies multiplying the measured time length $\Delta t$ by the factors 1, 5 and 10, keeping the first set of rates $\boldsymbol{\lambda}\times 1$.

We now approximate the PDF of a noisy location increment $\Delta y$, $\mathbb{P}(\Delta y)$, by considering both an up-to-one-switch approximation ($P_1(\Delta y)$; Section~\ref{Subsec:up-to-one-switch}) and an up-to-two-switch approximation ($P_2(\Delta y)$; Section~\ref{Subsec:up-to-two-switch}).

\subsection{Up-to-one-switch approximation for the probability distribution function of a location increment}\label{Subsec:up-to-one-switch}

\begin{figure*}[!ht]
    \centering
    \includegraphics[width=0.8\textwidth]{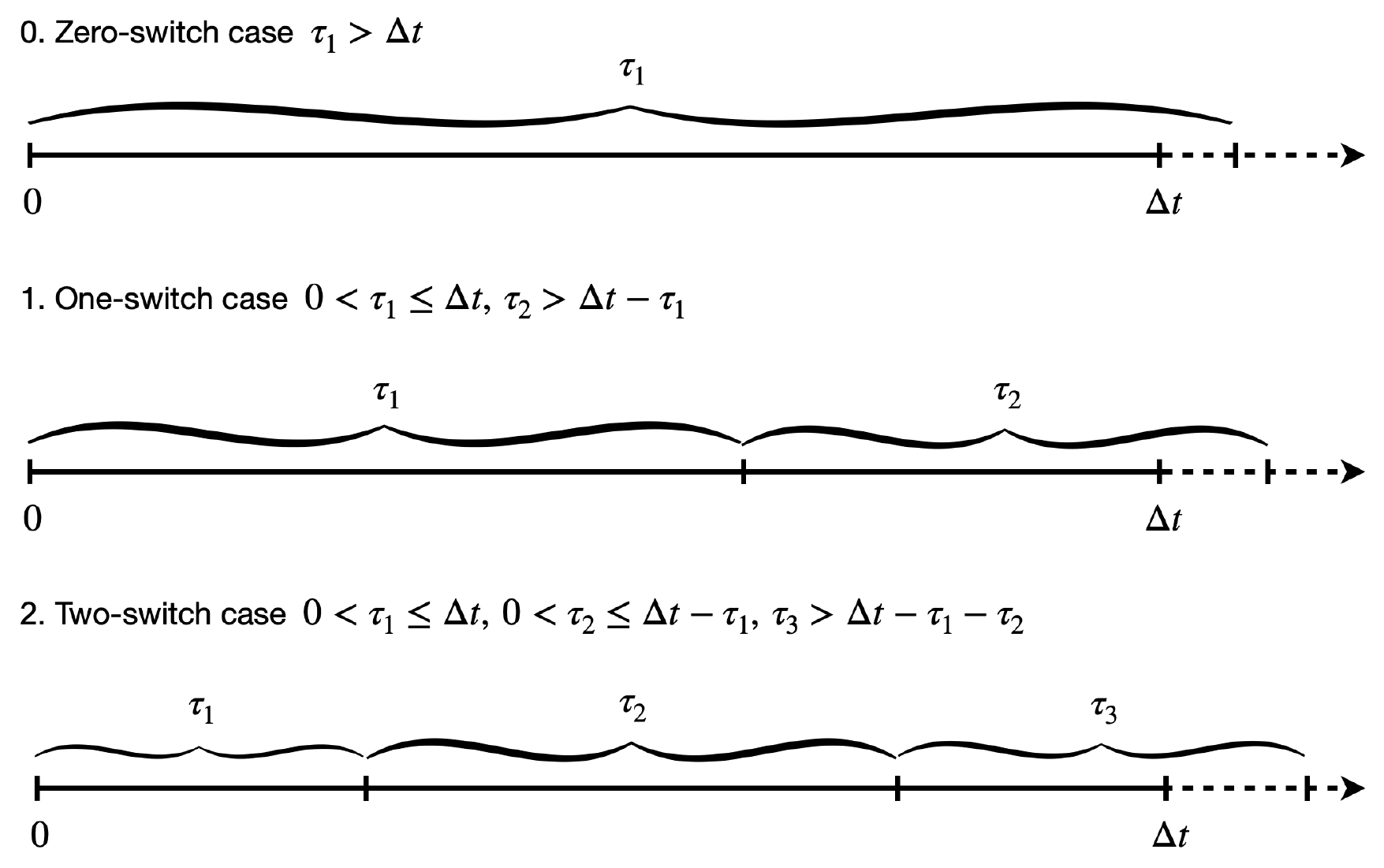}
    \caption{Diagrams representing the state switching times in the cases of zero switches, one switch and two switches. The variable $\tau_1>0$ indicates the time spent in state $S_1$, $\tau_2>0$ the time spent in state $S_2$ and $\tau_3>0$ the time spent in state $S_3$.}
    \label{Fig:switches_diagram}
\end{figure*}

We now consider the up-to-one-switch approximation
\begin{equation}\label{Eq:P_1(Delta y)}
    P_1(\Delta y) :=\mathbb{P}(\Delta y \,|\, W=0)\mathbb{P}(W=0)+\mathbb{P}(\Delta y \,|\, W=1)\mathbb{P}(W\ge1).
\end{equation}
We make progress by further conditioning each term in Equation~\eqref{Eq:P_1(Delta y)} on the state at the start of the interval, denoted as $S_{1}$. Namely, for $W = 0$, we see that
$$\mathbb{P}(\Delta y \,|\, W=0)\mathbb{P}(W=0) = \sum_{s=1}^n \mathbb{P}(\Delta y \,|\, W=0, S_{1}=s)\mathbb{P}(W=0\,|\,S_{1}=s) \mathbb{P}(S_{1}=s),$$
where $\mathbb{P}(S_{1}=s)=p_{s}$ is given by the equilibrium assumption (Equation~\eqref{Eq:P(S(t)=s)}). For $W = 0$, the amount of time spent in the first state is, trivially, given by $\Delta t$ (see Figure~\ref{Fig:switches_diagram} zero-switch case). Therefore we have
$$\mathbb{P}(W=0\,|\,S_{1}=s)=\exp{(-\lambda_{s}\Delta t)},$$
and
$$\mathbb{P}(\Delta y \,|\, W=0, S_{1}=s)=f_{\mathcal{N}(v_{s} \Delta t,2\sigma^2)}(\Delta y),$$
where the right hand side denotes the PDF of the normal distribution with mean $v_{s} \Delta t$ and variance $2\sigma^2$. For more details on these results see Supplementary Information Section~\ref{SI:Sec:P_1}.

To compute $\mathbb{P}(\Delta y \,|\, W=1)\mathbb{P}(W\ge1)$, we condition on both the first state and the second state visited within the interval, $S_{1}$ and $S_{2}$ (see one-switch case in Figure~\ref{Fig:switches_diagram}). This yields
$$\begin{aligned}
    \mathbb{P}(\Delta y \,|\, W=1)\mathbb{P}(W \ge1) =\sum_{\substack{s_{1}\\s_{2}\ne s_{1}}}
    &\mathbb{P}(\Delta y \,|\, W=1, S_{1}=s_{1}, S_{2}=s_{2})
    \\
    &\times \mathbb{P}(W\ge1\,|\,S_{1}=s_{1}, S_{2}=s_{2}) 
    \\
    &\times \mathbb{P}(S_{2}=s_{2}\,|\,S_{1}=s_{1})\mathbb{P}(S_{1}=s_{1}).
\end{aligned}$$
By definition, $\mathbb{P}(S_{2}=s_{2}\,|\, S_{1}=s_{1})=p_{s_{1}s_{2}}$ (Equation~\eqref{Eq:p_su}), and we have
$$\mathbb{P}(W\ge1\,|\,S_{1}=s_{1}, S_{2}=s_{2})=\mathbb{P}(W\ge 1\,|\,S_{1}=s_{1})=1-\exp{(-\lambda_{s_{1}}\Delta t)}$$
by considering that at least one switch occurs if $\tau_1<\Delta t$.

Computation of the PDF for $\Delta y$ if one switch occurs and given the two states visited, denoted as
$$\Tilde f_{s_{1},s_{2}}(\Delta y) := \mathbb{P}(\Delta y\,|\, W=1, S_{1}=s_{1}, S_{2}=s_{2}),$$
is significantly more involved. We proceed by considering the corresponding PDF for the exact increment 
$$\Tilde{g}_{s_{1},s_{2}}(\Delta x):= \mathbb{P}(\Delta x\,|\, W=1, S_{1}=s_{1}, S_{2}=s_{2}),$$
and then convoluting the result with the distribution for $\Delta \epsilon$ to obtain $\Tilde f_{s_{1},s_{2}}(\Delta y)$.

We construct the PDF $\Tilde{g}_{s_{1},s_{2}}(\Delta x)$ by first computing the distribution of the switching time $\tau_1$ during the time interval considered $[0,\Delta t]$, with distribution given by
$$\begin{aligned}
    F_{s_{1},s_{2}}(t_1):=&\ \mathbb{P}(\tau_1 \le t_1, W=1\,|\,S_{1}=s_{1}, S_{2}=s_{2})
    \\=&\ \mathbb{P}(0<\tau_1 \le t_1, \tau_2>\Delta t - \tau_1\,|\,S_{1}=s_{1}, S_{2}=s_{2})
    \\=& \ \int_0^{t_1}\int_{\Delta t-\tau_1}^{\infty}f_{s_{1},s_{2}}(\tau_1, \tau_2)\dd \tau_2 \dd \tau_1,
\end{aligned}$$
for $t_1\in (0, \Delta t]$,
where $f_{s_{1},s_{2}}(\tau_1, \tau_2)=f_{\text{Exp}(\lambda_{s_1})}(\tau_1)\cdot f_{\text{Exp}(\lambda_{s_2})}(\tau_2)$ is the joint distribution of the time spent in the first and second states which are independent (see two-switch case in Figure~\ref{Fig:switches_diagram}). Moreover, we obtain the cumulative distribution function (CDF) of the switching time as
$$\begin{aligned}
    G_{s_{1},s_{2}}(t_1):=&
    \ \mathbb{P}(\tau_1 \le t_1\,|\, W=1, S_{1}=s_{1}, S_{2}=s_{2})
    \\=&
    \ \mathbb{P}(\tau_1 \le t_1\,|\, 0\le\tau_1\le\Delta t, \tau_2>\Delta t - \tau_1, S_{1}=s_{1}, S_{2}=s_{2})
    \\
    =&\ \frac{F_{s_{1},s_{2}}(t_1)}{F_{s_{1},s_{2}}(\Delta t)}.
\end{aligned}$$
This CDF can be differentiated to obtain the PDF of the switching time $\tau_1=t_1$, from which we obtain the PDF of the exact location increment $\Delta x$, $\Tilde{g}_{s_{1},s_{2}}(\Delta x)$. We have that, for fixed $S_{1}$, $S_{2}$ and switching at time $\tau_1$,
$$\Delta x = h_{s_{1},s_{2}}(\tau_1) :=v_{s_{1}}\tau_1+v_{s_{2}}(\Delta t - \tau_1).$$

If $v_{s_{1}}=v_{s_{2}}$, then the velocity is constant for the whole interval; thus the exact increment is known, $\Delta x=h_{s_{1},s_{2}}(\Delta t)$, and the distribution for the measured increment is
$$\Tilde{f}_{s_{1},s_{2}}(\Delta y)=f_{\mathcal{N}({v_{s_{1}}\Delta t},2\sigma^2)}(\Delta y),$$
where the right hand side denotes the PDF of the normal distribution.
Otherwise, when $v_{s_{1}}\ne v_{s_{2}}$, the exact increment $\Delta x$ is determined by the time of the switch $\tau_1$. In this case, we note that $h_{s_{1},s_{2}}$ is a monotonic function in $\tau_1$, then its inverse exists and gives
$$\tau_1 = h_{s_{1},s_{2}}^{-1}(\Delta x);$$
thus the PDF of $\Delta x$ is given by
$$\Tilde{g}_{s_{1},s_{2}}(\Delta x)=g_{s_{1},s_{2}}\left(h_{s_{1},s_{2}}^{-1}(\Delta x)\right)\cdot\left|\frac{\dd}{\dd(\Delta x)}h_{s_{1},s_{2}}^{-1}(\Delta x)\right|,$$
where $g_{s_{1},s_{2}}(t_1)$ is the PDF obtained by differentiating the CDF $G_{s_{1},s_{2}}(t_1)$ in $t_1$.

Finally, we obtain the PDF for $\Delta y$, $\Tilde f_{s_{1},s_{2}}(\Delta y)$, by incorporating the Gaussian noise, computing it as the distribution of the sum of two independent random variables $\Delta x$ and $\Delta \epsilon$. Therefore, we integrate the product of the PDF for $\Delta x$ and the PDF for $\Delta \epsilon = \Delta y- \Delta x$ for all $\Delta x$ and obtain
\begin{equation}\label{Eq:tildef_s1,s2}
\Tilde f_{s_{1},s_{2}}(\Delta y) = \int_{a}^b \Tilde{g}_{s_{1},s_{2}}(\Delta x)f_{\mathcal{N}(0,2\sigma^2)}(\Delta y-\Delta x) \dd(\Delta x),
\end{equation}
where we define
$$a=a_{s_{1},s_{2}}:= \min \{h_{s_{1},s_{2}}(0), h_{s_{1},s_{2}}(\Delta t)\},$$
and
$$b=b_{s_{1},s_{2}}:= \max \{h_{s_{1},s_{2}}(0), h_{s_{1},s_{2}}(\Delta t)\},$$
since $\Tilde{g}_{s_{1},s_{2}}(\Delta x)=0$ for $\Delta x < a$ or $\Delta x > b$. The integral in Equation~\eqref{Eq:tildef_s1,s2} can be computed explicitly; this is provided in Supplementary Information Section~\ref{SI:Sec:P_1}.

\begin{figure*}[!ht]
    \includegraphics[width=1\textwidth]{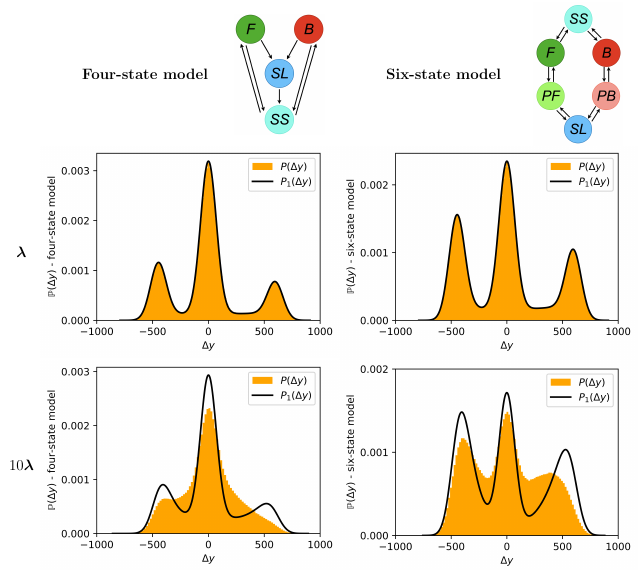}
    \caption{The empirical PDF for $\Delta y$, denoted $P(\Delta y)$, (orange histogram) is compared with its up-to-one-switch approximation $P_1(\Delta y)$ (black line). The left plots are obtained from a four-state model while the right ones from a six-state model. The top plots are obtained using the parameters as specified in Supplementary Information Figure~\ref{SI:Fig:networks}, while the bottom ones use the same parameters except the rates which are multiplied by 10.}
    \label{Fig:PDFs_4_6}
\end{figure*}

In Figure~\ref{Fig:PDFs_4_6} we present a comparison between the empirical PDF for $\Delta y$, denoted $P(\Delta y)$, and the up-to-one-switch approximation $P_1(\Delta y)$ for the four-state model and six-state model. To assess the performance of the approximation as the rate parameters increase for a fixed measurement interval, we also produce results with the rate parameters multiplied by 10. These results show that the up-to-one-switch approximation works well for infrequent switching relative to $\Delta t$ (low rates $\boldsymbol{\lambda}$), while the bottom plots indicate that the accuracy decreases as the rates are increased, as expected. This is expected since the \textit{in silico} data contains more increments with at least two switches as the rates are increased (for $\boldsymbol{\lambda}$ versus $10\boldsymbol{\lambda}$ for the four-state model the increments with at least two switches are 3\% versus 59\%, for the six-state model they are 6\% versus 72\%). Figure~\ref{Fig:PDFs_2_3} shows that for both the two-state model and the three-state model the accuracy of $P_1(\Delta y)$ decreases as the rates are increased. Finally, Supplementary Information Figure~\ref{SI:Fig:Error_PDF_4_6} shows the error of the up-to-one-switch PDF approximation, defined as $P_1(\Delta y)-P(\Delta y)$, for the panels shown in Figure~\ref{Fig:PDFs_4_6}.

To improve the accuracy of the approximation of $\mathbb{P}(\Delta y)$ for higher rates, one could use approximations that increase the maximum number of switches considered. In the next section, we compute an up-to-two-switch approximation; these steps could be adapted to produce approximations with an increased maximum number of switches.

\subsection{Up-to-two-switch approximation for the probability distribution function of a location increment}\label{Subsec:up-to-two-switch}

We now consider the up-to-two-switch approximation defined as in Equation~\eqref{Eq:P_m(Delta y)}
\begin{equation*}
\begin{aligned}
    P_2(\Delta y) :=&\ \mathbb{P}(\Delta y \,|\, W=0)\mathbb{P}(W=0)+\mathbb{P}(\Delta y \,|\, W=1)\mathbb{P}(W=1)
    \\
    &+\mathbb{P}(\Delta y \,|\, W=2)\mathbb{P}(W\ge2).
\end{aligned}
\end{equation*}
In Section~\ref{Subsec:up-to-one-switch} and in the Supplementary Information, we computed the terms needed to obtain $\mathbb{P}(\Delta y \,|\, W=0)\mathbb{P}(W=0)$
and $\mathbb{P}(\Delta y \,|\, W=1)\mathbb{P}(W=1)$.

In order to compute $\mathbb{P}(\Delta y \,|\, W=2)\mathbb{P}(W\ge2)$ we proceed as before by conditioning on all three subsequent states visited during an interval, $S_{1}$, $S_{2}$ and $S_{3}$ (two-switch case Figure~\ref{Fig:switches_diagram}), and use the Markov property and the fact that  $\mathbb{P}(S_{3}=s_{3}\,|\,S_{1}=s_{1},S_{2}=s_{2})= \mathbb{P}(S_{3}=s_{3}\,|\,S_{2}=s_{2})= p_{s_{2}s_{3}}$ (Equation~\eqref{Eq:p_su}), $\mathbb{P}(S_{2}=s_{2}\,|\,S_{1}=s_{1})=p_{s_{1}s_{2}}$ (Equation~\eqref{Eq:p_su}), and $\mathbb{P}(S_{1}=s_{1})=p_{s_{1}}$ (Equation~\eqref{Eq:P(S(t)=s)}), to obtain
$$\begin{aligned}
\mathbb{P}(\Delta y \,|\, W=2)\mathbb{P}(W\ge2)
=\sum_{\substack{s_{1}\\ s_{2}\ne s_{1}\\ s_{3}\ne s_{2}}}& \mathbb{P}(\Delta y\,|\,W=2, S_{1}=s_{1}, S_{2}=s_{2}, S_{3}=s_{3})
\\
&\times \mathbb{P}(W\ge2\,|\, S_{1}=s_{1}, S_{2}=s_{2}, S_{3}=s_{3})
\\
&
\times \mathbb{P}(S_{3}=s_{3}\,|\,S_{1}=s_{1},S_{2}=s_{2})
\\
&
\times\mathbb{P}(S_{2}=s_{2}\,|\,S_{1}=s_{1})\mathbb{P}(S_{1}=s_{1}).
\end{aligned}$$
Hence, we only need to compute the probability of having two switches $\mathbb{P}(W\ge2\,|\, S_{1}=s_{1}, S_{2}=s_{2}, S_{3}=s_{3})$ (see Supplementary Information Section~\ref{SI:Sec:P_2}).

As for $W=1$, the PDF for $\Delta y$ if two switches occur and given the three states visited, denoted as
$$\Tilde f_{s_{1},s_{2},s_{3}}(\Delta y) := \mathbb{P}(\Delta y\,|\,W=2, S_{1}=s_{1}, S_{2}=s_{2}, S_{3}=s_{3}),$$
is obtained by first considering the PDF for an exact increment
$$\Tilde{g}_{s_{1},s_{2},s_{3}}(\Delta x):=\mathbb{P}(\Delta x\,|\, W=2, S_{1}=s_{1}, S_{2}=s_{2}, S_{3}=s_{3}),$$
and then convoluting with the distribution for $\Delta \epsilon$. We proceed as before, although note that we must deal with the case $s_{3}= s_{1}$ separately, since it leads to different time and increment distributions (see Supplementary Information Section~\ref{SI:Sec:P_2}). \textcolor{black}{In this section, we only focus on the case $s_{3}\ne s_{1}$, while the case $s_{3}= s_{1}$ is investigated in Supplementary Information Section~\ref{SI:Sec:P_2}.}

We first determine the joint distribution of the time $\tau_i$ spent in each state $S_{i}$, $i=1,2,3$, within the interval $[0,\Delta t]$ for the case of two switches ($W=2$). We note that the number of switches is two if and only if $0<\tau_1\le\Delta t$, $0<\tau_2\le\Delta t-\tau_1$ and $\tau_3> \Delta t-\tau_1-\tau_2$. Following a similar approach to before, we condition on the three subsequent states attained. For $t_1,t_2\in(0, \Delta t]$ with $t_2\le \Delta t - t_1$, we compute
$$\begin{aligned}
    F_{s_{1},s_{2},s_{3}}(t_1,t_2):=
    & \ \mathbb{P}(\tau_1\le t_1, \tau_2 \le t_2, W=2 \,|\, S_{1}=s_{1}, S_{2}=s_{2}, S_{3}=s_{3})\\
    =& \ \mathbb{P}\left(\ \begin{aligned}
        &0<\tau_1 \le t_1, 0<\tau_2\le t_2, 
        \\
        &\tau_3>\Delta t-\tau_1-\tau_2
    \end{aligned} \;\middle|\; 
    \begin{aligned}
        S_{1}=s_{1}, S_{2}=s_{2}, S_{3}=s_{3}
    \end{aligned}\right)
    \\
    =&\ \int_0^{t_1}\left(\int_0^{t_2}\left(\int_{\Delta t-\tau_1-\tau_2}^{\infty}\prod_{i=1}^3 f_{\text{Exp}(\lambda_{s_i})}(\tau_i)
    \dd \tau_3\right)\dd \tau_2\right)\dd \tau_1,
\end{aligned}$$
by integrating the joint distribution of the time spent in the first, second and third states that are again independent. Hence, we can compute the joint CDF of the times when the two switches occur given the three states attained as follows
$$\begin{aligned}
G_{s_{1},s_{2},s_{3}}(t_1, t_2):=
    & \ \mathbb{P}(\tau_1\le t_1, \tau_2 \le t_2\,|\, W=2, S_{1}=s_{1}, S_{2}=s_{2}, S_{3}=s_{3})\\
    =&\ \mathbb{P}\left(\ \begin{aligned}
    &\tau_1 \le t_1,\\& \tau_2\le t_2
\end{aligned}  \;\middle|\; \begin{aligned}
    &0<\tau_1\le\Delta t, 0<\tau_2\le\Delta t-\tau_1, \tau_3>\Delta t - \tau_1-\tau_2,\\
    & S_{1}=s_{1}, S_{2}=s_{2}, S_{3}=s_{3}
\end{aligned}\right)
\\
=&\ \dfrac{F_{s_{1},s_{2},s_{3}}(t_1,t_2)}{\mathbb{P}\left(
\begin{aligned}
    &0<\tau_1\le\Delta t, 0<\tau_2\le\Delta t-\tau_1,
    \\ & \tau_3>\Delta t - \tau_1-\tau_2
\end{aligned}  \;\middle|\; S_{1}=s_{1}, S_{2}=s_{2}, S_{3}=s_{3} \right)}
\\
=&\ \dfrac{F_{s_{1},s_{2},s_{3}}(t_1,t_2)}{\int_0^{\Delta t}\left(\int_0^{\Delta t - \tau_1}\left(\int_{\Delta t-\tau_1-\tau_2}^{\infty}f_{s_{1},s_{2},s_{3}}(\tau_1, \tau_2,\tau_3)\dd \tau_3\right)\dd \tau_2\right)\dd \tau_1}.
\end{aligned}$$

Now, we compute the PDF for an exact increment $\Delta x$, integrating over all possible switching times 
$$
\begin{aligned}
&\Tilde{g}_{s_{1},s_{2}, s_{3}}(\Delta x)=
\\
&=\int_0^{\Delta t}\Bigg( \int_0^{\Delta t - t_1} \mathbb{P}\left(\Delta x \,|\, \tau_1 = t_1, \tau_2=t_2, W=2, S_{1}=s_{1}, S_{2}=s_{2}, S_{3}=s_{3} \right)
\\
&\qquad \qquad \qquad \qquad \times \mathbb{P}\left(t_1, t_2
\,|\, W=2, S_{1}=s_{1}, S_{2}=s_{2}, S_{3}=s_{3} \right)\dd t_2\Bigg)\dd t_1
\\
&=\int_0^{\Delta t} \left(\int_0^{\Delta t - t_1} \delta\Big(v_{s_{1}}t_1 +v_{s_{2}}t_2 +v_{s_{3}}(\Delta t- t_1 -t_2) -\Delta x\Big)g_{s_{1}, s_{2}, s_{3}}(t_1, t_2) \dd t_2 \right)\dd t_1,
\end{aligned}$$
where $\delta(\cdot)$ denotes the Dirac delta function and $g_{s_{1},s_{2},s_{3}}(t_1, t_2)$ is the PDF obtained differentiating the CDF $G_{s_{1},s_{2},s_{3}}(t_1, t_2)$ in $t_1$ and $t_2$. Using the sifting property of the Dirac delta function we obtain
$$\begin{aligned}
\Tilde{g}_{s_{1},s_{2}, s_{3}}(\Delta x)=\int_{0}^{\Delta t}\frac{1}{|v_{s_{3}}-v_{s_{2}}|} \ g_{s_{1}, s_{2}, s_{3}}\left(t_1, \frac{(v_{s_{1}}-v_{s_{3}})t_1 +v_{s_{3}}\Delta t -\Delta x}{v_{s_{3}}-v_{s_{2}}}\right)\dd t_1,
\end{aligned}$$
where $g_{s_{1}, s_{2}, s_{3}}$ is defined to be zero for $(t_1, t_2)$ outside of $[0, \Delta t]\times [0, \Delta t-t_1]$.

In order to obtain $\Tilde{g}_{s_{1},s_{2}, s_{3}}(\Delta x)$ we need to compute the indefinite integral in $t_1$
$$\begin{aligned}
I(t_1):=\int_{t_1}\frac{1}{|v_{s_{3}}-v_{s_{2}}|} \ g_{s_{1}, s_{2}, s_{3}}\left(t_1, \frac{(v_{s_{1}}-v_{s_{3}})t_1 +v_{s_{3}}\Delta t -\Delta x}{v_{s_{3}}-v_{s_{2}}}\right)\dd t_1,
\end{aligned}$$
and define the domain of integration for $t_1$
$$A:=\left\{t_1 \;\middle|\; t_1\in[0,\Delta t], \frac{(v_{s_{1}}-v_{s_{3}})t_1 +v_{s_{3}}\Delta t -\Delta x}{v_{s_{3}}-v_{s_{2}}}\in[0, \Delta t-t_1]\right\}.$$

We denote $v_{\min}, v_{\text{int}}, v_{\max} \in \{v_{s_{1}}, v_{s_{2}}, v_{s_{3}}\} $ such that $v_{\min} <v_{\text{int}} <v_{\max}$. We note that for $\Delta x\in[v_{\min}\Delta t, v_{\max}\Delta t]$, $A$ is an interval, $A:=[E_0,E_1]$. In Supplementary Information Section~\ref{SI:Sec:P_2}, we obtain that $E_0$ and $E_1$ are functions of $\Delta x$, piecewise linear in the intervals $[v_{\min} \Delta t, v_{\text{int}} \Delta t]$ and $[v_{\text{int}} \Delta t, v_{\max} \Delta t]$. The PDF for an exact increment is obtained as
$$\Tilde{g}_{s_{1}, s_{2}, s_{3}}(\Delta x)=I(E_1(\Delta x))-I(E_0(\Delta x)).$$

Finally, we compute the PDF of a noisy increment $\Delta y$, again by convoluting the PDFs for $\Delta x$ and $\Delta \epsilon$, to obtain
$$\Tilde{f}_{s_{1},s_{2},s_{3}}(\Delta y) =\ \int_{v_{\min}\Delta t}^{v_{\max}\Delta t} \Tilde{g}_{s_{1},s_{2}, s_{3}}(\Delta x)f_{\mathcal{N}(0,2\sigma^2)}(\Delta y-\Delta x) \dd(\Delta x),$$ 
whose analytical expression is presented in Supplementary Information Section~\ref{SI:Sec:P_2}.

\begin{figure*}[!ht]
    \includegraphics[width=1\textwidth]{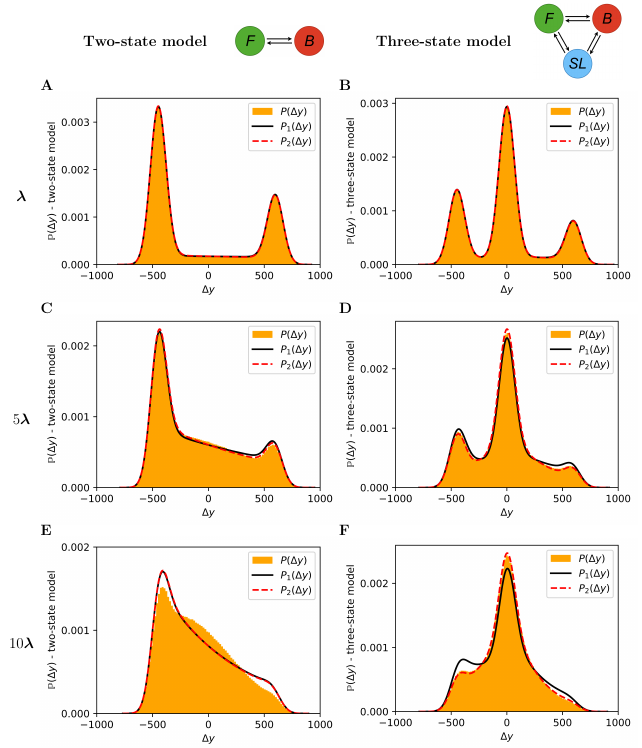}
    \caption{The empirical PDF for $\Delta y$, denoted as $P(\Delta y)$, (orange histogram) is compared with its up-to-one-switch approximation $P_1(\Delta y)$ (black line) and its up-to-two-switch approximation $P_2(\Delta y)$ (red dashed line). The left plots are obtained from a two-state model while the right ones from a three-state model. The panels \textbf{A}-\textbf{B} are obtained using the parameters as specified in Supplementary Information Figure~\ref{SI:Fig:networks}, while the panels \textbf{C}-\textbf{D} and \textbf{E}-\textbf{F} are obtained with the same parameters except the rates which are multiplied by 5 and 10, respectively.}
    \label{Fig:PDFs_2_3}
\end{figure*}

In Figure~\ref{Fig:PDFs_2_3} we present a comparison between the empirical PDF for $\mathbb{P}(\Delta y)$, the up-to-one-switch approximation $P_1(\Delta y)$, and the up-to-two-switch approximation $P_2(\Delta y)$, for the two-state model and three-state model. We do this in order to assess the improvement made by incorporating an extra switch in the approximation. From the plots in panels \textbf{A} and \textbf{B} we can see that the up-to-one-switch and up-to-two-switch approximations are comparable for infrequent switching (low rates $\boldsymbol{\lambda}$), while looking at panels \textbf{C}, \textbf{D} and \textbf{F} we notice a higher accuracy given by the two-switch approximation as the rates increase. This is expected since the probability of having more than two switches $\mathbb{P}(W>2)$ is higher for $\boldsymbol{\lambda}\times 10$ than for $\boldsymbol{\lambda}\times 1$ and $\boldsymbol{\lambda}\times 5$ (see $\mathbb{P}(W=w)$ in Figure~\ref{Fig:P(K_w)}). 

\textcolor{black}{Panel \textbf{E} in Figure~\ref{Fig:PDFs_2_3} shows that going from an up-to-one-switch to an up-to-two-switch approximation does not always lead to an appreciably improved approximation. In particular, comparing panel \textbf{E} with panel \textbf{F} in Figure~\ref{Fig:PDFs_2_3}, we note that the accuracy for the two-state model for rates $10\boldsymbol{\lambda}$ is not significantly improved since the probability of having more than two switches is still quite high ($\mathbb{P}(W>2)\approx 0.32$ for the two-state model, while $\mathbb{P}(W>2)\approx 0.19$ for the three-state model). Moreover, another reason for the difference between the accuracy of the up-to-two-switch approximation in panels \textbf{E} and \textbf{F} is that the approximation used ($\mathbb{P}(\Delta y\,|\,W>2)\approx \mathbb{P}(\Delta y\,|\,W=2)$) is less accurate for the two-state model in this parameter regime, compared to the three-state model (see Supplementary Information Figure~\ref{SI:Fig:newfig}). Finally, Supplementary Information Figure~\ref{SI:Fig:Comparison_error_PDFs_2_3} shows a comparison between the error of the up-to-one-switch PDF approximation, defined as $P_1(\Delta y)-P(\Delta y)$, and the error of the up-to-two-switch approximation, defined as $P_2(\Delta y)-P(\Delta y)$, in the panels shown in Figure~\ref{Fig:PDFs_2_3}.}

The approximations for $\mathbb{P}(\Delta y)$ give the distribution of single location increments. In the next section, we use the results obtained to approximate the probability distribution of a track by considering a set of subsequent location increments.

\section{Approximation for the probability distribution function of a set of subsequent location increments}\label{Sec:4}

Now, we aim to compute an approximation for the joint PDF of a set of $N$ subsequent location increments. We compute an up-to-one-switch approximation using the results obtained in Section~\ref{Subsec:up-to-one-switch}, as it is simpler and quicker compared to the up-to-two-switch approximation. We start by computing the PDF of $N$ exact subsequent location increments
$$\boldsymbol{\Delta x}_N = [\Delta x_1, \Delta x_2, \ldots, \Delta x_{N-1}, \Delta x_N],$$
for any $N\in\{1,2,\ldots\}$. Moreover, $\boldsymbol{\Delta x}_i$ denotes the first $i$ elements of $\boldsymbol{\Delta x}_N$, while $\Delta x_i$ its element in position $i$. We denote the PDF of $N$ exact subsequent location increments with $\mathbb{P}(\boldsymbol{\Delta x}_N)$, and a similar notation is used for $\boldsymbol{\Delta y}$.

\begin{figure*}[!ht]
    \centering
    \begin{minipage}{1\textwidth}
    \includegraphics[width=1\textwidth]{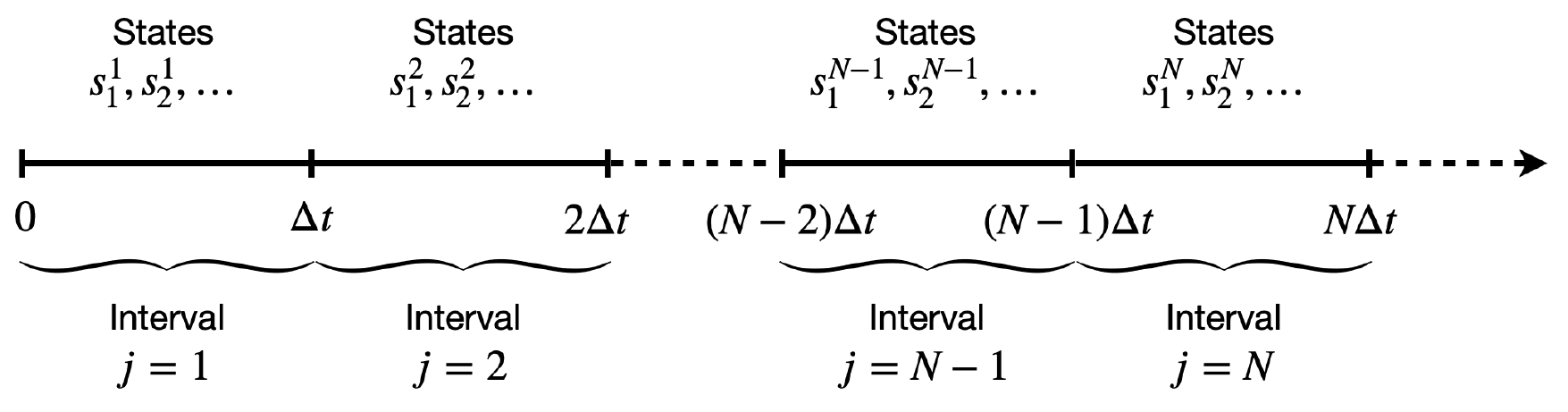}
    \end{minipage}
    \caption{Diagram representing the notation used for measured intervals of time length $\Delta t$ such that the agent location is measured at the beginning and end of such interval. We note that the state at the end of a measured interval coincides with the first state attained in the following measured interval.}
    \label{Fig:states_diagram_track}
\end{figure*}

We start by rewriting the PDF in terms of conditional distributions taking into account that the state at the end of a measured interval corresponds to the state at the beginning of the following measured interval (see Figure~\ref{Fig:displacement_increments}). We obtain
\begin{equation}\label{Eq:P(exact_track)}
    \mathbb{P}(\boldsymbol{\Delta x}_N) =\mathbb{P}(\Delta x_N \,|\, \boldsymbol{\Delta x}_{N-1})\ldots \mathbb{P}(\Delta x_3\,|\,\boldsymbol{\Delta x}_2)\mathbb{P}(\Delta x_2\,|\,\Delta x_1) \mathbb{P}(\Delta x_1),
\end{equation}
where $\mathbb{P}(\Delta x_1)$ corresponds to the marginal distribution for a single increment $\Delta x_1$, approximated in Section~\ref{Subsec:up-to-one-switch} and Section~\ref{Subsec:up-to-two-switch}. In this section, we first compute an approximation for the PDF of a $N$ exact location increments $\mathbb{P}(\boldsymbol{\Delta x}_N)$. We show that for any $N\ge 2$ we can construct the joint PDF of a set of exact subsequent location increments $\boldsymbol{\Delta x}_N$, $\mathbb{P}(\boldsymbol{\Delta x}_N)$, using a recursive method. From now on, we denote by $S^{j}_{i}$ the $i$-th state attained and $W^j$ the number of switches that occur during the $j$-th interval (as in Figure~\ref{Fig:states_diagram_track}).

We compute an iterative formula for
$\mathbb{P}(\Delta x_N \,|\, \boldsymbol{\Delta x}_{N-1})$, which does not only depend on $\mathbb{P}(\Delta x_{N-1} \,|\, \boldsymbol{\Delta x}_{N-2})$, but also on the probabilities of attaining a state or having a number of switches at the previous measured intervals. In particular, we compute an iterative formula for
$\mathbb{P}(\Delta x_N \,|\, \boldsymbol{\Delta x}_{N-1})$
that involves the approximations for
$$\begin{aligned}
    \mathbb{P}(\Delta x_{N} \,|\,S^{N}_{1}=s^{N}_{1}),  &\ \mathbb{P}(W^{N}=0 \,|\,S^{N}_{1}=s^{N}_{1}),
    \\ \mathbb{P}(W^{N}\ge 1 \,|\,S^{N}_{1}=s^{N}_{1}), &\ \mathbb{P}(S^{N}_{1}=s^{N}_{1} \,|\, \boldsymbol{\Delta x}_{N-1}),
\end{aligned}$$
using $\mathbb{P}(\Delta x_{N-1} \,|\, \boldsymbol{\Delta x}_{N-2})$, and 
$$\begin{aligned}
    \mathbb{P}(\Delta x_{N-1} \,|\,S^{N-1}_{1}=s^{N-1}_{1}), &\ \mathbb{P}(W^{N-1}=0 \,|\,S^{N-1}_{1}=s^{N-1}_{1}),\\ \mathbb{P}(W^{N-1}\ge 1 \,|\,S^{N-1}_{1}=s^{N-1}_{1}), &\ \mathbb{P}(S^{N-1}_{1}=s^{N-1}_{1} \,|\, \boldsymbol{\Delta x}_{N-2}).
\end{aligned}$$
We note that, for $N=2$ these terms above are again the marginal PDFs, for which we have computed approximations in the previous sections. For simplicity, we use the up-to-one-switch approximation presented in Section~\ref{Subsec:up-to-one-switch} which we denote by $P_1$, however, the result could be extended to incorporate more switches.

We note that conditioning on the state at the beginning of the $N$-th interval, $S^N_{1}$, makes the process at the $N$-th interval independent of $\boldsymbol{\Delta x}_{N-1}$. Thus, we obtain
\begin{equation}\label{Eq:Property}
    \mathbb{P}(\Delta x_N \,|\, S^N_{1}=s^{N}_{1}, \boldsymbol{\Delta x}_{N-1})=\mathbb{P}(\Delta x_N \,|\, S^N_{1}=s^{N}_{1}),
\end{equation}
and, similarly, we obtain
\begin{equation}\label{Eq:Property_2}
\mathbb{P}(S^N_{1}=s^{N}_{1} \,|\,S^{N-1}_{1}=s^{N-1}_{1}, \boldsymbol{\Delta x}_{N-1})=\mathbb{P}(S^N_{1}=s^{N}_{1} \,|\,S^{N-1}_{1}=s^{N-1}_{1}, \Delta x_{N-1}).
\end{equation}

Using Equation~\eqref{Eq:Property} we write 
$$\begin{aligned}
    \mathbb{P}(\Delta x_N \,|\, \boldsymbol{\Delta x}_{N-1})=&\ \sum_{s^{N}_{1}}\mathbb{P}(\Delta x_N \,|\, S^N_{1}=s^{N}_{1}, \boldsymbol{\Delta x}_{N-1})\mathbb{P}(S^N_{1}=s^{N}_{1} \,|\, \boldsymbol{\Delta x}_{N-1})
    \\
    =&\ \sum_{s^{N}_{1}}\mathbb{P}(\Delta x_N \,|\, S^N_{1}=s^{N}_{1})\mathbb{P}(S^N_{1}=s^{N}_{1} \,|\, \boldsymbol{\Delta x}_{N-1}).
\end{aligned}$$
Moreover, we can compute $\mathbb{P}(S^N_{1}=s^{N}_{1} \,|\, \boldsymbol{\Delta x}_{N-1})$ by considering the state at the beginning of the previous interval $S^{N-1}_{1}$, 
\begin{align}
    &\mathbb{P}(S^N_{1}=s^{N}_{1} \,|\, \boldsymbol{\Delta x}_{N-1})\nonumber
    \\ &= \sum_{s^{N-1}_{1}}\mathbb{P}(S^N_{1}=s^{N}_{1} \,|\,S^{N-1}_{1}=s^{N-1}_{1}, \boldsymbol{\Delta x}_{N-1})\mathbb{P}(S^{N-1}_{1}=s^{N-1}_{1} \,|\, \boldsymbol{\Delta x}_{N-1}) \nonumber
    \\
    \label{Eq:P(track)_terms}
    &=\sum_{s^{N-1}_{1}}\frac{\mathbb{P}(\Delta x_{N-1}, S^N_{1}=s^{N}_{1} \,|\,S^{N-1}_{1}=s^{N-1}_{1})}{\mathbb{P}(\Delta x_{N-1} \,|\,S^{N-1}_{1}=s^{N-1}_{1})}\mathbb{P}(S^{N-1}_{1}=s^{N-1}_{1} \,|\, \boldsymbol{\Delta x}_{N-1}),
\end{align}
where we use Equation~\eqref{Eq:Property_2}.

The final term in Equation~\eqref{Eq:P(track)_terms} can be obtained using Bayes' theorem, such that
$$
\begin{aligned}
    \mathbb{P}(S^{N-1}_{1}=s^{N-1}_{1} \,|\, \boldsymbol{\Delta x}_{N-1}) =& \ \mathbb{P}( \Delta x_{N-1}\,|\, S^{N-1}_{1}=s^{N-1}_{1}, \boldsymbol{\Delta x}_{N-2})
    \\
    &\times \frac{\mathbb{P}(S^{N-1}_{1}=s^{N-1}_{1} \,|\, \boldsymbol{\Delta x}_{N-2})}{\mathbb{P}( \Delta x_{N-1}\,|\, \boldsymbol{\Delta x}_{N-2})}
    \\
    =&\ \mathbb{P}( \Delta x_{N-1}\,|\, S^{N-1}_{1}=s^{N-1}_{1})\frac{\mathbb{P}(S^{N-1}_{1}=s^{N-1}_{1} \,|\, \boldsymbol{\Delta x}_{N-2})}{\mathbb{P}( \Delta x_{N-1}\,|\, \boldsymbol{\Delta x}_{N-2})},
\end{aligned}$$
where the second equality is obtained using the property in Equation~\eqref{Eq:Property} at the $(N-1)$-th step.
We substitute this result into Equation~\eqref{Eq:P(track)_terms}, and, by simplifying $\mathbb{P}(\Delta x_{N-1} \,|\,S^{N-1}_{1}=s^{N-1}_{1})$, we obtain
$$\begin{aligned}
    &\mathbb{P}(S^N_{1}=s^{N}_{1} \,|\, \boldsymbol{\Delta x}_{N-1}) 
    \\
    &= \sum_{s^{N-1}_{1}}\mathbb{P}(\Delta x_{N-1}, S^N_{1}=s^{N}_{1} \,|\,S^{N-1}_{1}=s^{N-1}_{1})\frac{\mathbb{P}(S^{N-1}_{1}=s^{N-1}_{1} \,|\, \boldsymbol{\Delta x}_{N-2})}{\mathbb{P}( \Delta x_{N-1}\,|\, \boldsymbol{\Delta x}_{N-2})}.
\end{aligned}$$
Here, both the numerator and denominator of the fraction are computed at the $(N-1)$-th induction step.

The final term that we must compute is $\mathbb{P}(\Delta x_{N-1}, S^N_{1}=s^{N}_{1} \,|\,S^{N-1}_{1}=s^{N-1}_{1})$. To do this, we condition on the number of switches during the $(N-1)$-th interval, $W^{N-1}$, and obtain
$$\begin{aligned}
    &\mathbb{P}(\Delta x_{N-1}, S^N_{1}=s^{N}_{1} \,|\,S^{N-1}_{1}=s^{N-1}_{1})
    \\
    &= \sum_{w=0}^{\infty}
    \mathbb{P}(\Delta x_{N-1}\,|\, S^N_{1}=s^{N}_{1}, S^{N-1}_{1}=s^{N-1}_{1}, W^{N-1}=w)
    \\& \qquad \quad \times \mathbb{P}(S^N_{1}=s^{N}_{1}\,|\,S^{N-1}_{1}=s^{N-1}_{1}, W^{N-1}=w)\mathbb{P}(W^{N-1}=w \,|\,S^{N-1}_{1}=s^{N-1}_{1}).
\end{aligned}$$
For simplicity, we apply the up-to-one-switch approximation in Section~\ref{Sec:3} which yields
$$\begin{aligned}
    &\mathbb{P}(\Delta x_{N-1}, S^N_{1}=s^{N}_{1} \,|\,S^{N-1}_{1}=s^{N-1}_{1})\\
    &\approx P_1(\Delta x_{N-1}, S^N_{1}=s^{N}_{1} \,|\,S^{N-1}_{1}=s^{N-1}_{1})
    \\
    &
    =\begin{dcases}
    \begin{aligned}
        &\mathbb{P}(\Delta x_{N-1}\,|\, S^N_{1}=s^{N}_{1}, S^{N-1}_{1}=s^{N-1}_{1}, W^{N-1}=0)
        \\
        &\ \times\mathbb{P}(W^{N-1}=0 \,|\,S^{N-1}_{1}=s^{N-1}_{1}), 
    \end{aligned}
    & \text{if } s^{N}_{1}=s^{N-1}_{1},
    \\
    \begin{aligned}
        &\mathbb{P}(\Delta x_{N-1}\,|\, S^N_{1}=s^{N}_{1}, S^{N-1}_{1}=s^{N-1}_{1}, W^{N-1}=1)
        \\
        &\ \times \mathbb{P}(S^{N-1}_{2}=s^{N}_{1}\,|\,S^{N-1}_{1}=s^{N-1}_{1})\mathbb{P}(W^{N-1}\ge 1 \,|\,S^{N-1}_{1}=s^{N-1}_{1}),
    \end{aligned}
    & \text{if } s^{N}_{1}\ne s^{N-1}_{1},
\end{dcases}
\end{aligned}$$
since if $W^{N-1}=0$, then $S^{N}_{1}=S^{N-1}_{1}$, and this gives a unitary probability of keeping the same state
$$\mathbb{P}(S^N_{1}=s^{N}_{1}\,|\,S^{N-1}_{1}=s^{N-1}_{1}, W^{N-1}=0)=1.$$
Otherwise, if $W^{N-1}=1$, then we have $S^{N}_{1}=S^{N-1}_{2}\ne S^{N-1}_{1}$ and therefore
$$\mathbb{P}(S^N_{1}=s^{N}_{1}\,|\,S^{N-1}_{1}=s^{N-1}_{1}, W^{N-1}=1)=\mathbb{P}(S^{N-1}_{2}=s^{N}_{1}\,|\,S^{N-1}_{1}=s^{N-1}_{1}).$$

This result can be extended to compute an approximation for the PDF of $N$ noisy subsequent location increments $\boldsymbol{\Delta y}_N = [\Delta y_1, \Delta y_2, \ldots, \Delta y_{N-1}, \Delta y_N].$ The PDF of $N$ noisy subsequent location increments can be written in terms of conditional distributions as
\begin{equation}\label{Eq:P(track)}
    \mathbb{P}(\boldsymbol{\Delta y}_N) =\mathbb{P}(\Delta y_N \,|\, \boldsymbol{\Delta y}_{N-1}) \ldots \mathbb{P}(\Delta y_3\,|\,\boldsymbol{\Delta y}_2)\mathbb{P}(\Delta y_2\,|\,\Delta y_1) \mathbb{P}(\Delta y_1),
\end{equation}
where we note that $\mathbb{P}(\Delta y_1)$ is the marginal distribution for a single increment $\Delta y_1$.

\textcolor{black}{The distributions of the $\Delta y_j$ are obtained by convoluting the distributions for the $\Delta x_j$ with the appropriate Gaussian distribution. The steps proposed in this section thus far are also valid for the $\Delta y_j$, with the exception being application of the Markov property in Equations~\eqref{Eq:Property} and~\eqref{Eq:Property_2}, as successive noisy increments, $\Delta y_{j-1}$ and $\Delta y_{j}$, are correlated. In particular, $\Delta y_j|\Delta x_j$ would be obtained by conditioning on $\Delta \epsilon_{j}$, which depends on $\Delta y_{j-1}$ and subsequently on $\Delta \epsilon_{j-1}$, which similarly depends on all the previous $\Delta y_{i}$ and $\Delta \epsilon_i$. Indeed, it is non-trivial to obtain an analytical expression that incorporates a convolution of the measurement noise correlated with all the previous location and noise increments. To make progress, we assume the correlation between the noise increments to be negligible. In practice, this corresponds to assuming that $\Delta \epsilon_N:=\epsilon_N-\epsilon_{N-1}$ and $\Delta \epsilon_{N-1}:=\epsilon_{N-1}-\epsilon_{N-2}$ are independent. In Supplementary Information Section~\ref{SI:Sec:P(track)} we present the full approximate PDF derivation for noisy subsequent location increments.}

\begin{figure*}
    \centering\includegraphics[width=1\textwidth]{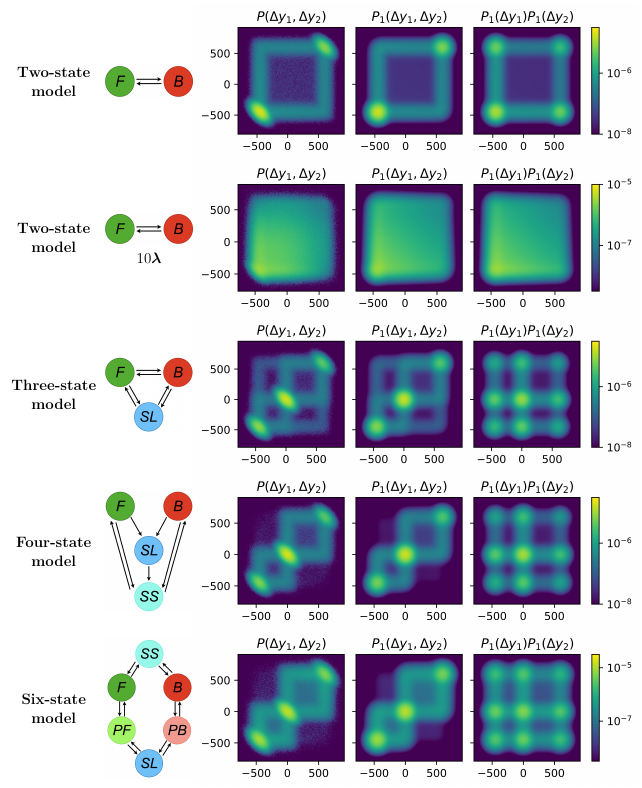}
    \caption{\textbf{Comparison of the empirical PDF for two subsequent measured increments $\Delta y_1$ and $\Delta y_2$, denoted as $P(\Delta y_1, \Delta y_2)$ (left column) with its up-to-one-switch approximation of the probability of two noisy subsequent location increments ($P_{1}(\Delta y_1, \Delta y_2)$) (middle column) and with the independent up-to-one-switch approximation for each increment $P_1(\Delta y_1) P_1(\Delta y_2)$ (right column)}. The plots are obtained in a logarithmic color scale, using the parameters specified in Supplementary Information Figure~\ref{SI:Fig:networks}. $\Delta y_1$ is represented on the vertical axis, while $\Delta y_2$ is on the horizontal axis.}
    \label{Fig:P(Delta y_1, Delta y_2)}
\end{figure*}

We visualise and demonstrate the up-to-one-switch PDF approximation for a two noisy subsequent location increments, $P_1(\Delta y_1, \Delta y_2)$, with the empirical PDF for $\mathbb{P}(\Delta y_1, \Delta y_2)$, denoted as $P(\Delta y_1, \Delta y_2)$, however, the result can be applied to any number of subsequent increments. We also compare $P_1(\Delta y_1, \Delta y_2)$ with the naive approximation $P_1(\Delta y_1) P_1(\Delta y_2)$ that assumes that successive increments are approximately independent, to test the impact of neglecting these correlations. Figure~\ref{Fig:P(Delta y_1, Delta y_2)} shows that, as expected, $P_{1}(\Delta y_1,\Delta y_2)$ is a better approximation for $\mathbb{P}(\Delta y_1, \Delta y_2)$ than $P_{1}(\Delta y_1) P_1(\Delta y_2)$ for low switching rates (infrequent switches). In contrast, the panels representing the two-state model with $\boldsymbol{\lambda}\times 10$ show no noticeable difference between $P_{1}(\Delta y_1,\Delta y_2)$ and $P_{1}(\Delta y_1) P_1(\Delta y_2)$. This suggests that for high switching frequency the correlation between two subsequent increments is negligible.

Comparing the two-state model $P(\Delta y_1, \Delta y_2)$ (left column) with $P_1(\Delta y_1, \Delta y_2)$ (middle column) in Figure~\ref{Fig:P(Delta y_1, Delta y_2)} we notice peaks around $\Delta y_1 = \Delta y_2 = v_1\Delta t$ and $\Delta y_1 = \Delta y_2 = v_2\Delta t$. Moreover, the three, four and six-state models have a peak at zero ($0\cdot \Delta t$), since their networks include one or more stationary states. These peaks represent a measured interval with no change in velocity, corresponding to the noisy increment $\Delta y_1$, followed by another measured interval that keeps the same velocity as before, corresponding to the noisy increment $\Delta y_2$. On the other hand, the right panels representing $P_1(\Delta y_1) P_1(\Delta y_2)$ include extra peaks which correspond to the probability of having a measured interval with no change in velocity, corresponding to the noisy increment $\Delta y_1$, followed by another measured interval with no change in velocity, corresponding to the noisy increment $\Delta y_2$, but each with a distinct velocity and therefore state. Having two subsequent measured intervals with two different states attained is not a likely scenario, as the state switch would have to occur exactly at the time between the two measured increments ($t_1$) which is a zero-probability event. We also notice that these peaks are more oval in the left panel while round in the middle (and right) panels, which is due to the assumption used in the approximation that neglects the correlation of the noise of subsequent increments (see Equation~\eqref{Eq:Approx_no_noise_1}).

To summarise, Figure~\ref{Fig:P(Delta y_1, Delta y_2)} shows that, for low switching rates (infrequent switches), it is important to take into account the correlation of subsequent measured increments to approximate $\mathbb{P}(\Delta y_1, \Delta y_2)$. Indeed, $P_{1}(\Delta y_1,\Delta y_2)$ is a better approximation for $\mathbb{P}(\Delta y_1, \Delta y_2)$ than $P_{1}(\Delta y_1)P_1(\Delta y_2)$ for all model networks presented. Finally, Supplementary Information Figure~\ref{SI:Fig:Err_P(Delta y_1, Delta y_2)} compares the error of the joint up-to-one-switch PDF approximation for two noisy subsequent location increments, defined as $|P_1(\Delta y_1,\Delta y_2)-P(\Delta y_1,\Delta y_2)|$, with the error of the product of the marginal approximate PDFs, defined as $|P_1(\Delta y_1)P_1(\Delta y_2)-P(\Delta y_1,\Delta y_2)|$, for the panels shown in Figure~\ref{Fig:P(Delta y_1, Delta y_2)}.

\section{Discussion}\label{Sec:5}

\textcolor{black}{In this manuscript, we present a general $n$-state one-dimensional stochastic velocity-jump model and provide approximate solutions to the model subject to discrete-time noisy observations.} The state evolution is a CTMC, and each state is associated with a defined velocity, rate of switching, and probabilities of transitioning to every other state. Single-agent tracking data is intrinsically noisy and collected at discrete time points, which impedes the determination of state-switching times. \textcolor{black}{The methods we propose to approximate the solutions apply to the velocity-jump model proposed for any network of $n$ states. The approximate solutions can be used to obtain fast forwards predictions, to inform experimental design and to carry out parameter inference and model selection.}

\textcolor{black}{We compute the approximate PDFs by conditioning on the states attained between two subsequent measurements, limiting the number to at most one or two state switches, as for frequent enough image collection we expect the majority of the increments to include few state transitions.} Hence, we derive an up-to-one-switch and an up-to-two-switch approximation for the PDF of a single increment, and compare them to give insights on the parameter regimes at which they are valid. Comparing these approximations to the empirical PDFs we note that both approximations agree with the empirical distributions if the switching between states occurs at sufficiently low frequencies, and they become less precise as the switching rates increase. As expected, the up-to-two-switch approximation is more accurate than the up-to-one-switch approximation, especially as the switching rates increase. We note that the up-to-two-switch approximation was explicitly computed for states with distinct velocities and rates, however, it is possible to extend the up-to-two-switch approximation to models in which different states have the same velocities or rates by distinguishing the cases, as in the up-to-one-switch approximation. Moreover, the methods proposed are applicable for any number of switches per interval, but the possible sets of ordered visited states are combinatorially explosive, which leads to a correspondingly large number of terms to compute.

We also derive an up-to-one-switch approximation for the joint PDF of a set of subsequent location increments, to extend the results of the marginal PDF previously obtained, considering the correlations between successive location increments. Comparing the joint PDF approximation to the empirical PDF and to the product of the marginal PDFs for each increment shows that the joint PDF approximation agrees with the empirical PDF, highlighting the importance of considering the correlation between successive location increments, not captured by multiplying the marginal PDFs. Again, the methods are valid with any up-to-$m$-switch marginal PDF approximation. \textcolor{black}{In order to be able to obtain an analytical expression for the joint PDF we make the assumption that the correlation between the noise increments in a track is negligible. To improve the approximation of the joint PDF, the correlation between the noise increments could be incorporated; however, it may be necessary to use numerical approximations to compute the convolution of the exact subsequent increments with the subsequent noise increments explicitly.}

\textcolor{black}{Describing the dynamics of similar stochastic processes and providing frameworks to estimate model parameters has long been an area of mathematical interest~\citep{ liptser2013statisticsI, liptser2013statisticsII, doucet2001sequential}. Here, we extend the results for a two-state model provided in~\citep{harrison2018impact, rosser2013novel} to a more general setting. Existing frameworks for the calibration of stochastic models are often based on likelihood-free approaches, such as particle filtering pseudomarginal methods~\citep{simpson2022reliable, warne2020practical, king2016statistical, andrieu2010particle}. In contrast, our approach can provide analytically tractable approximations of the likelihood that can be used with any likelihood-based inference method.}

\textcolor{black}{We now discuss extensions of the model presented and its approximate solutions, which may come at the cost of analytical tractability of the PDF approximations. The focus of our work is 
motion in one spatial dimension (Figure~\ref{Fig:introduction}\textbf{A}). However, in different biological settings motion is often in two or three spatial dimensions (Figure~\ref{Fig:introduction}\textbf{B}-\textbf{C}), which motivates a natural extension of the work to higher dimensions. In particular, our model can be extended to higher dimensions by describing the agent velocities as vectors, and considering multivariate normal for the noise. If each states considered is characterised by a fixed velocity vector, the methods proposed to approximate the PDFs of exact increments are carried over by noting that the exact location increments become vectors, and the noise is added by convoluting a multivariate normal distribution. In contrast, having states with fixed velocity modulus but varying direction would be a more complex extension. In this case, one could include angular reorientations following a state switch, extending the work of \cite{harrison2018impact} with a set of distinct velocities, but this may lead to analytically intractable integrals for which we may need to resort to numerical methods.}

\textcolor{black}{For the data collection model we take a standard approach and use Gaussian noise with zero mean and a fixed variance, which is then incorporated into the analysis by a convolution. We expect our approach to be suitable for other noise models provided that the convolution integral can be carried out. Thus, the methods can be adapted for use with more general noise distributions. However, it may not be trivial to convolve the exact increment distributions with other noise distributions. As before, numerical methods may allow the approximation of the integrals resulting from these convolutions.}

\textcolor{black}{In some contexts, motion may be punctuated by intervals in which the individuals also undergo diffusion, which may be characterised by normally distributed space jumps. The approximations for a single increment can be modified by convolving the noise distribution with the space-jump distribution. For the track likelihood, the same modification applies, though it may be difficult to distinguish between measurement noise and diffusion in practice. Moreover, motion may be characterised by a set of velocity distributions rather than a set of constant velocities. This also motivates an extension to a model in which each state has a velocity distribution rather than a fixed velocity. In this case, while modifying the model is relatively simple, approximating the model solutions may involve convolutions which lead to integrals that are not solvable analytically. Other future challenges include extending the results obtained for non-Markovian switching time distributions, or for time-dependent parameters or self-exciting states.}

\textcolor{black}{In conclusion, in this paper we approximate the solutions to an $n$-state velocity-jump model designed to describe single-agent motion in one dimension. We characterise the PDFs of noisy location increments, overcoming the challenges posed by the experimental constraints of data collection, which is noisy and discrete. These PDF approximations may be used to obtain forwards predictions, inform experimental design and as likelihoods in any likelihood-based framework to carry out parameter inference and model selection. Moreover, this work paves the way for providing solutions and calibration frameworks for more general velocity-jump models, in potentially higher spatial dimensions.}

\backmatter


\bmhead*{Acknowledgements}
A.C. is supported by an EPSRC/UKRI Doctoral Training Award. A.P.B. thanks the Mathematical Institute, Oxford for a Hooke Research Fellowship. R.E.B. is supported by a grant from the Simons Foundation (MP-SIP-00001828).

\section*{Declarations}


\bmhead*{Conflict of interest}
The authors declare that they have no conflict of interests.


\bmhead*{Consent for publication}
All the authors approved the final version of the manuscript.

\bmhead*{Data availability}
The authors confirm that all data generated or analysed during this study can be generated through the code in the Github repository\\ \href{https://github.com/a-ceccarelli/distributions_n-state_VJ_model}{a-ceccarelli/distributions\_n-state\_VJ\_model}.


\bmhead*{Code availability}
The code is available on GitHub in the repository\\ \href{https://github.com/a-ceccarelli/distributions_n-state_VJ_model}{a-ceccarelli/distributions\_n-state\_VJ\_model}.

\bmhead*{Author contribution}
A.C., R.E.B. and A.P.B. conceived the original idea. A.C. computed the approximations, created the code, carried out the analysis, and wrote the manuscript. A.P.B. and R.E.B. gave additional inputs to the manuscript and supervised the work. All authors gave final approval for publication.


\bibliography{sn-bibliography_rev}

\end{document}


\title{\textbf{Supplementary Information} - Approximate solutions of a general stochastic velocity-jump \textcolor{black}{model} subject to discrete-time noisy observations}

\author*[1]{\fnm{Arianna} \sur{Ceccarelli}\orcidlink{0000-0002-9598-8845}}\email{ceccarelli@maths.ox.ac.uk}

\author[1]{\fnm{Alexander P.} \sur{Browning}\orcidlink{0000-0002-8753-1538}}

\author[1]{\fnm{Ruth E.} \sur{Baker}\orcidlink{0000-0002-6304-9333}}

\affil[1]{\orgdiv{Mathematical Institute}, \orgname{University of Oxford}, \orgaddress{\city{Oxford}, \postcode{OX2 6GG}, \state{Oxfordshire}, \country{UK}}}

\newgeometry{margin=2.5cm}

\maketitle

\tableofcontents

\makeatletter
\renewcommand \thesection{S\@arabic\c@section}
\renewcommand\thetable{S\@arabic\c@table}
\renewcommand \thefigure{S\@arabic\c@figure}
\renewcommand\theequation{S\@arabic\c@equation}
\makeatother

\section{Irreducibility hypothesis}\label{SI:Sec:Irreducibility}

We now present the irreducibility hypothesis for a CTMC. The Markov chain, or equivalently its matrix $\boldsymbol{Q}$, needs to be irreducible in order for $\mathbb{P}(s)$ to be constant at any time $t\ge0$. $\boldsymbol{Q}$ is irreducible if and only if it leads to a single communicating class of states \citep{norris1998markov_SI}. In other words, for any states $s$ and $u$ there exists an integer $n\ge0$ such that the probability of going from a state $s$ to a state $u$ in $n$ switches is strictly positive, or equivalently, the probability of going from a state $s$ to a state $u$ after a time $t>0$ is strictly positive \citep{norris1998markov_SI}.
Proposition~2.59 in \cite{liggett2010continuous_SI} can be used to prove the existence of a stationary probability distribution, but it requires the chain to be recurrent in addition to it being irreducible. In fact, the irreducible chain considered is also recurrent, meaning that all its states are recurrent. 

For an irreducible chain, Proposition~2.52 in \cite{liggett2010continuous_SI} states that a state is recurrent if and only if all states are recurrent. Moreover, if the chain is not recurrent it is transient, which would imply that the expected amount of time spent in $u$ of a chain starting at $s$, denoted with $G(s,u)$, is finite for all $s, u\in \{1,2,\ldots,n\}$. Hence, if by contradiction the chain was not recurrent, then, for a finite number of states, for any $s$, we would have $\sum_u G(s,u)<\infty$. But this cannot be true since $$\sum_u G(s,u)=\sum_u \lim_{t\to\infty}\int_0^t p_y(s,u)\dd t = \lim_{t\to\infty}\int_0^t \sum_u 
p_y(s,u)\dd t=\lim_{t\to\infty}t=\infty.$$
Thus, we are guaranteed existence and uniqueness of a stationary probability distribution and therefore we can compute $\boldsymbol{\pi}$. We note that having sets of classes that do not communicate would lead to a set of disconnected models rather than a single one, which could be studied separately using the methods that follow.

\section{Method to compute the initial probability distribution}\label{SI:Sec:P(s)}

Here, we propose a method to compute the initial probability distribution $\boldsymbol{\pi}$ \citep{ross2014introduction_SI}, assumed to be the stationary probability distribution of the Markov chain considered. 
The irreducibility assumption presented in Supplementary Information Section~\ref{SI:Sec:Irreducibility} guarantees the existence and uniqueness of $\boldsymbol{\pi}$.

For $\boldsymbol{Q}$ regular, $\boldsymbol{\pi}$ is a solution of the corresponding chemical master equation \citep{kuntz2021stationary}.
In other words, we aim to find $\boldsymbol{\pi}$ such that
$$\boldsymbol{\pi} \boldsymbol{Q} = 0,$$
where
$$\boldsymbol{Q} =
\begin{bmatrix}
    -\lambda_1 & \lambda_1 p_{12} & \lambda_1p_{13} & \hdots & \lambda_1p_{1n}
    \\
    \lambda_2p_{21} & -\lambda_2 & \lambda_2p_{23} & \hdots & \lambda_2p_{2n}
    \\
    \lambda_3p_{31} & \lambda_3p_{32} & -\lambda_3 & \hdots & \lambda_3p_{3n}
    \\
    \vdots & \vdots & \vdots & \ddots & \vdots
    \\
    \lambda_np_{n1} & \lambda_np_{n2} & \lambda_np_{n3} & \hdots & -\lambda_n
\end{bmatrix}.$$
We now compute an expression for a vector $\boldsymbol{\omega}=[\omega_1, \omega_2, \omega_3, \ldots, \omega_n]^T\ne 0$ such that $\boldsymbol{Q}^T\boldsymbol{\omega} = \boldsymbol{0}$. The kernel of $\boldsymbol{Q}^T$ is not $\{\boldsymbol{0}\}$ since, by definition, the rows of $\boldsymbol{Q}$ are linearly dependent. Moreover, the irreducibility hypothesis guarantees that the dimension of the kernel is exactly one.

We can use inverse power method to find such eigenvector with zero eigenvalue \citep{ford2014numerical}. 
In particular, we obtain that for all states $s=1,2,\ldots,n$,
$$\lambda_s \omega_s = \sum_{u\ne s}\lambda_u p_{us}\omega_u.$$
For all $s,u=1,2,\ldots,n$, $u\ne s$, $\lambda_s>0$ and $p_{su}\ge0$ are non-negative coefficients. Hence, if there exists $\omega_s>0$ then all the entries of $\boldsymbol{\omega}$ must be non-negative. Therefore, by defining $\Omega:=\sum_{s=1,\ldots,n}\omega_s$ and $p_s:=\omega_i/\Omega$, we must have that $p_s\in[0,1]$, for all states $s=1,2,\ldots,n$.

\section{Networks with parameters}

\begin{figure*}[!ht]
    \centering
    \begin{minipage}{\textwidth}
    \begin{minipage}{0.5\textwidth}
    \small \centering
    \textbf{Two-state model}
    \end{minipage}%
    \begin{minipage}{0.5\textwidth}
    \small \centering 
    \textbf{Three-state model}
    \end{minipage}
    \begin{minipage}{0.5\textwidth}
    \includegraphics[width=\textwidth]{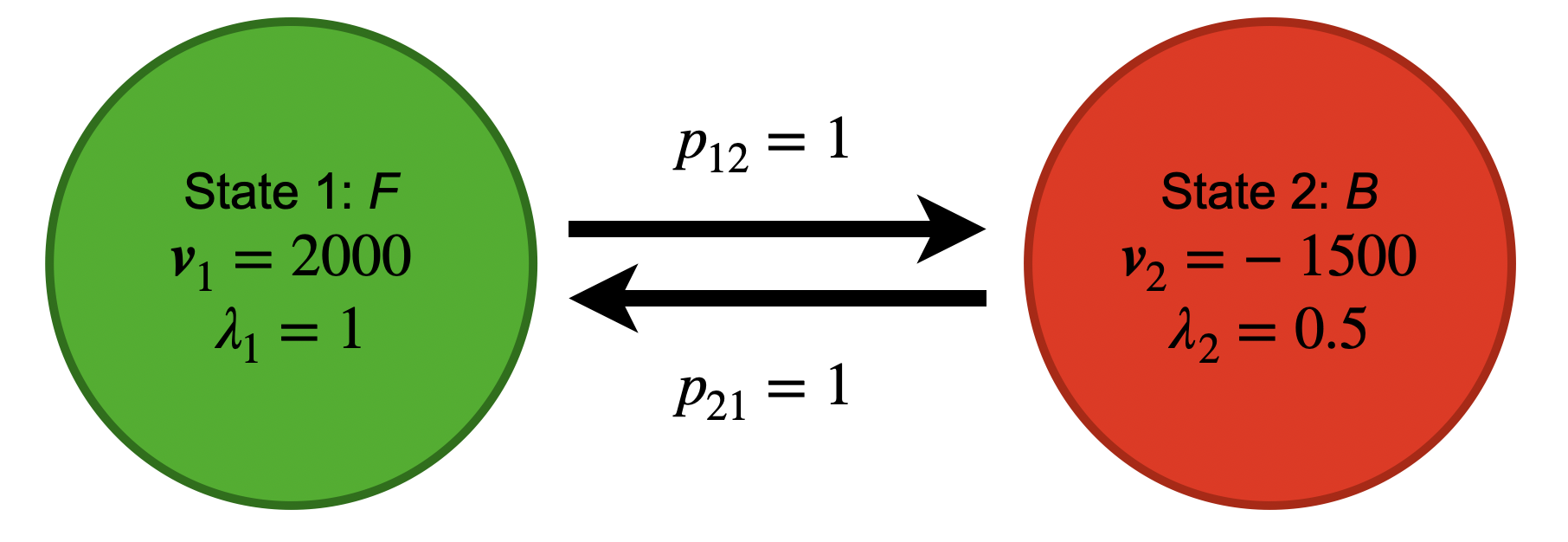}
    \end{minipage}%
    \begin{minipage}{0.5\textwidth}
    \includegraphics[width=\textwidth]{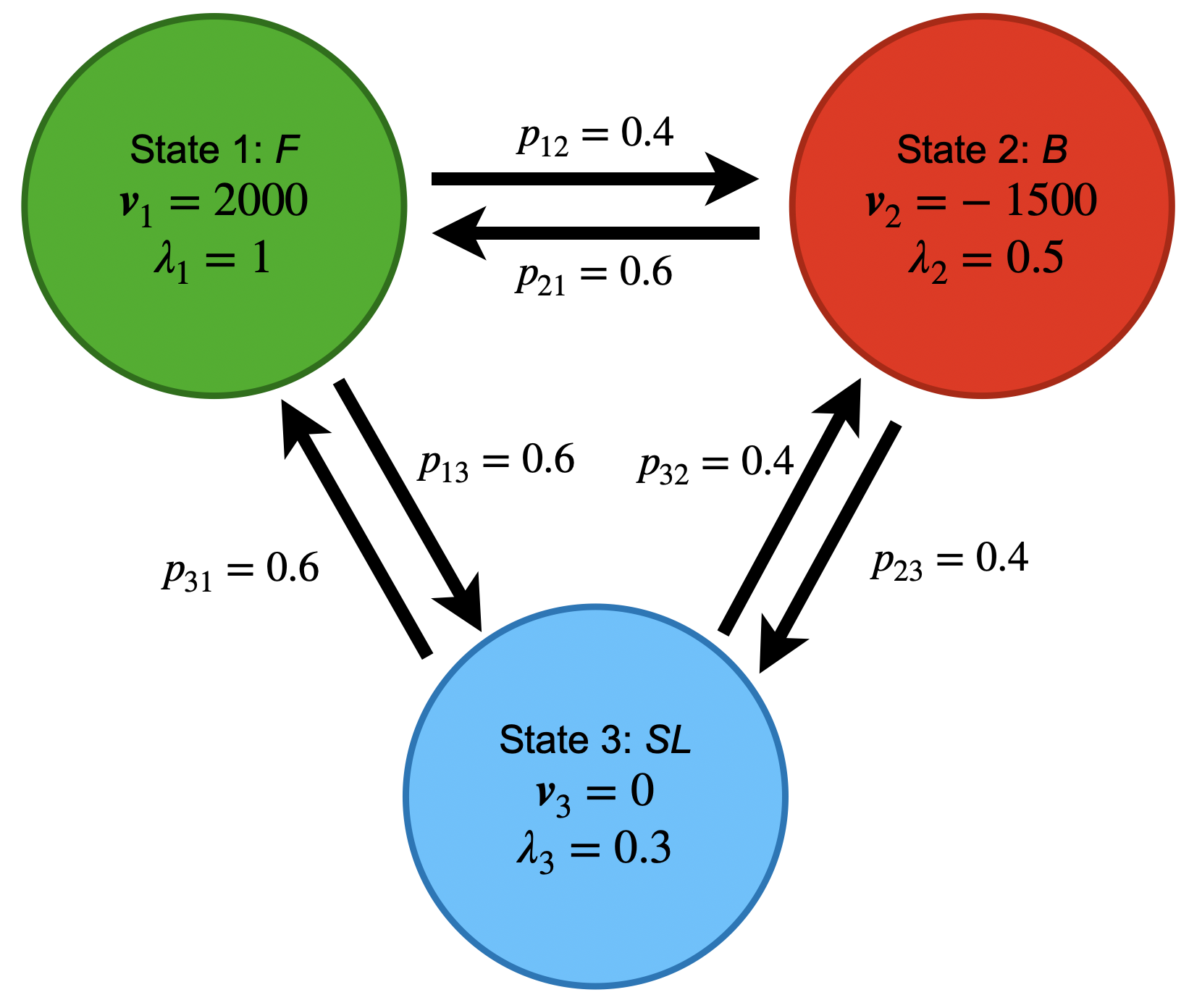}
    \end{minipage}
    \begin{minipage}{0.5\textwidth}
    \small \centering
    \textbf{Four-state model}
    \end{minipage}%
    \begin{minipage}{0.5\textwidth}
    \small \centering 
    \textbf{Six-state model}
    \end{minipage}
    \begin{minipage}{0.5\textwidth}
    \includegraphics[width=\textwidth]{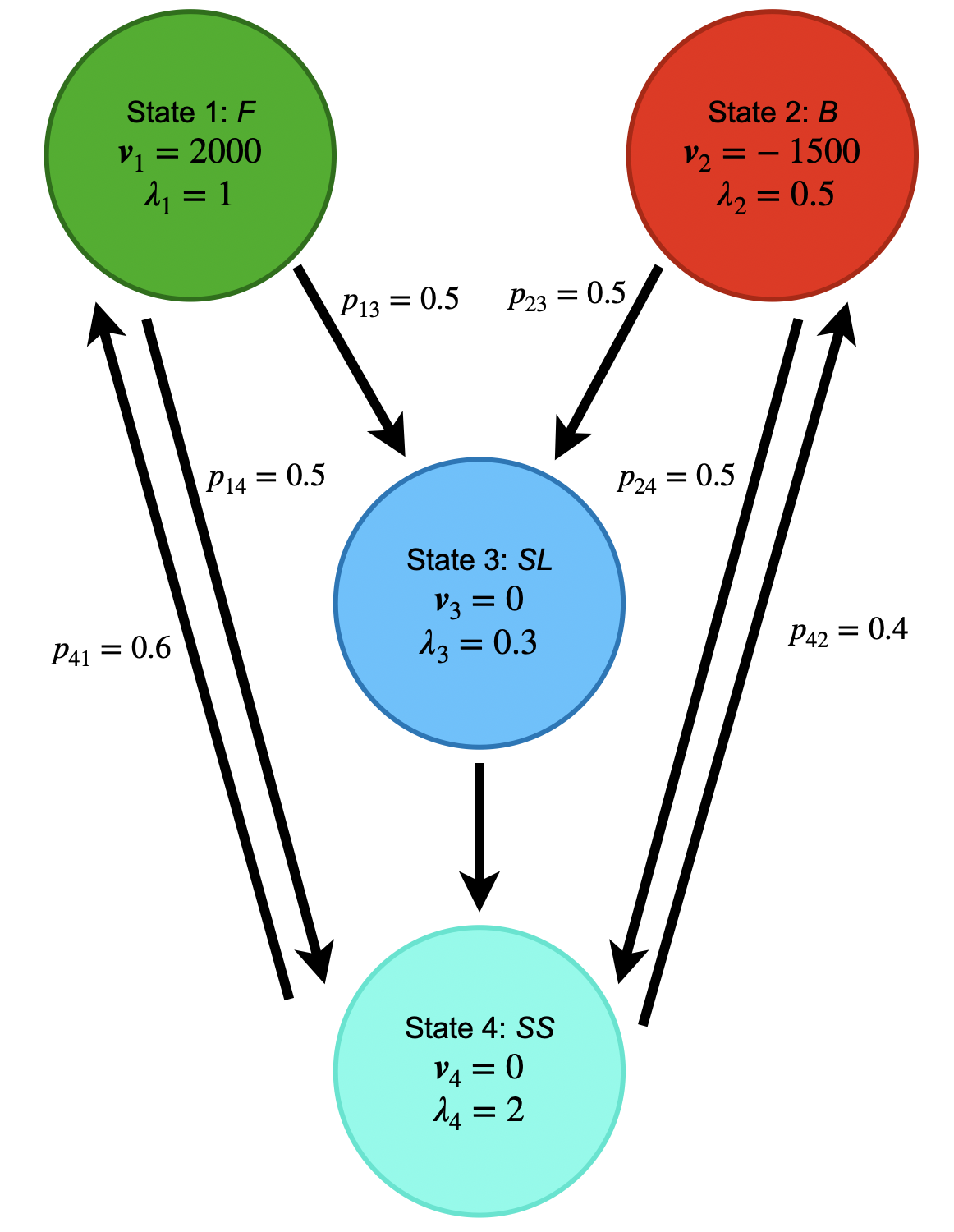}
    \end{minipage}%
    \begin{minipage}{0.5\textwidth}
    \includegraphics[width=\textwidth]{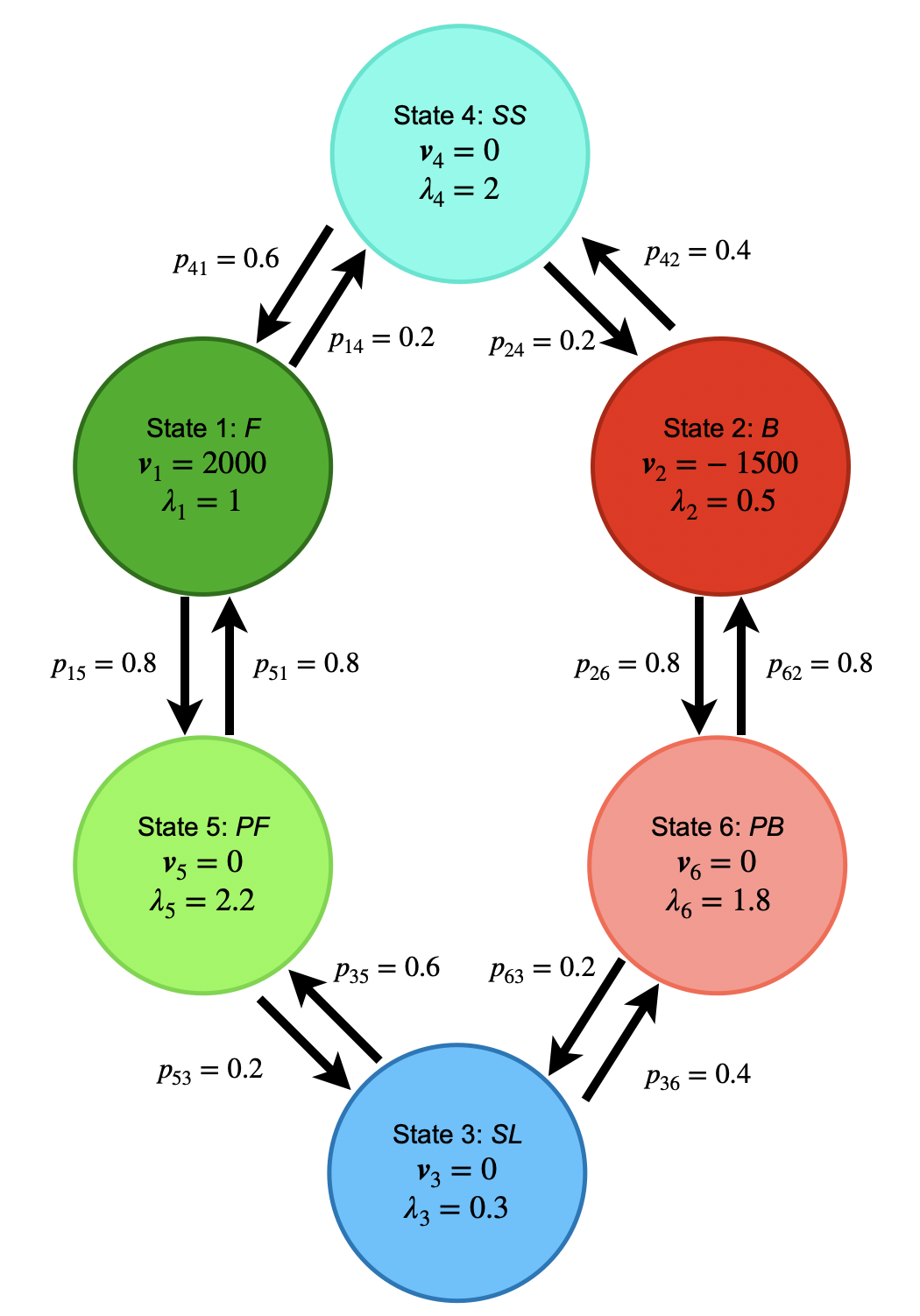}
    \end{minipage}
    \end{minipage}\caption{\textbf{Examples of $n$-state models with specified networks and parameters.} A two-state, three-state, four-state and six-state model are presented. \textit{F} stands for forward state, \textit{B} for backward state,  \textit{SL} for stationary state with long average permanence, and \textit{SS} for stationary state with short average permanence, \textit{PF} for pause in forward movement and \textit{PB} for pause in backward movement. Velocities and values of the transition matrix $P=(p_{ij})$ are fixed, while the switching rates $\lambda$ are varied for testing purposes, but their ratios are fixed. Moreover, we fix $\Delta t = 0.3$ and $\sigma = 50$.}
    \label{SI:Fig:networks}
\end{figure*}

Figure~\ref{SI:Fig:networks} shows the full networks with parameters presented as examples. All the parameters are fixed, except the set of rates $\boldsymbol{\lambda} = [\lambda_1, \lambda_2, \ldots, \lambda_n]$ which are often multiplied by a factor of 10 (or integers from 1 to 10) to obtain comparisons between data collection scenarios. The parameters are chosen to reflect the motion of molecular motors along microtubules since the movement is often captured in kymographs, which are intrinsically one-dimensional (see Figure~\ref{Fig:introduction}\textbf{A}). The time step is fixed to be $\Delta t = 0.3$. Moreover,
\cite{maday2014axonal} estimate the velocities of molecular motors to reach the order of $1000$, thus we define $v_F = 2000$ and $v_B = -1500$, and we set all stationary and pausing state velocities to zero. Finally, the noise is fixed to $\sigma = 50$, approximately an order of magnitude lower than the maximal exact increment magnitude, $v_F \Delta t = 600$.

\section{Up-to-one-switch approximation for the probability distribution function of a noisy location increment}\label{SI:Sec:P_1}

Here, we compute explicitly all the results obtained in Section~\ref{Subsec:up-to-one-switch}. The up-to-one-switch approximation of a single location increment is provided as a Python code in the function \textit{approx\_pdf\_up\_to\_1\_switch} in the file \textit{functions.py}.

We now consider the up-to-one-switch approximation for the probability density function (PDF) of $\Delta y$, defined as
\begin{equation}\label{SI:Eq:P_1(Delta y)}
    P_1(\Delta y) :=\mathbb{P}(\Delta y \,|\, W=0)\mathbb{P}(W=0)+\mathbb{P}(\Delta y \,|\, W=1)\mathbb{P}(W\ge1).
\end{equation}
We make progress by further by conditioning each term in Equation~\eqref{SI:Eq:P_1(Delta y)} on the state at the start of the interval, denoted as $S_1$. Namely, for $W = 0$, we see that
$$\mathbb{P}(\Delta y \,|\, W=0)\mathbb{P}(W=0) = \sum_{s=1}^n \mathbb{P}(\Delta y \,|\, W=0, S_1=s)\mathbb{P}(W=0\,|\,S_1=s) \mathbb{P}(S_1=s),$$
where $\mathbb{P}(S_1=s)=p_{s}$ is given by the equilibrium assumption (Equation~\eqref{Eq:P(S(t)=s)}). For $W = 0$, the amount of time spent in the first state is, trivially, given by $\Delta t$ (see Figure~\ref{Fig:switches_diagram} zero-switch case). Therefore we obtain
$$\mathbb{P}(W=0\,|\,S_1=s)=\exp{(-\lambda_{s}\Delta t)}.$$

To compute $\mathbb{P}(\Delta y \,|\, W=1)\mathbb{P}(W\ge1)$, we condition on both the first state and the second state visited within the interval, $S_1$ and $S_2$ (see one-switch case in Figure~\ref{Fig:switches_diagram}). This yields
$$\begin{aligned}
    \mathbb{P}(\Delta y \,|\, W=1)\mathbb{P}(W \ge1) =\sum_{\substack{s_1\\s_2\ne s_1}}
    &\mathbb{P}(\Delta y \,|\, W=1, S_1=s_1, S_2=s_2)
    \\
    &\times \mathbb{P}(W\ge1\,|\,S_1=s_1, S_2=s_2) 
    \\
    &\times \mathbb{P}(S_2=s_2\,|\,S_1=s_1)\mathbb{P}(S_1=s_1).
\end{aligned}$$
By definition, $\mathbb{P}(S_2=s_2\,|\, S_1=s_1)=p_{s_1s_2}$ (Equation~\eqref{Eq:p_su}), by the equilibrium assumption $\mathbb{P}(S_1=s_1)=p_{s_1}$ (Equation~\eqref{Eq:P(S(t)=s)}), and we obtain
$$\mathbb{P}(W\ge1\,|\,S_1=s_1, S_2=s_2)=\mathbb{P}(W\ge 1\,|\,S_1=s_1)=1-\exp{(-\lambda_{s_1}\Delta t)}$$
by considering that at least one switch occurs if $\tau_1<\Delta t$.

\subsection{Computing \texorpdfstring{$\mathbb{P}(\Delta y \,|\, W=0, S_1=s)$}{}}\label{SI:Subsec:P(Delta y|W=0)}

For $W=0$ and given the initial state of the interval $S_1$, we note that
$$(\Delta x \,|\, W=0, S_1=s)=v_{s} \Delta t.$$
Therefore,
$$\mathbb{P}(\Delta x \,|\, W=0, S_1=s)=\delta_{v_{s} \Delta t}(\Delta x),$$
where $\delta_{v_{s} \Delta t}$ denotes the Dirac delta function centered at $v_{s} \Delta t$.
Adding the noise $\Delta \epsilon \sim \mathcal{N}(0, 2\sigma^2)$, we obtain 
$$\mathbb{P}(\Delta y \,|\, W=0, S_1=s)=f_{\mathcal{N}(v_{s} \Delta t,2\sigma^2)}(\Delta y),$$
where $f_{\mathcal{N}(v_{s} \Delta t,2\sigma^2)}$ denotes the PDF of a normal distribution with mean $v_{s}\Delta t$ and variance $2\sigma^2$.

\subsection{Computing \texorpdfstring{$\mathbb{P}(W=1\,|\,S_1=s_1, S_2=s_2)$}{}}

For $t_1\in(0, \Delta t]$, we define
$$F_{s_1,s_2}(t_1):=\mathbb{P}(\tau_1 \le t_1, \tau_1>0, \tau_2>\Delta t - \tau_1\,|\,S_1=s_1, S_2=s_2).$$
Then
$$G_{s_1,s_2}(t_1):=\mathbb{P}(\tau_1 \le t_1\,|\, 0\le\tau_1\le\Delta t, \tau_2>\Delta t - \tau_1, S_1=s_1, S_2=s_2)=\frac{F_{s_1,s_2}(t)}{F_{s_1,s_2}(\Delta t)}.$$

We can compute $F_{s_1,s_2}$ integrating the joint distribution $f_{s_1,s_2}(\tau_1, \tau_2)$ for $\tau_1 \in (0, t]$ and $\tau_2 > \Delta t-\tau_1$, thus
$$\begin{aligned}
    F_{s_1,s_2}(t_1):=& \ \mathbb{P}(\tau_1 \le t_1, \tau_1>0, \tau_2>\Delta t - \tau_1\,|\,S_1=s_1,S_2=s_2)
    \\
    =& \ \int_0^{t_1}\left(\int_{\Delta t-\tau_1}^{\infty}f_{s_1,s_2}(\tau_1, \tau_2)\dd \tau_2\right)\dd \tau_1.
\end{aligned}$$
Note that $F_{s_1,s_2}(\Delta t)=\mathbb{P}(W=1\,|\,S_1=s_1,S_2=s_2)$.
The joint distribution of two independent random variables is the product of the two distributions, and we note that
$f_{s}(\tau)= \lambda_{s}e^{-\lambda_{s}t}$, which yields
$$f_{s_1,s_2}(\tau_1, \tau_2) = f_{s_1}(\tau_1) f_{s_2}(\tau_2) = \lambda_{s_1} \exp\left({-\lambda_{s_1} \tau_1}\right) \lambda_{s_2}  \exp\left({-\lambda_{s_2} \tau_2}\right) .$$
By substitution, we obtain
$$F_{s_1,s_2}(t_1)=\int_0^t\left(\int_{\Delta t-\tau_1}^{\infty}\lambda_{s_1} \exp\left({-\lambda_{s_1} \tau_1}\right) \lambda_{s_2}  \exp\left({-\lambda_{s_2} \tau_2}\right)\dd \tau_2\right)\dd \tau_1,$$
thus,
$$ F_{s_1,s_2}(t)= 
\begin{dcases}
    \lambda_{s_1} \exp\left(-\lambda_{s_1}\Delta t\right)t & \text{if } \lambda_{s_1}= \lambda_{s_2},
    \\
    \frac{\lambda_{s_1} \exp\left({-\lambda_{s_2} \Delta t}\right)}{-\lambda_{s_1} + \lambda_{s_2}}\left( \exp\left({(-\lambda_{s_1} + \lambda_{s_2})t}\right)-1 \right) & \text{if } \lambda_{s_1} \ne \lambda_{s_2}.
\end{dcases}$$
It follows that
$$G_{s_1,s_2}(t) = 
\begin{dcases}
    \frac{t}{\Delta t} & \text{if } \lambda_{s_1}= \lambda_{s_2},
    \\
    \frac{\exp\left({(-\lambda_{s_1} + \lambda_{s_2})t}\right)-1}{\exp\left({(-\lambda_{s_1} + \lambda_{s_2})\Delta t}\right)-1} & \text{if } \lambda_{s_1}\ne \lambda_{s_2}.
\end{dcases}$$
Finally, we note that
$\mathbb{P}(W=1\,|\,S_1=s_1,S_2=s_2)=F_{s_1,s_2}(\Delta t)$.

\subsection{Computing \texorpdfstring{$\mathbb{P}(\Delta x \,|\, W=1, S_1=s_1, S_2=s_2)$}{}}\label{SI:Subsec:P(Delta x | W=1, S_1=s_1, S_2=s_2)}

For fixed $S_1=s_1$, $S_2=s_2$ and a time $\tau_1$ spent in state $S_1$ during the time interval considered, we get
$$(\Delta x \,|\, W=1, S_1=s_1, S_2=s_2)=v_{s_1}\tau_1+v_{s_2}(\Delta t - \tau_1).$$
We define
$$h_{s_1,s_2}(\tau_1) := v_{s_1}\tau_1+v_{s_2}(\Delta t - \tau_1).$$
We note that $h_{s_1,s_2}(t)$ is monotonically increasing if $v_{s_1}-v_{s_2}>0$, and decreasing if $v_{s_1}-v_{s_2}<0$. 

If $v_{s_1}=v_{s_2}$, then the velocity is constant for the whole interval; thus the exact location increment is known $\Delta x=h_{s_1,s_2}(\Delta t)$ and
$$\mathbb{P}(\Delta y \,|\, W=1, S_1=s_1,S_2=s_2)=\mathbb{P}(\Delta y \,|\, W=0, S_1=s_1)=f_{\mathcal{N}({v_{s_1}\Delta t},2\sigma^2)}(\Delta y).$$

Otherwise, the exact increment $\Delta x$ is determined by the time of the switch, and we compute
$$\begin{aligned}
g_{s_1,s_2}(t_1):=&\ \mathbb{P}(t_1\,|\, W=1, S_1=s_1, S_2=s_2)= \frac{\dd}{\dd t}G_{s_1,s_2}(t_1)
\\
=&\begin{dcases}
    \frac{1}{\Delta t} & \text{if } \lambda_{s_1}= \lambda_{s_2},
    \\
    (-\lambda_{s_1} + \lambda_{s_2})\frac{\exp\left({(-\lambda_{s_1} + \lambda_{s_2})t}\right)-1}{\exp\left({(-\lambda_{s_1} + \lambda_{s_2})\Delta t}\right)-1} & \text{if } \lambda_{s_1}\ne \lambda_{s_2}.
\end{dcases}
\end{aligned}$$
Since $h_{s_1,s_2}:\mathbb{R} \to \mathbb{R}$ is a monotonic function and $\Delta x=h_{s_1,s_2}(t)$, the density function of $\Delta x$ is
$$\begin{aligned}
    \Tilde{g}_{s_1,s_2}(\Delta x)&=\mathbb{P}(\Delta x\,|\, W=1, S_1=s_1, S_2=s_2)
    \\
    & =g_{s_1,s_2}\left(h_{s_1,s_2}^{-1}(\Delta x)\right)\left|\frac{\dd}{\dd(\Delta x)}h_{s_1,s_2}^{-1}(\Delta x)\right|.
\end{aligned}$$

Hence, we compute $t$ in terms of $\Delta x$ as
$$t = h_{s_1,s_2}^{-1}(\Delta x) = \frac{\Delta x - v_{s_2} \Delta t}{v_{s_1} - v_{s_2}}$$
and its derivative
$$\frac{\dd}{\dd(\Delta x)}h_{s_1,s_2}^{-1}(\Delta x) = \frac{1}{v_{s_1} - v_{s_2}}.$$
Hence,
$$\left|\frac{\dd}{\dd(\Delta x)}h_{s_1,s_2}^{-1}(\Delta x)\right| = \frac{1}{|v_{s_1} - v_{s_2}|}.$$

\subsection{Computing \texorpdfstring{$\mathbb{P}(\Delta y \,|\, W=1, S_1=s_1, S_2=s_2)$}{}}\label{SI:Subsec:P(Delta y | W=1, S_1=s_1, S_2=s_2)}

Now, we incorporate noise
$$\begin{aligned}
    \Tilde{f}_{s_1,s_2}(\Delta y) :=&\ \mathbb{P}(\Delta y\,|\, W=1, S_1=s_1, S_2=s_2 ) 
    \\ =& \int_{a}^b \Tilde{g}_{s_1,s_2}(\Delta x)f_{\mathcal{N}(0,2\sigma^2)}(\Delta y-\Delta x) \dd(\Delta x),
\end{aligned}$$
where 
$$a=a_{s_1,s_2}:= \min \{h_{s_1,s_2}(0), h_{s_1,s_2}(\Delta t)\},$$
$$b=b_{s_1,s_2}:= \max \{h_{s_1,s_2}(0), h_{s_1,s_2}(\Delta t)\},$$
and 
$$f_{\mathcal{N}(0,2\sigma^2)}(\Delta \epsilon):= \frac{1}{\sqrt{2}\sigma \sqrt{2\pi}}\exp\left({-\frac{1}{2}\frac{(\Delta\epsilon)^2}{2\sigma^2}}\right)=\frac{1}{2\sigma \sqrt{\pi}}\exp\left({-\frac{(\Delta\epsilon)^2}{4\sigma^2}}\right).$$

First, we consider the case $\lambda_{s_1}\ne\lambda_{s_2}$. Then,
$$\Tilde{g}_{s_1,s_2}(\Delta x)= \frac{-\lambda_{s_1}+\lambda_{s_2}}{|v_{s_1} - v_{s_2}|}\frac{\exp{\left((-\lambda_{s_1} +\lambda_{s_2})h^{-1}_{s_1,s_2}(\Delta x)\right)}-1}{\exp\left({(-\lambda_{s_1} + \lambda_{s_2})\Delta t}\right)-1}.$$
To obtain $\Tilde{f}_{s_1,s_2}(\Delta y)$ we integrate over $\Delta x$. We note that
\begin{equation*}
    \int_{a}^{b} \exp{\left( -c (\Delta x)^2 + r \Delta x + \hat{r} \right)} \dd(\Delta x) = \frac{\sqrt{\pi}}{2\sqrt{c}}\exp{\left(\frac{r^2}{4c}+\hat{r}\right)}\left[\erf\left(\frac{2c \Delta x -r}{2\sqrt{c}} \right) \right]_{\Delta x = a}^{\Delta x = b},
\end{equation*}
where $\erf$ is the error function, and write
$$\Tilde{f}_{s_1,s_2}(\Delta y) =   k\left[\erf\left(\frac{2c \Delta x-r}{2\sqrt{c}} \right) \right]_{\Delta x = a}^{\Delta x = b}.
$$
Here, $a$ and $b$ are as previously defined,
$$c:=\frac{1}{4\sigma^2},$$
while, $r$, $\hat{r}$ and $k$ vary depending on the states $s_1,s_2$ and on $\Delta y$:
$$
r=r_{s_1,s_2}(\Delta y) := 2c\Delta y + \frac{-\lambda_{s_1} + \lambda_{s_2}}{v_{s_1} - v_{s_2}},$$
$$\hat{r} =\hat{r}_{s_1,s_2}(\Delta y):= -c(\Delta y)^2 + \frac{-\lambda_{s_1} + \lambda_{s_2}}{|v_{s_1} - v_{s_2}|}|v_{s_2}| \Delta t,$$
and
$$k=k_{s_1,s_2} := \frac{-\lambda_{s_1} + \lambda_{s_2}}{2|v_{s_1} - v_{s_2}|} \frac{\exp\left({\frac{r^2}{4c}+\hat{r}}\right)}{\exp\left({(-\lambda_{s_1} +\lambda_{s_2})\Delta t}\right)-1}.$$

If $\lambda_{s_1}= \lambda_{s_2}$, then
$$\Tilde{g}_{s_1,s_2}(\Delta x)=  \frac{1}{|v_{s_1} - v_{s_2}|\Delta t},$$
and we note that here $\Tilde{g}_{s_1,s_2}(\Delta x)$ is the length of the interval $[a,b]$, as previously defined.
Thus
$$\Tilde{g}_{s_1,s_2}(\Delta x)=f_{\mathcal{U}_{(a,b)}}(\Delta x),$$
where $f_{\mathcal{U}_{(a,b)}}$ denotes the PDF of a uniform distribution in the interval $[a,b]$.

Therefore,
$$\Tilde{f}_{s_1,s_2}(\Delta y) =\int_a^b f_{\mathcal{U}(a,b)}(\Delta x) f_{\mathcal{N}(0, 2\sigma^2)}(\Delta y-\Delta x)\dd(\Delta x) = \frac{1}{b-a}\int_{\Delta y-a}^{\Delta y-b}-f_{\mathcal{N}(0, 2\sigma^2)}(w)\dd w,$$
where $f_{\mathcal{U}(a,b)}=1/(b-a)$ is the PDF of the uniform distribution in $[a,b]$ and the change of variable $ W=\Delta y-\Delta x$. Hence,
$$\Tilde{f}_{s_1,s_2}(\Delta y) = \frac{1}{b-a}\left(F_{\mathcal{N}(0, 2\sigma^2)}(\Delta y-a) - F_{\mathcal{N}(0, 2\sigma^2)}(\Delta y -b)\right),$$
where $F_{\mathcal{N}(0, 2\sigma^2)}$ represents the CDF of the normal distribution with mean $\mu=0$ and variance $\hat{\sigma}^2=2\sigma^2$,
defined as
$$F_{\mathcal{N}(\mu, \hat{\sigma})}(w)=\frac{1}{2}\left[1+\erf\left(\frac{w-\mu}{\hat{\sigma}\sqrt{2}}\right)\right].$$
We note that $\Tilde{f}_{s_1,s_2}(\Delta y)$ for $\lambda_1=\lambda_2$ can also be obtained by taking the limit for $\lambda_1-\lambda_2\to 0$ of the function computed for $\lambda_1\ne\lambda_2$.

\subsection{Error of the up-to-one-switch approximation}

\begin{figure*}[!ht]
    \begin{minipage}{0.06\textwidth}
    \textcolor{white}{-}
    \end{minipage}%
    \begin{minipage}{0.47\textwidth}
    \centering
    \begin{minipage}{0.6\textwidth}
    \small \centering
    \textbf{Four-state model}
    \end{minipage}%
    \begin{minipage}{0.3\textwidth}
    \includegraphics[width=1\textwidth]{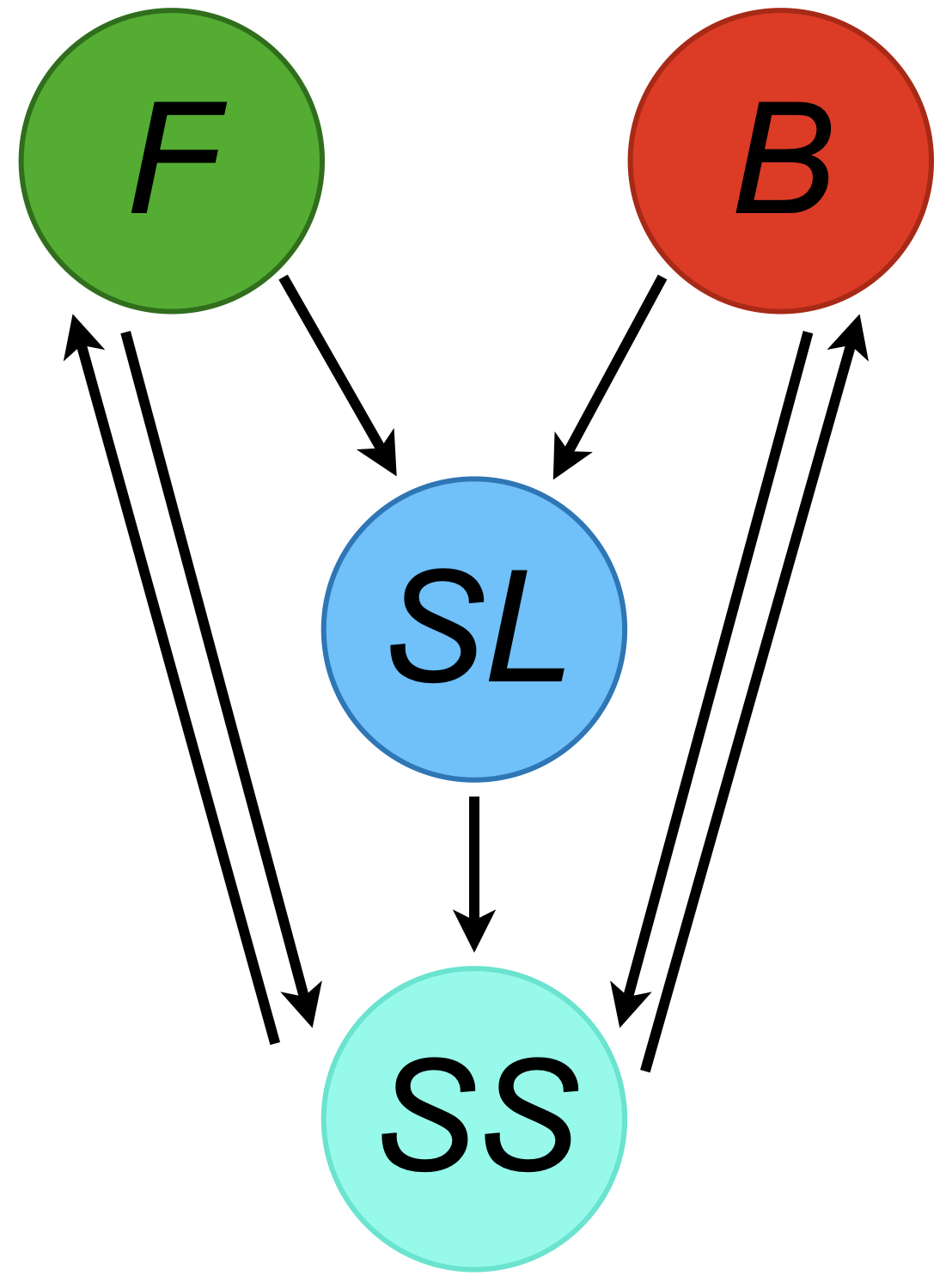}
    \end{minipage}
    \end{minipage}%
    \begin{minipage}{0.47\textwidth}
    \centering
    \begin{minipage}{0.6\textwidth}
    \small \centering 
    \textbf{Six-state model}
    \end{minipage}%
    \begin{minipage}{0.3\textwidth}
    \includegraphics[width=1\textwidth]{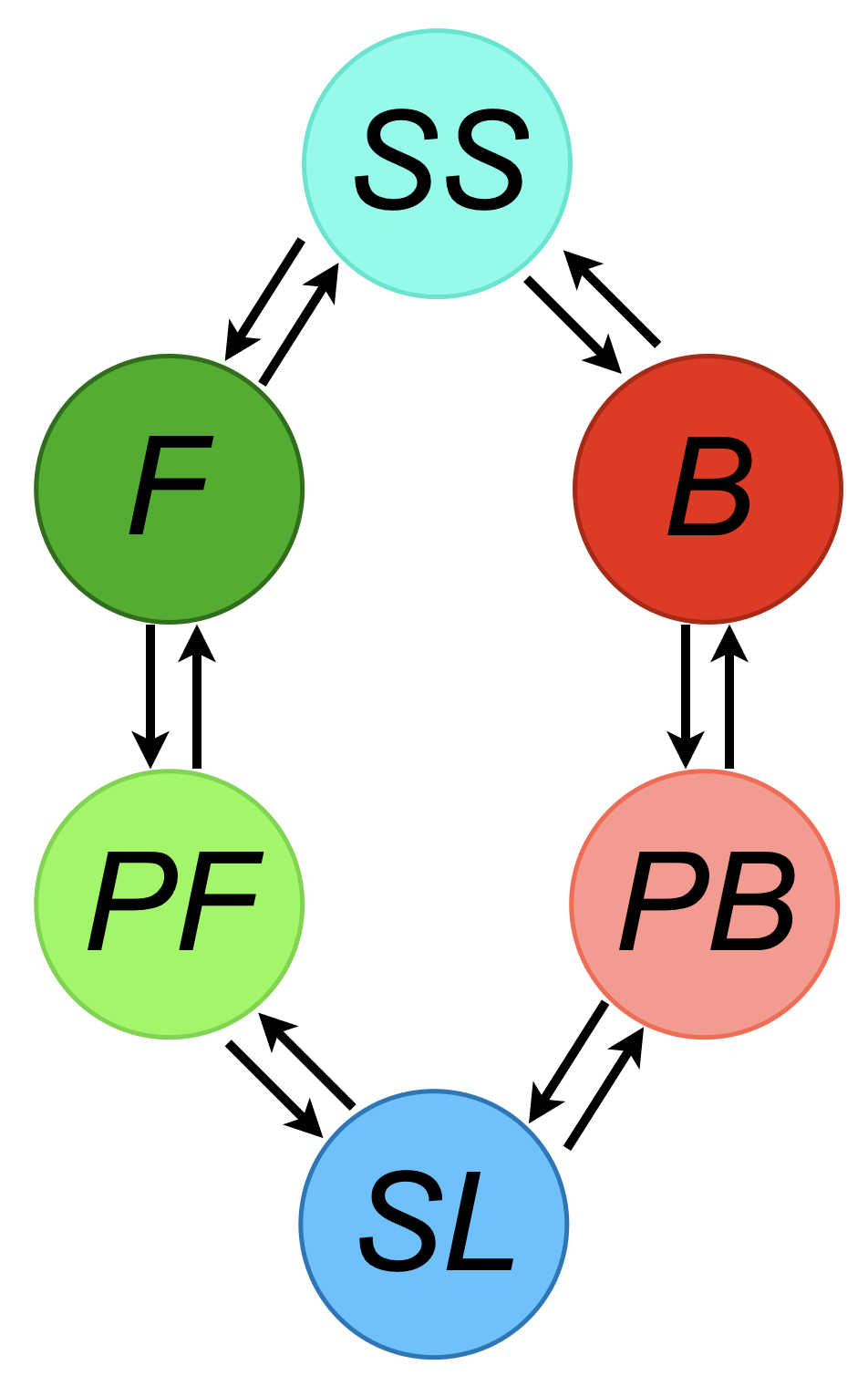}
    \end{minipage}
    \end{minipage}
    \begin{minipage}{0.06\textwidth}
    \centering
    $\boldsymbol{\lambda}$
    \end{minipage}%
    \begin{minipage}{0.47\textwidth}
    \includegraphics[width=1\textwidth]{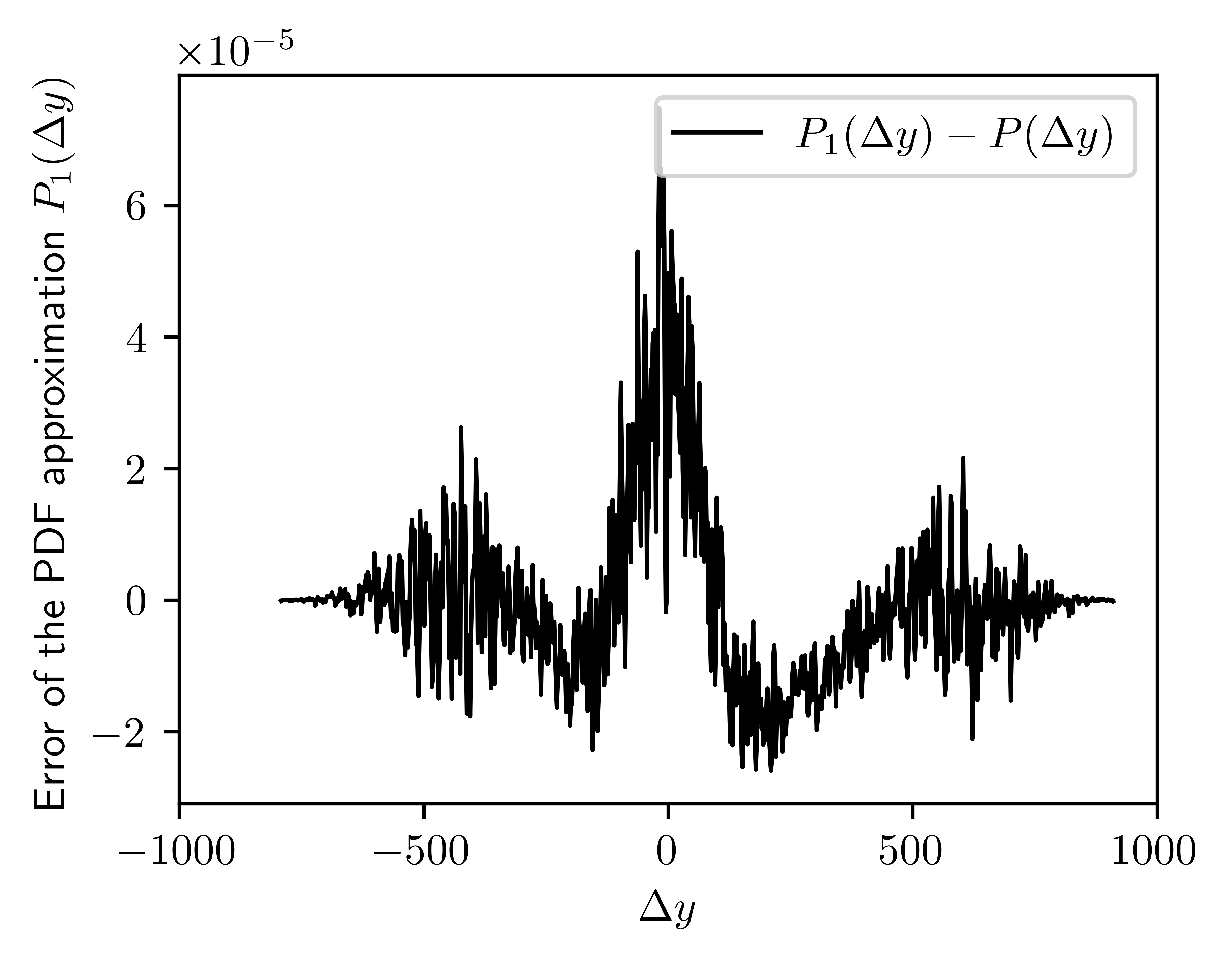}
    \end{minipage}%
    \begin{minipage}{0.47\textwidth}
    \includegraphics[width=1\textwidth]{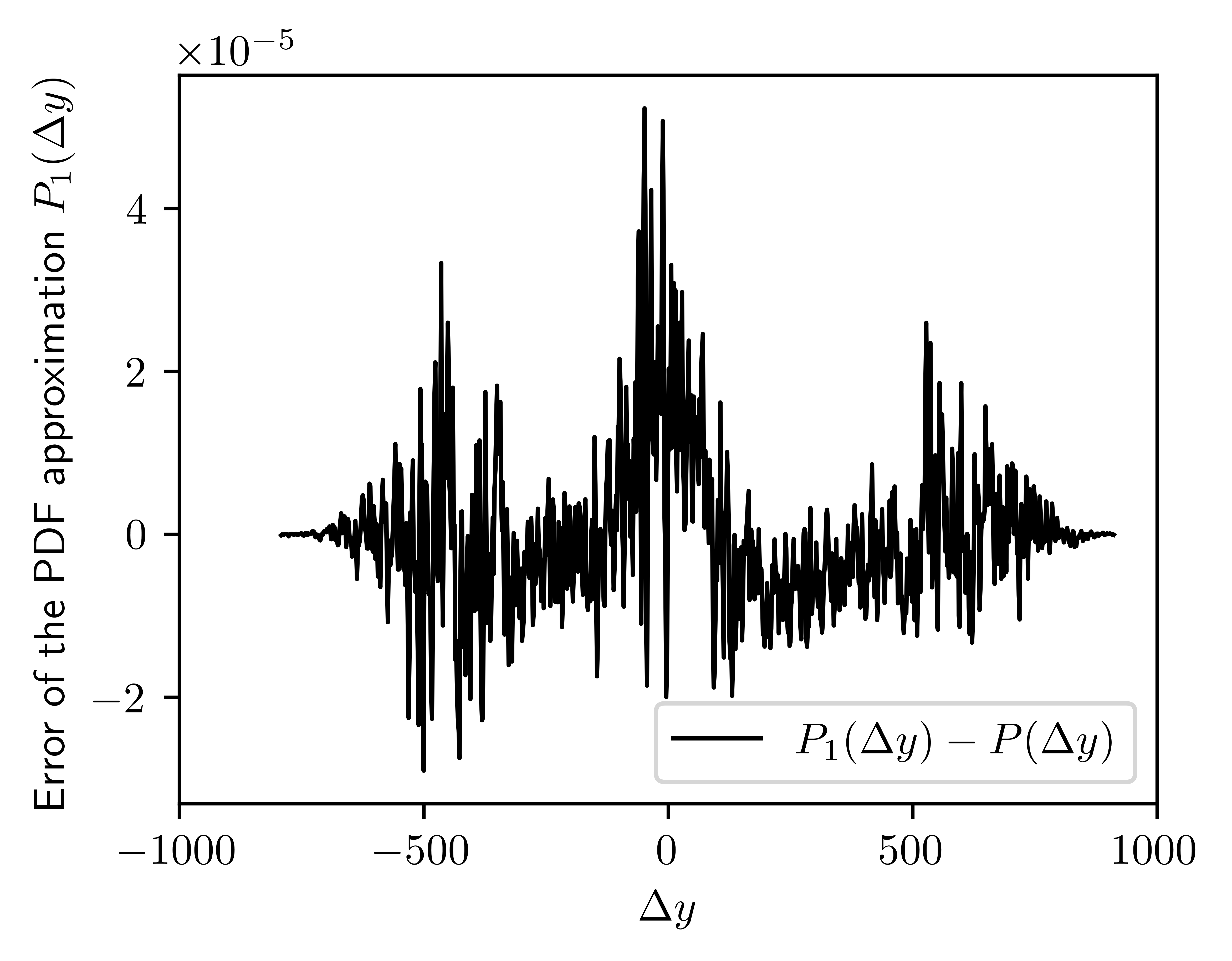}
    \end{minipage}
    \begin{minipage}{0.06\textwidth}
    \centering
    $10\boldsymbol{\lambda}$
    \end{minipage}%
    \begin{minipage}{0.47\textwidth}
    \includegraphics[width=1\textwidth]{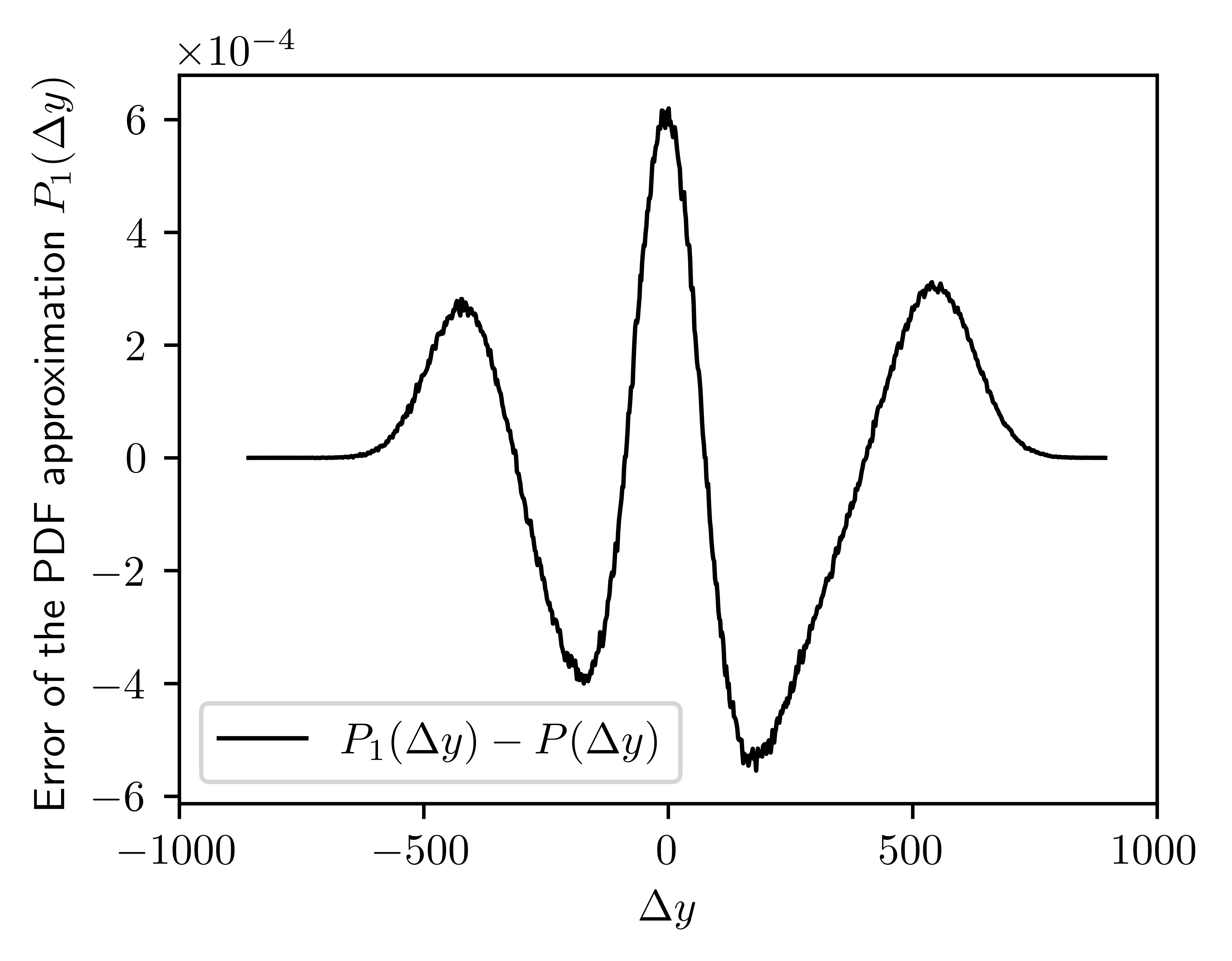}
    \end{minipage}%
    \begin{minipage}{0.47\textwidth}
    \includegraphics[width=1\textwidth]{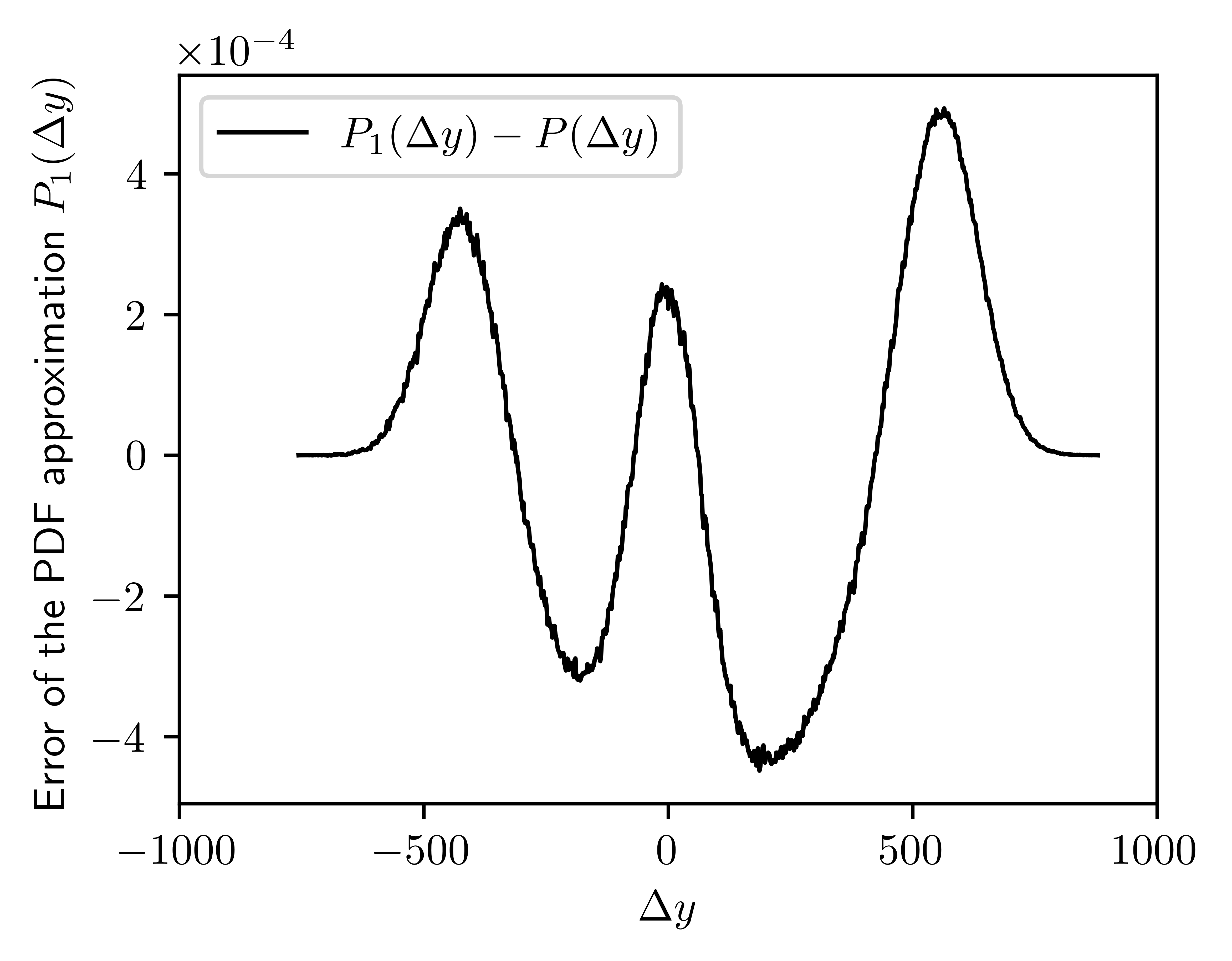}
    \end{minipage}
    \caption{The error of the approximation $P_1(\Delta y)$, defined as $P_1(\Delta y)-P(\Delta y)$ where $P(\Delta y)$ is the
    empirical PDF for $\Delta y$ is shown. The left plots are obtained from a four-state model while the right ones from a six-state model. The top plots are obtained using the parameters as specified in the Supplementary Information Figure~\ref{SI:Fig:networks}, while the bottom ones use the same parameters except the rates which are multiplied by 10.}
    \label{SI:Fig:Error_PDF_4_6}
\end{figure*}

Figure~\ref{SI:Fig:Error_PDF_4_6} shows the error of the up-to-one-switch approximation, $P_1(\Delta y)-P(\Delta y)$, for the four-state model and the six-state model.

\section{Up-to-two-switch approximation for the probability distribution function of a noisy location increment}\label{SI:Sec:P_2}

Here, we compute explicitly all the results obtained in Section~\ref{Subsec:up-to-two-switch}. The up-to-two-switch approximation of a single location increment is provided as a Python code in the function \textit{approx\_pdf\_up\_to\_2\_switch} in the file \textit{functions.py}.

We now consider the up-to-two-switch approximation for the PDF of $\Delta y$ defined as
\begin{equation*}
    P_2(\Delta y) :=\mathbb{P}(\Delta y \,|\, W=0)\mathbb{P}(W=0)+\mathbb{P}(\Delta y \,|\, W=1)\mathbb{P}(W=1)+\mathbb{P}(\Delta y \,|\, W=2)\mathbb{P}(W\ge2).
\end{equation*}
We have already computed $\mathbb{P}(\Delta y \,|\, W=0) \mathbb{P}(W=0)$ and $\mathbb{P}(\Delta y \,|\, W=1)$ by conditioning on the states attained, therefore we need to compute $\mathbb{P}(W=1)$ and $\mathbb{P}(\Delta y \,|\, W=2) \mathbb{P}(W\ge2).$

We note that
$$\mathbb{P}(W\ge2) = 1 -\mathbb{P}(W=0)-\mathbb{P}(W=1),$$
which also holds after conditioning on the states attained. And in particular,
\begin{align}\label{SI:Eq:P(W=2|S)}
    \mathbb{P}(W\ge2\,|\, S_1=s_1, S_2=s_2, S_3=s_3) &= 1 -\mathbb{P}(W=0\,|\,S_1=s_1)-\mathbb{P}(W=1\,|\,S_1=s_1, S_2=s_2)
    \\
    &=\mathbb{P}(W\ge2\,|\, S_1=s_1, S_2=s_2), \nonumber
\end{align}
where $\mathbb{P}(W=0\,|\,S_1=s_1)$ and $\mathbb{P}(W=1\,|\,S_1=s_1, S_2=s_2)$ have been computed in the previous section.

Moreover, as in the one-switch case, we need to condition on the states. For two switches, we have three subsequent states to consider
$$\begin{aligned}
\mathbb{P}(\Delta y \,|\, W=2) \mathbb{P}(W\ge2)=\sum_{s_1}\sum_{s_2\ne s_1}\sum_{s_3\ne s_2}&\mathbb{P}(\Delta y\,|\,W=2, S_1=s_1, S_2=s_2, S_3=s_3)
\\
&\quad \times \mathbb{P}(W\ge2\,|\, S_1=s_1, S_2=s_2, S_3=s_3)
\\
&\quad \times \mathbb{P}(S_3=s_3\,|\,S_1=s_1,S_2=s_2)
\\& \quad \times \mathbb{P}(S_2=s_2\,|\,S_1=s_1)\mathbb{P}(S_1=s_1),
\end{aligned}$$
where $\mathbb{P}(S_3=s_3\,|\,S_1=s_1,S_2=s_2)= \mathbb{P}(S_3=s_3\,|\,S_2=s_2)= p_{s_2,s_3}$, $\mathbb{P}(S_2=s_2\,|\,S_1=s_1)=p_{s_1,s_2}$ and $\mathbb{P}(S_1=s_1)=p_{s_1}$.
Hence, we only need to compute $\mathbb{P}(\Delta y\,|\,W=2, S_1=s_1, S_2=s_2, S_3=s_3)$ and, as before, we start by computing $\mathbb{P}(W=2\,|\,S_1=s_1, S_2=s_2, S_3=s_3)$.

\subsection{Computing \texorpdfstring{$\mathbb{P}(W=2\,|\,S_1=s_1, S_2=s_2, S_3=s_3),\ s_3\ne s_1$}{}}\label{Subsec:st3!=st1}

Here we compute $\mathbb{P}(W=2\,|\,S_1=s_1, S_2=s_2, S_3=s_3)$ when $s_3\ne s_1$. We consider the case $s_3= s_1$ separately, in Section~\ref{SI:Subsec:s3=s1}. Moreover, we assume that the velocities $v_{s_i}$ are all distinct and the rates $\lambda_{s_i}$ are all distinct for $i=1,2,3$. We denote $\tau_i$ the time spent in state $S_i$ during the $\Delta t$ time interval considered. For $W=2$, we need to condition on the three subsequent states attained ($S_1, S_2, S_3$) to have all the necessary information to compute $\mathbb{P}(W=2\,|\,S_1=s_1, S_2=s_2, S_3=s_3)$. The number of switches is $W=2$ if and only if $0<\tau_1\le\Delta t$, $0<\tau_2\le\Delta t-\tau_1$ and $\tau_3\ge \Delta t-\tau_1-\tau_2$. 

For $t_1,t_2\in(0, \Delta t]$ with $t_2\le \Delta t - t_1$, we define
$$\begin{aligned}
    F_{s_1,s_2,s_3}(t_1,t_2):=&\ \mathbb{P}\left(\begin{aligned}
        &\tau_1 \le t_1, \tau_2\le t_2, \tau_1>0, 
        \\
        &0<\tau_2\le\Delta t - \tau_1, \tau_3>\Delta t-\tau_1-\tau_2
    \end{aligned}\,\middle|\,S_1=s_1, S_2=s_2,S_3=s_3\right)
    \\
    =&\ \mathbb{P}(0<\tau_1 \le t_1, 0<\tau_2\le t_2, \tau_3>\Delta t-\tau_1-\tau_2\,|\,S_1=s_1, S_2=s_2,S_3=s_3),
\end{aligned}$$
since $t_2\le \Delta t - t_1\le \Delta t-\tau_1$. We can compute $F_{s_1,s_2,s_3}$ by integrating the joint distribution $f_{s_1,s_2,s_3}(\tau_1, \tau_2, \tau_3)$ for $\tau_1 \in (0, t_1]$ and $\tau_2 \in (0, t_2]$ and $\tau_3 > \Delta t-\tau_1-\tau_2$, thus
$$F_{s_1,s_2,s_3}(t_1,t_2):=\int_0^{t_1}\left(\int_0^{t_2}\left(\int_{\Delta t-\tau_1-\tau_2}^{\infty}f_{s_1,s_2,s_3}(\tau_1, \tau_2,\tau_3)\dd \tau_3\right)\dd \tau_2\right)\dd \tau_1.$$

The joint distribution of three independent random variables is the product of the three distributions, and we note that
$f_{s}(\tau)= \lambda_{s}e^{-\lambda_{s}\tau}$; hence,
$$\begin{aligned}
f_{s_1,s_2,s_3}(\tau_1, \tau_2, \tau_3)& = f_{s_1}(\tau_1) f_{s_2}(\tau_2) f_{s_2}(\tau_3) 
\\
&=\lambda_{s_1} \exp\left({-\lambda_{s_1} \tau_1}\right) \lambda_{s_2}  \exp\left({-\lambda_{s_2} \tau_2}\right) \lambda_{s_3}  \exp\left({-\lambda_{s_3} \tau_3}\right).
\end{aligned}$$
Hence, in the assumption that $\lambda_{s_i}$ are all distinct, we obtain
$$\begin{aligned}
F_{s_1,s_2, s_3}(t_1,t_2)=\lambda_{s_1}\lambda_{s_2}\exp{\left(-\lambda_{s_3}\Delta t\right)}\ \frac{\exp{\left((-\lambda_{s_2}+\lambda_{s_3}) t_2\right)}-1}{{-\lambda_{s_2}+\lambda_{s_3}}}\frac{\exp{\left((-\lambda_{s_1}+\lambda_{s_3}) t_1\right)}-1}{-\lambda_{s_1}+\lambda_{s_3}}.
\end{aligned}$$

We define
$$
G_{s_1,s_2,s_3}(t_1, t_2):=\mathbb{P}\left(\tau_1 \le t_1, \tau_2\le t_2\,\middle|\,\begin{aligned}&0<\tau_1\le\Delta t, 0<\tau_2\le\Delta t-\tau_1, \tau_3>\Delta t - \tau_1-\tau_2, \\& S_1=s_1, S_2=s_2,S_3=s_3\end{aligned}\right),
$$
which can be computed as
$$G_{s_1,s_2,s_3}(t_1, t_2):=\frac{F_{s_1,s_2,s_3}(t_1,t_2)}{\mathbb{P}\left(0<\tau_1\le\Delta t, 0<\tau_2\le\Delta t-\tau_1, \tau_3>\Delta t - \tau_1-\tau_2\,\middle|\, S_1=s_1, S_2=s_2,S_3=s_3\right)},$$
where
$$\begin{aligned}
&\mathbb{P} (0<\tau_1\le\Delta t, 0<\tau_1\le\Delta t-\tau_2, \tau_3>\Delta t - \tau_1-\tau_2\,|\, S_1=s_1, S_2=s_2,S_3=s_3)
\\
&=\int_0^{\Delta t}\left(\int_0^{\Delta t - \tau_1}\left(\int_{\Delta t-\tau_1-\tau_2}^{\infty}f_{s_1,s_2,s_3}(\tau_1, \tau_2,\tau_3)\dd \tau_3\right)\dd \tau_2\right)\dd \tau_1
\\
&=\frac{\lambda_{s_1}\lambda_{s_2}}{-\lambda_{s_2}+\lambda_{s_3}}\left( \frac{\exp{\left(-\lambda_{s_1}\Delta t\right)}-\exp{\left(-\lambda_{s_2}\Delta t\right)}}{-\lambda_{s_1}+\lambda_{s_2}}-\frac{\exp{\left(-\lambda_{s_1}\Delta t\right)}-\exp{\left(-\lambda_{s_3}\Delta t\right)}}{-\lambda_{s_1}+\lambda_{s_3}}\right).
\end{aligned}$$
Therefore, simplifying we obtain
$$G_{s_1,s_2,s_3}(t_1, t_2)=\frac{\exp{\left(-\lambda_{s_3}\Delta t\right)} \left(\exp{\left((-\lambda_{s_2}+\lambda_{s_3}) t_2\right)}-1\right)\left(\exp{\left((-\lambda_{s_1}+\lambda_{s_3}) t_1\right)}-1\right)}{\left( \frac{\exp{\left(-\lambda_{s_1}\Delta t\right)}-\exp{\left(-\lambda_{s_2}\Delta t\right)}}{-\lambda_{s_1}+\lambda_{s_2}}-\frac{\exp{\left(-\lambda_{s_1}\Delta t\right)}-\exp{\left(-\lambda_{s_3}\Delta t\right)}}{-\lambda_{s_1}+\lambda_{s_3}}\right)(-\lambda_{s_1}+\lambda_{s_3})}.$$
Finally, we note that
$$\mathbb{P}(W=2\,|\,S_1=s_1,S_2=s_2, S_3=s_3)=F_{s_1,s_2, s_3}(\Delta t,\Delta t).$$

\subsection{Computing \texorpdfstring{$\mathbb{P}(\Delta x\,|\,W=2, S_1=s_1, S_2=s_2, S_3=s_3),\ s_3\ne s_1$}{}}
From the CDF $G_{s_1,s_2,s_3}(t_1, t_2)$ we can obtain the PDF
$$
g_{s_1,s_2,s_3}(t_1, t_2):= \mathbb{P}(t_1, t_2\,|\, W=2, S_1=s_1,S_2=s_2, S_3=s_3) =\frac{\partial^2}{\partial t_1 \partial t_2} G_{s_1,s_2,s_3}(t_1, t_2).$$ 
Simplifying we obtain
$$g_{s_1,s_2,s_3}(t_1, t_2):=\frac{N(t_1,t_2)}{D},$$
where we define
$D:=D_{1,2}+D_{2,3}+D_{3,1}$,
$$D_{i,j}:=(\lambda_{s^{(i)}}-\lambda_{s^{(j)}})\exp{(\Delta t (\lambda_{s^{(i)}}+\lambda_{s^{(j)}}))},$$
and
$$N(t_1,t_2):=D_{1,2}\cdot (\lambda_{s_3}-\lambda_{s_1})\exp{(t_1 (\lambda_{s_3}-\lambda_{s_1}))} \cdot (\lambda_{s_3}-\lambda_{s_2})\exp{(t_2 (\lambda_{s_3}-\lambda_{s_2}))}.$$
We define
$$\begin{aligned}
\Tilde{g}_{s_1,s_2, s_3}(\Delta x):=&\ \mathbb{P}(\Delta x\,|\, W=2, S_1=s_1, S_2=s_2, S_3=s_3)=
\\
=&\int_0^{\Delta t} (\int_0^{\Delta t - t_1} \mathbb{P}(\Delta x\,|\, \tau_1=t_1, \tau_2=t_2, W=2, S_1=s_1, S_2=s_2, S_3=s_3)
\\
&\qquad \qquad \qquad \quad \times \mathbb{P}(t_1, t_2\,|\, W=2, S_1=s_1, S_2=s_2, S_3=s_3)\dd t_2 )\dd t_1
\\
=& \ \int_0^{\Delta t}\left( \int_0^{\Delta t - t_1} \delta\left(v_{s_1}t_1 +v_{s_2}t_2 +v_{s_3}(\Delta t- t_1 -t_2) -\Delta x\right)g_{s_1, s_2, s_3}(t_1, t_2) \dd t_2\right)\dd t_1.
\end{aligned}$$

Now, we use the sifting property of the Dirac delta function: for any continuous function $f$,
$$\int_{-\infty}^{\infty}\delta(x-s)f(s)\dd s=f(x).$$
Since the Dirac delta function is evaluated at
$$[(v_{s_1}-v_{s_3})t_1 +v_{s_3}\Delta t -\Delta x] - [(v_{s_3}-v_{s_2})t_2],$$
for $v_{s_3}\ne v_{s_2}$, we use the changes of variable  $x=x(t_1):=(v_{s_1}-v_{s_3})t_1 +v_{s_3}\Delta t -\Delta x$ and 
$z=z(t_2):=(v_{s_3}-v_{s_2})t_2$, and we obtain
$$\begin{aligned}
&\int_{0}^{\Delta t - t_1} \delta\left(v_{s_1}t_1 +v_{s_2}t_2 +v_{s_3}(\Delta t- t_1 -t_2) -\Delta x\right)g_{s_1, s_2, s_3}(t_1, t_2) \dd t_2
\\
&=\int_{z^{-1}(0)}^{z^{-1}(\Delta t-t_1)} \delta(x-z)\ g_{s_1, s_2, s_3}\left(t_1, \frac{z}{v_{s_3}-v_{s_2}}\right)\frac{1}{v_{s_3}-v_{s_2}}\dd z.
\end{aligned}$$
We note that $z^{-1}(0)=0$ and $z^{-1}(\Delta t-t_1)= (\Delta t-t_1)/(v_{s_3}-v_{s_2})$ and for all $t_1\in[0, \Delta t]$ 
$$\begin{dcases}
    z^{-1}(0)\le z^{-1}(\Delta t-t_1) & \text{if } v_{s_3}-v_{s_2}>0,
    \\
    z^{-1}(0)\ge z^{-1}(\Delta t-t_1) & \text{if } v_{s_3}-v_{s_2}<0.
\end{dcases}$$
Hence, $\sign\left(z^{-1}(\Delta t-t_1)-z^{-1}(0)\right)=\sign\left(v_{s_3}-v_{s_2}\right)$.
Moreover, since $g_{s_1, s_2, s_3}(t_1, t_2)$ is only defined for $t_1\in[0,\Delta t]$ and $t_2\in[0, \Delta t-t_1]$, we extend its definition to $\mathbb{R}$ using the function
$$\Bar{g}_{s_1, s_2, s_3}(t_1, t_2):=
\begin{dcases}
g_{s_1, s_2, s_3}(t_1, t_2) & \text{if } t_1\in[0, \Delta t], t_2\in[0, \Delta t-t_1],
\\
0 & \text{otherwise}.
\end{dcases}$$

Hence, we write the integral
$$\begin{aligned}
&\int_{z^{-1}(0)}^{z^{-1}(\Delta t-t_1)} \delta(x-z)\ g_{s_1, s_2, s_3}\left(t_1, \frac{z}{v_{s_3}-v_{s_2}}\right)\frac{1}{v_{s_3}-v_{s_2}}\dd z
\\
&=\frac{\sign\left(v_{s_3}-v_{s_2}\right)}{v_{s_3}-v_{s_2}}\int_{-\infty}^{\infty} \delta(x-z)\ \Bar{g}_{s_1, s_2, s_3}\left(t_1, \frac{z}{v_{s_3}-v_{s_2}}\right)\dd z
\\
&=\frac{1}{|v_{s_3}-v_{s_2}|} \ \Bar{g}_{s_1, s_2, s_3}\left(t_1, \frac{x}{v_{s_3}-v_{s_2}}\right)
\\
&=\frac{1}{|v_{s_3}-v_{s_2}|} \ \Bar{g}_{s_1, s_2, s_3}\left(t_1, \frac{(v_{s_1}-v_{s_3})t_1 +v_{s_3}\Delta t -\Delta x}{v_{s_3}-v_{s_2}}\right).
\end{aligned}$$
We also see that
$$g_{s_1, s_2, s_3}\left(t_1, \frac{(v_{s_1}-v_{s_3})t_1 +v_{s_3}\Delta t -\Delta x}{v_{s_3}-v_{s_2}}\right)=\frac{1}{D} \cdot  N\left(t_1,\frac{(v_{s_1}-v_{s_3})t_1 +v_{s_3}\Delta t -\Delta x}{v_{s_3}-v_{s_2}}\right),$$
where we define
$$\begin{aligned}
    N(t_1,t_2):= (\lambda_{s_1}-\lambda_{s_2})\exp{(\Delta t (\lambda_{s_1}+\lambda_{s_2}))}(\lambda_{s_3}-\lambda_{s_1})\exp{(t_1 (\lambda_{s_3}-\lambda_{s_1}))}(\lambda_{s_3}-\lambda_{s_2})\exp{(t_2 (\lambda_{s_3}-\lambda_{s_2}))}.
\end{aligned}$$
Thus, by substitution we get
$$\begin{aligned}
N\left(t_1, \frac{(v_{s_1}-v_{s_3})t_1 +v_{s_3}\Delta t -\Delta x}{v_{s_3}-v_{s_2}}\right)
=&\ (\lambda_{s_1}-\lambda_{s_2}) (\lambda_{s_3}-\lambda_{s_1})(\lambda_{s_3}-\lambda_{s_2})\exp{(\Delta t (\lambda_{s_1}+\lambda_{s_2}))}
\\
&\cdot\exp{\left(t_1 (\lambda_{s_3}-\lambda_{s_1})+\frac{(v_{s_1}-v_{s_3})t_1 +v_{s_3}\Delta t -\Delta x}{v_{s_3}-v_{s_2}} (\lambda_{s_3}-\lambda_{s_2})\right)}
\\
=&\ k_{1,2,3} \cdot\Tilde{M}(\Delta x)\cdot \Tilde{N}(t_1),
\end{aligned}$$
which was written denoting
$$k_{1,2,3}:=-(\lambda_{s_1}-\lambda_{s_2})(\lambda_{s_2}-\lambda_{s_3}) (\lambda_{s_3}-\lambda_{s_1})\cdot\exp{\left( \frac{v_{s_2}(\lambda_{s_1}+\lambda_{s_2})-v_{s_3}(\lambda_{s_1}+\lambda_{s_3})}{v_{s_2}-v_{s_3}}\Delta t \right)},$$
$$\Tilde{M}(\Delta x):=\exp{\left(\frac{-(\lambda_{s_3}-\lambda_{s_2})}{v_{s_3}-v_{s_2}}\Delta x\right)},$$
and
$$\begin{aligned}
\Tilde{N}(t_1):=&\exp{\left( \left(\lambda_{s_3}-\lambda_{s_1}+\frac{(v_{s_1}-v_{s_3})}{v_{s_3}-v_{s_2}} (\lambda_{s_3}-\lambda_{s_2})\right)t_1\right)}
\\
=&\exp{\left(\frac{v_{s_1}(\lambda_{s_3}-\lambda_{s_2})
+v_{s_2}(\lambda_{s_1}-\lambda_{s_3})
+v_{s_3}(\lambda_{s_2}-\lambda_{s_1})}{v_{s_3}-v_{s_2}} t_1\right)}
\\
=&\exp{\left(\frac{\nu_{1,2,3}}{v_{s_2}-v_{s_3}} t_1\right)},
\end{aligned}$$
where
$$\nu_{1,2,3} := v_{s_1}(\lambda_{s_2}-\lambda_{s_3})
+v_{s_2}(\lambda_{s_3}-\lambda_{s_1})
+v_{s_3}(\lambda_{s_1}-\lambda_{s_2}).$$

Now, we define and compute the indefinite integral in $t_1$
$$\begin{aligned} I(t_1):=
&\ \int_{t_1} \frac{1}{|v_{s_3}-v_{s_2}|} \ \Bar{g}_{s_1, s_2, s_3}\left(t_1, \frac{(v_{s_1}-v_{s_3})t_1 +v_{s_3}\Delta t -\Delta x}{v_{s_3}-v_{s_2}}\right)\dd t_1 
\\
=&\ \frac{1}{|v_{s_3}-v_{s_2}|} \frac{k_{1,2,3}}{D}\Tilde{M}(\Delta x) \int_{t_1} \Tilde{N}(t_1)\dd t_1
\\
=&\ \frac{k_{1,2,3}\Tilde{M}(\Delta x)}{|v_{s_3}-v_{s_2}|\cdot D}\frac{v_{s_2}-v_{s_3}}{\nu_{1,2,3}}\Tilde{N}(t_1)
\\
= &\ \frac{k_{1,2,3}\Tilde{M}(\Delta x)}{D}\frac{\sign(v_{s_2}-v_{s_3})}{\nu_{1,2,3}}\Tilde{N}(t_1),
\end{aligned}$$
in which the result of the indefinite integral is obtained up to a constant.

We now need to determine area of integration, which depend on $\Delta x$ and on the increasing order assumed by the three velocities. Hence, we first denote $v_{\min}, v_{\text{int}}, v_{\max} \in \{v_{s_1}, v_{s_2}, v_{s_3}\} $ such that $v_{\min} <v_{\text{int}} <v_{\max}$. Now, the probability density of $\Delta x$ is non-zero for $\Delta x \in (v_{\min} \Delta t, v_{\max} \Delta t)$.

We denote the domain of integration for $t_1$
$$A:=\left\{t_1 \;\middle|\; t_1\in[0,\Delta t], \frac{(v_{s_1}-v_{s_3})t_1 +v_{s_3}\Delta t -\Delta x}{v_{s_3}-v_{s_2}}\in[0, \Delta t-t_1]\right\},$$
and we note that if $A$ is not empty, it is an interval $A=[E_0,E_1]$. Since $t_1\in[0,\Delta t]$ then $E_0\ge 0$ and $E_1\le\Delta t$. Moreover,
$$t_2=\frac{\Delta x-(v_{s_1}-v_{s_3})t_1 -v_{s_3}\Delta t }{v_{s_2}-v_{s_3}}\in[0, \Delta t-t_1].$$
gives other conditions on the area of integration:
\begin{equation}\label{SI:Eq:conditions}
    \begin{dcases}
    (v_{s_1}-v_{s_3})t_1 +v_{s_3}\Delta t \le \Delta x \le (v_{s_1}-v_{s_2})t_1 +v_{s_2}\Delta t, & \text{if } v_{s_2}>v_{s_3},
    \\
    (v_{s_1}-v_{s_2})t_1 +v_{s_2}\Delta t \le \Delta x \le (v_{s_1}-v_{s_3})t_1 +v_{s_3}\Delta t, & \text{if } v_{s_2}<v_{s_3}.
\end{dcases}
\end{equation}
We define 
$$a_1:=\frac{\Delta x- v_{s_2}\Delta t}{v_{s_1}-v_{s_2}},$$
and
$$b_1:=\frac{\Delta x- v_{s_3}\Delta t}{v_{s_1}-v_{s_3}},$$
and we notice that we can write the conditions in Equation~\eqref{SI:Eq:conditions} as
$$\begin{dcases}
\text{if } v_{s_2}>v_{s_3}, &
\begin{dcases}
    t_1 \le b_1 & \text{if } v_{s_1}>v_{s_3},
    \\
    t_1 \ge b_1 & \text{if } v_{s_1}<v_{s_3},
\end{dcases}
\text{ and }
\begin{dcases}
    t_1 \ge a_1 & \text{if } v_{s_1}>v_{s_2},
    \\
    t_1 \le a_1 & \text{if } v_{s_1}<v_{s_2},
\end{dcases}
\\
\text{if } v_{s_2}<v_{s_3}, &
\begin{dcases}
    t_1 \ge b_1 & \text{if } v_{s_1}>v_{s_3},
    \\
    t_1 \le b_1 & \text{if } v_{s_1}<v_{s_3},
\end{dcases}
\text{ and }
\begin{dcases}
    t_1 \le a_1 & \text{if } v_{s_1}>v_{s_2},
    \\
    t_1 \ge a_1 & \text{if } v_{s_1}<v_{s_2}.
\end{dcases}
\end{dcases}$$
We apply the lower boundaries for $t_1$ to determine $E_0$ and upper boundaries to determine $E_1$, which depend on whether $\Delta x\in[v_{\min} \Delta t, v_{\text{int}} \Delta t]$ or $\Delta x\in[v_{\text{int}} \Delta t, v_{\max} \Delta t]$. In particular, we obtain
$$\begin{dcases}
    \text{if } v_{s_2}>v_{s_1}>v_{s_3},
    & 
    \begin{dcases}
    E_0=0,\ E_1=b_1, & \text{if }\Delta x\le v_{\text{int}} \Delta t,
    \\
    E_0=0,\ E_1=a_1, & \text{if }\Delta x\ge v_{\text{int}} \Delta t,
    \end{dcases}
    \\
    \text{if } v_{s_3}>v_{s_1}>v_{s_2},
    & 
    \begin{dcases}
    E_0=0,\ E_1=a_1, & \text{if }\Delta x\le v_{\text{int}} \Delta t,
    \\
    E_0=0,\ E_1=b_1, & \text{if }\Delta x\ge v_{\text{int}} \Delta t,
    \end{dcases}
    \\
    \text{if } v_{s_1}>v_{s_2}>v_{s_3},
    & 
    \begin{dcases}
    E_0=0,\ E_1=a_1, & \text{if }\Delta x\le v_{\text{int}} \Delta t,
    \\
    E_0=b_1,\ E_1=a_1, & \text{if }\Delta x\ge v_{\text{int}} \Delta t,
    \end{dcases}
    \\
    \text{if } v_{s_3}>v_{s_2}>v_{s_1},
    & 
    \begin{dcases}
    E_0=a_1,\ E_1=b_1, & \text{if }\Delta x\le v_{\text{int}} \Delta t,
    \\
    E_0=0,\ E_1=b_1, & \text{if }\Delta x\ge v_{\text{int}} \Delta t,
    \end{dcases}
    \\
    \text{if } v_{s_1}>v_{s_3}>v_{s_2},
    & 
    \begin{dcases}
    E_0=0,\ E_1=a_1, & \text{if }\Delta x\le v_{\text{int}} \Delta t,
    \\
    E_0=b_1,\ E_1=a_1, & \text{if }\Delta x\ge v_{\text{int}} \Delta t,
    \end{dcases}
    \\
    \text{if } v_{s_2}>v_{s_3}>v_{s_1},
    & 
    \begin{dcases}
    E_0=b_1,\ E_1=a_1, & \text{if }\Delta x\le v_{\text{int}} \Delta t,
    \\
    E_0=0,\ E_1=a_1, & \text{if }\Delta x\ge v_{\text{int}} \Delta t.
    \end{dcases}
\end{dcases}$$
In conclusion,
$$\Tilde{g}_{s_1, s_2, s_3}(\Delta x):=\mathbb{P}(\Delta x\,|\, W=2, s_1, s_2, s_3)=I(E_1)-I(E_0).$$

\subsection{Computing \texorpdfstring{$\mathbb{P}(\Delta y\,|\,W=2, S_1=s_1, S_2=s_2, S_3=s_3),\ s_3\ne s_1$}{}}
By writing $\Tilde{g}$ explicitly, it can be verified that
$$\Tilde{g}_{s_1, s_2, s_3}(\Delta x)=\Tilde{g}_{s^{(i_1)}, s^{(i_2)}, s^{(i_3)}}(\Delta x),$$
for any $i_1, i_2, i_3\in\{1,2,3\}$ all distinct. In other words, once three distinct states (with distinct rates and velocities) are fixed, the probability distribution of obtaining a $\Delta x$ with two switches does not depend on the order of the states are attained.
Hence, we can compute 
$$\begin{aligned}
\Tilde{f}_{s_1,s_2,s_3}(\Delta y) &:= \mathbb{P}(\Delta y\,|\,W=2, S_1=s_1, S_2=s_2, S_3=s_3)
    \\
    &\ = \int_{v_{\min}\Delta t}^{v_{\max}\Delta t} \Tilde{g}_{s_1,s_2, s_3}(\Delta x)f_{\mathcal{N}(0,2\sigma^2)}(\Delta y-\Delta x) \dd(\Delta x),
\end{aligned}$$ 
by considering the simplest form for 
$$\Tilde{g}_{s_1,s_2, s_3}(\Delta x)=I_{s_1,s_2, s_3}(E_1)-I_{s_1,s_2, s_3}(0),$$
where we fix $v_{s_2}=v_{\max}$, $v_{s_1}=v_{\text{int}}$, $v_{s_3}=v_{\min}$, and rearrange the rates $\lambda_{s_i}$ accordingly, and
$$\begin{dcases}
    E_1=b_1=\frac{\Delta x- v_{s_3}\Delta t}{v_{s_1}-v_{s_3}}, & \text{if }\Delta x\le v_{\text{int}} \Delta t,
    \\
    E_1=a_1=\frac{\Delta x- v_{s_2}\Delta t}{v_{s_1}-v_{s_2}}, & \text{if }\Delta x\ge v_{\text{int}} \Delta t.
\end{dcases}$$
Therefore,
$$\begin{aligned}
    \Tilde{f}_{s_1,s_2,s_3}(\Delta y) =& \int_{v_{\min}\Delta t}^{ v_{\text{int}}\Delta t} \Tilde{g}_{s_1,s_2, s_3}(\Delta x)f_{\mathcal{N}(0,2\sigma^2)}(\Delta y-\Delta x) \dd(\Delta x)
    \\
    &+\int_{v_{\text{int}}\Delta t}^{v_{\max}\Delta t} \Tilde{g}_{s_1,s_2, s_3}(\Delta x)f_{\mathcal{N}(0,2\sigma^2)}(\Delta y-\Delta x) \dd(\Delta x)
    \\
    =& \int_{v_{\min}\Delta t}^{v_{\text{int}}\Delta t} (I(b_1)-I(0))f_{\mathcal{N}(0,2\sigma^2)}(\Delta y-\Delta x) \dd(\Delta x)
    \\
    &+\int_{v_{\text{int}}\Delta t}^{v_{\max}\Delta t} (I(a_1)-I(0))f_{\mathcal{N}(0,2\sigma^2)}(\Delta y-\Delta x) \dd(\Delta x)
    \\
    =&\int_{v_{\min}\Delta t}^{v_{\max}\Delta t} -I(0)f_{\mathcal{N}(0,2\sigma^2)}(\Delta y-\Delta x) \dd(\Delta x),
    \\
    &+\int_{v_{\min}\Delta t}^{v_{\text{int}}\Delta t} I(b_1)f_{\mathcal{N}(0,2\sigma^2)}(\Delta y-\Delta x) \dd(\Delta x)
    \\
    &+\int_{v_{\text{int}}\Delta t}^{v_{\max}\Delta t} I(a_1)f_{\mathcal{N}(0,2\sigma^2)}(\Delta y-\Delta x) \dd(\Delta x)
    \\
    =&: J_0 + J_{b_1} + J_{a_1},
\end{aligned}$$ 
where
$$f_{\mathcal{N}(0,2\sigma^2)}(\Delta \epsilon):= \frac{1}{2\sigma \sqrt{\pi}}\exp\left({-\frac{(\Delta\epsilon)^2}{4\sigma^2}}\right),$$
and
$$I(t_1)=\frac{k_{1,2,3}\Tilde{M}(\Delta x)}{D}\frac{\sign(v_{s_2}-v_{s_3})}{\nu_{1,2,3}}\Tilde{N}(t_1),$$
$$\Tilde{N}(t_1)=\exp{\left(\frac{\nu_{1,2,3}}{v_{s_2}-v_{s_3}} t_1\right)},$$
$$\Tilde{M}(\Delta x)=\exp{\left(\frac{\lambda_{s_2}-\lambda_{s_3}}{v_{s_3}-v_{s_2}}\Delta x\right)}.$$
By defining
$$\Tilde{k}:=\frac{k_{1,2,3}}{D}\frac{\sign(v_{s_2}-v_{s_3})}{\nu_{1,2,3}}\frac{1}{2\sigma\sqrt{\pi}},$$
then we write
$$\begin{aligned}
J_0:=&\ \int_{v_{\min}\Delta t}^{v_{\max}\Delta t} -I(0)f_{\mathcal{N}(0,2\sigma^2)}(\Delta y-\Delta x) \dd(\Delta x)
\\
=&\ \int_{v_{\min}\Delta t}^{v_{\max}\Delta t} -\Tilde{k}\exp{\left(\frac{\lambda_{s_2}-\lambda_{s_3}}{v_{s_3}-v_{s_2}}\Delta x\right)}\exp\left({-\frac{(\Delta y-\Delta x)^2}{4\sigma^2}}\right)\dd(\Delta x)
\\
=&\ k_0\left[\erf\left(\frac{2c(\Delta x)-r_0}{2\sqrt{c}} \right) \right]_{\Delta x = v_{\min}\Delta t}^{\Delta x = v_{\max}\Delta t},
\end{aligned}$$
where
$$c:=\frac{1}{4\sigma^2},$$
$$r_0=r_0(\Delta y):= 2c\Delta y + \frac{\lambda_{s_2} - \lambda_{s_3}}{v_{s_3} - v_{s_2}},$$
$$\hat{r}_0=\hat{r}_0(\Delta y) := -c(\Delta y)^2,$$
and
$$\begin{aligned}
k_0:=& -\Tilde{k}\frac{\sqrt{\pi}}{2\sqrt{c}}\exp{\left(\frac{r_0^2}{4c}+\hat{r}_0 \right)}
\\
=& -\frac{k_{1,2,3}}{D}\frac{\sign(v_{s_2}-v_{s_3})}{\nu_{1,2,3}}\frac{1}{2}\exp{\left(\frac{r_0^2}{4c}+\hat{r}_0 \right)}
\\
=& \ \frac{k_{1,2,3}}{D}\frac{\sign(v_{s_3}-v_{s_2})}{\nu_{1,2,3}}\frac{1}{2}\exp{\left(\frac{r_0^2}{4c}+\hat{r}_0 \right)}.
\end{aligned}$$

Similarly,
$$\begin{aligned}
J_{b_1}:=&\int_{v_{\min}\Delta t}^{v_{\text{int}}\Delta t} I(b_1)f_{\mathcal{N}(0,2\sigma^2)}(\Delta y-\Delta x) \dd(\Delta x)
\\
=& \int_{v_{\min}\Delta t}^{ v_{\text{int}}\Delta t} \Tilde{k}\Tilde{N}\left(\frac{\Delta x- v_{s_3}\Delta t}{v_{s_1}-v_{s_3}}\right)\exp{\left(\frac{-(\lambda_{s_3}-\lambda_{s_2})}{v_{s_3}-v_{s_2}}\Delta x\right)}\exp\left({-\frac{(\Delta y-\Delta x)^2}{4\sigma^2}}\right)\dd(\Delta x)
\\
=& \ k_{b_1}\left[\erf\left(\frac{2c(\Delta x)-r_{b_1}}{2\sqrt{c}} \right) \right]_{\Delta x = v_{\min}\Delta t}^{\Delta x = v_{\text{int}}\Delta t},
\end{aligned}$$
but now
$$\begin{aligned}
    r_{b_1}=r_{b_1}(\Delta y):=&\ 2c\Delta y + \frac{\lambda_{s_2} - \lambda_{s_3}}{v_{s_3} - v_{s_2}}+\frac{\nu_{1,2,3}}{(v_{s_2}-v_{s_3})(v_{s_1}-v_{s_3})} 
\\
=&\ r_0 + \frac{\nu_{1,2,3}}{(v_{s_2}-v_{s_3})(v_{s_1}-v_{s_3})},
\end{aligned}$$
$$\begin{aligned}
\hat{r}_{b_1} :=& -c(\Delta y)^2-\frac{\nu_{1,2,3}\cdot v_{s_3}\Delta t}{(v_{s_2}-v_{s_3})(v_{s_3}-v_{s_1})}
\\
=&\ \hat{r}_0-\frac{\nu_{1,2,3}\cdot v_{s_3}\Delta t}{(v_{s_3}-v_{s_2})(v_{s_3}-v_{s_1})},
\end{aligned}$$
and
$$\begin{aligned}
k_{b_1}:=&\ \Tilde{k}\frac{\sqrt{\pi}}{2\sqrt{c}}\exp{\left(\frac{r_{b_1}^2}{4c}+\hat{r}_{b_1} \right)}
\\
=&\ \frac{k_{1,2,3}}{D}\frac{\sign(v_{s_2}-v_{s_3})}{\nu_{1,2,3}}\frac{1}{2}\exp{\left(\frac{r_{b_1}^2}{4c}+\hat{r}_{b_1} \right)}.
\end{aligned}$$

Finally,
$$\begin{aligned}
J_{a_1}:=&\int_{v_{\text{int}}\Delta t}^{v_{\max}\Delta t} I(a_1)f_{\mathcal{N}(0,2\sigma^2)}(\Delta y-\Delta x) \dd(\Delta x)
\\
=&\int_{v_{\text{int}}\Delta t}^{v_{\max}\Delta t} \Tilde{k}\Tilde{N}\left(\frac{\Delta x- v_{s_2}\Delta t}{v_{s_1}-v_{s_2}}\right)\exp{\left(\frac{-(\lambda_{s_3}-\lambda_{s_2})}{v_{s_3}-v_{s_2}}\Delta x\right)}\exp\left({-\frac{(\Delta y-\Delta x)^2}{4\sigma^2}}\right)\dd(\Delta x)
\\
=&\ k_{a_1}\left[\erf\left(\frac{2c(\Delta x)-r_{a_1}}{2\sqrt{c}} \right) \right]_{\Delta x = v_{\text{int}}\Delta t}^{\Delta x = v_{\max}\Delta t},
\end{aligned}$$
but now
$$\begin{aligned}
    r_{a_1}=r_{a_1}(\Delta y):=&\ 2c\Delta y + \frac{\lambda_{s_2} - \lambda_{s_3}}{v_{s_3} - v_{s_2}}+\frac{\nu_{1,2,3}}{(v_{s_2}-v_{s_3})(v_{s_1}-v_{s_2})} 
    \\
    =&\ r_0 +\frac{\nu_{1,2,3}}{(v_{s_2}-v_{s_3})(v_{s_1}-v_{s_2})} ,
\end{aligned}$$
$$\begin{aligned}
\hat{r}_{a_1} =\hat{r}_{a_1}(\Delta y):=& -c(\Delta y)^2 -\frac{\nu_{1,2,3}\cdot v_{s_2}\Delta t}{(v_{s_2}-v_{s_3})(v_{s_1}-v_{s_2})} 
    \\
    =&\ \hat{r}_0  -\frac{\nu_{1,2,3}\cdot v_{s_2}\Delta t}{(v_{s_2}-v_{s_3})(v_{s_1}-v_{s_2})},
\end{aligned}$$
and
$$\begin{aligned}
k_{a_1}:=&\ \Tilde{k}\frac{\sqrt{\pi}}{2\sqrt{c}}\exp{\left(\frac{r_{a_1}^2}{4c}+\hat{r}_{a_1} \right)}
\\
=&\ \frac{k_{1,2,3}}{D}\frac{\sign(v_{s_2}-v_{s_3})}{\nu_{1,2,3}}\frac{1}{2}\exp{\left(\frac{r_{a_1}^2}{4c}+\hat{r}_{a_1} \right)}.
\end{aligned}$$

\subsection{Computing \texorpdfstring{$\mathbb{P}(W=2\,|\,S_1=s_1, S_2=s_2, S_3=s_3)$}{} for \texorpdfstring{$s_3=s_1$}{}}\label{SI:Subsec:s3=s1}
Here, we compute $\mathbb{P}(W=2\,|\,S_1=s_1, S_2=s_2, S_3=s_1)$. Since $S_3=S_1$, we only need to define a function of  $t_2\in(0, \Delta t]$
$$\begin{aligned}
    F_{s_1,s_2,s_1}(t_2) :=&\ \mathbb{P}(\tau_2\le t_2, 0<\tau_1 \le \Delta t, 0<\tau_2\le \Delta t -\tau_1, \tau_3>\Delta t-\tau_1-\tau_2\,|\,S_1=s_1, S_2=s_2,S_3=s_1) 
    \\
    &= \mathbb{P}(0\le\tau_2\le t_2, 0<\tau_1 \le \Delta t-\tau_2, \tau_3>\Delta t-\tau_1-\tau_2\,|\,S_1=s_1, S_2=s_2,S_3=s_1) 
    \\
    &= \int_0^{t_2}\left(\int_0^{\Delta t - \tau_2}\left(\int_{\Delta t-\tau_1-\tau_2}^{\infty}f_{s_1,s_2,s_1}(\tau_1, \tau_2,\tau_3)\dd \tau_3\right)\dd \tau_1\right)\dd \tau_2 
    \\
    &
    \begin{aligned}
        = \frac{\lambda_{s_1}\lambda_{s_2}\exp{(-\lambda_{s_1}\Delta t)}}{(\lambda_{s_1}-\lambda_{s_2})^2}\Big(&((\lambda_{s_1}-\lambda_{s_2})\Delta t (\exp{((\lambda_{s_1}-\lambda_{s_2})t_2)}-1))
        \\
        &- ((\lambda_{s_1}-\lambda_{s_2})t_2 - 1)\exp{((\lambda_{s_1}-\lambda_{s_2})t_2)}-1\Big),
    \end{aligned}
\end{aligned}$$
in the assumption $\lambda_{s_1}\ne\lambda_{s_2}$.
Therefore, in the previous notation we obtain
$$\begin{aligned}
    G_{s_1,s_2,s_1}(t_2):=&\  \mathbb{P}(\tau_2\le t_2\,|\, 0<\tau_1 \le \Delta t, 0<\tau_2\le \Delta t -\tau_1, \tau_3>\Delta t-\tau_1-\tau_2,S_1=s_1,S_2=s_2,S_3=s_1) 
    \\
    =&\ \frac{F_{s_1,s_2,s_1}(t_2)}{F_{s_1, s_2,s_1}(\Delta t)}
    \\
    =&\ \frac{(\lambda_{s_1}-\lambda_{s_2})\Delta t (\exp{((\lambda_{s_1}-\lambda_{s_2})t_2)}-1) - ((\lambda_{s_1}-\lambda_{s_2})t_2 - 1)\exp{((\lambda_{s_1}-\lambda_{s_2})t_2)}-1}{\exp{((\lambda_{s_1}-\lambda_{s_2})\Delta t)}-1-(\lambda_{s_1}-\lambda_{s_2})\Delta t}.
\end{aligned}$$
Finally, 
$$\begin{aligned}
    \mathbb{P}(W=1\,|\,S_1=s_1,S_2=s_2, S_3=s_1)&=F_{s_1,s_2,s_1}(\Delta t)
    \\
    & = \frac{\lambda_{s_1}\lambda_{s_2}\exp{(-\lambda_{s_1}\Delta t)}}{(\lambda_{s_1}-\lambda_{s_2})^2}(\exp{((\lambda_{s_1}-\lambda_{s_2})\Delta t)}-1-(\lambda_{s_1}-\lambda_{s_2})\Delta t).
\end{aligned}$$

\subsection{Computing \texorpdfstring{$\mathbb{P}(\Delta x\,|\,W=2, S_1=s_1, S_2=s_2, S_3=s_3)$}{} for \texorpdfstring{$s_3=s_1$}{}}
From the CDF $G_{s_1,s_2,s_1}(t_2)$ we can obtain the PDF
$$\begin{aligned}
g_{s_1,s_2,s_1}(t_2):=&\ \mathbb{P}(t_2\,|\, W=2, S_1=s_1, S_2=s_2,S_3=s_1) =\frac{\dd}{\dd t_2} G_{s_1,s_2,s_1}(t_2)
\\
=&\ \frac{(\lambda_{s_1}-\lambda_{s_2})^2(\Delta t-t_2)\exp{((\lambda_{s_1}-\lambda_{s_2})t_2)}}{\exp{((\lambda_{s_1}-\lambda_{s_2})\Delta t)}-1-(\lambda_{s_1}-\lambda_{s_2})\Delta t}.
\end{aligned}$$ 
Now, we approach the problem similarly to Section~\ref{SI:Subsec:P(Delta x | W=1, S_1=s_1, S_2=s_2)}.

We define
$$h_{s_1,s_2,s_1}(t_2) := v_{s_2}t_2+v_{s_1}(\Delta t - t_2).$$
We note that $h_{s_1,s_2,s_1}:\mathbb{R} \to \mathbb{R}$ is a monotonic function in $t_2$ and $\Delta x=h_{s_1,s_2,s_1}(t_2)$. Thus the PDF of $\Delta x$ is
$$\begin{aligned}
    \Tilde{g}_{s_1,s_2,s_1}(\Delta x)&=\mathbb{P}(\Delta x\,|\, W=2, S_1=s_1, S_2=s_2, S_3=s_1)
    \\
    &=g_{s_1,s_2,s_1}\left(h_{s_1,s_2,s_1}^{-1}(\Delta x)\right)\left|\frac{\dd}{\dd(\Delta x)}h_{s_1,s_2,s_1}^{-1}(\Delta x)\right|,
\end{aligned}$$
where
$$h_{s_1,s_2,s_1}^{-1}(\Delta x) = \frac{\Delta x - v_{s_1} \Delta t}{v_{s_2} - v_{s_1}},$$
and its derivative is
$$\frac{\dd}{\dd(\Delta x)}h_{s_1,s_2,s_1}^{-1}(\Delta x) = \frac{1}{v_{s_2} - v_{s_1}}.$$
Hence,
$$\begin{aligned}
& \Tilde{g}_{s_1,s_2,s_1}(\Delta x) =\frac{(\lambda_{s_1}-\lambda_{s_2})^2}{|v_{s_2} - v_{s_1}|}\cdot
\frac{-(\Delta x -v_{s_2} \Delta t)}{v_{s_2} - v_{s_1}}\cdot \frac{\exp{\left((\lambda_{s_1}-\lambda_{s_2})(\Delta x - v_{s_1} \Delta t)/(v_{s_2} - v_{s_1})\right)}}{\exp{((\lambda_{s_1}-\lambda_{s_2})\Delta t)}-1-(\lambda_{s_1}-\lambda_{s_2})\Delta t}.
\end{aligned}$$

\subsection{Computing \texorpdfstring{$\mathbb{P}(\Delta y\,|\,W=2, S_1=s_1, S_2=s_2, S_3=s_3)$}{} for \texorpdfstring{$s_3=s_1$}{}}
By convoluting with the PDF for $\Delta\epsilon$ we now incorporate noise
$$\begin{aligned}
    \Tilde{f}_{s_1,s_2,s_1}(\Delta y) :=&\ \mathbb{P}(\Delta y\,|\, W=1, S_1=s_1, S_2=s_2, S_3=s_1 ) 
    \\
    =&\ \int_{a}^b \Tilde{g}_{s_1,s_2, s_1}(\Delta x)f_{\mathcal{N}(0,2\sigma^2)}(\Delta y-\Delta x) \dd(\Delta x),
\end{aligned}$$
where 
$$a=a_{s_1,s_2,s_1}:= \min \{h_{s_1,s_2,s_1}(0), h_{s_1,s_2,s_1}(\Delta t)\},$$
$$b=b_{s_1,s_2,s_1}:= \max \{h_{s_1,s_2,s_1}(0), h_{s_1,s_2,s_1}(\Delta t)\},$$
and 
$$f_{\mathcal{N}(0,2\sigma^2)}(\Delta \epsilon):= \frac{1}{2\sigma \sqrt{\pi}}\exp\left({-\frac{(\Delta\epsilon)^2}{4\sigma^2}}\right).$$
We obtain
$$\begin{aligned}
\Tilde{f}_{s_1,s_2,s_1}(\Delta y)& = 
\hat{k}\int_a^b (-v_{s_2}\Delta t + \Delta x)\exp{(-c(\Delta x)^2 +r \Delta x +\hat{r})}\dd(\Delta x)
\\
& = \frac{-\hat{k} e^g}{4 c^{3/2}}\left[ \sqrt{\pi} e ^{\frac{r^2}{4c}}(2c(-v_{s_2}\Delta t)+r) \erf\left(\frac{2c\Delta x-r}{2\sqrt{c}}\right)-2\sqrt{c}e^{\Delta x (r-c\Delta x)}  \right]_{\Delta x=a}^{\Delta x = b}
\\
&=k \left[ \sqrt{\pi} e ^{\frac{r^2}{4c}}(2c(-v_{s_2}\Delta t)+r) \erf\left(\frac{2c\Delta x-r}{2\sqrt{c}}\right)-2\sqrt{c}e^{\Delta x (r-c\Delta x)}  \right]_{\Delta x=a}^{\Delta x = b},
\end{aligned}
$$
where
$$c:=\frac{1}{4\sigma^2},$$
$$
r=r_{s_1,s_2,s_1}(\Delta y) := 2c\Delta y + \frac{\lambda_{s_1} - \lambda_{s_2}}{v_{s_2} - v_{s_1}},$$
$$\hat{r} =\hat{r}_{s_1,s_2,s_1}(\Delta y):= -c(\Delta y)^2 + \frac{\lambda_{s_1} - \lambda_{s_2}}{|v_{s_2} - v_{s_1}|}|v_{s_1}| \Delta t,$$
$$\hat{k}=\hat{k}_{s_1,s_2,s_1} :=\frac{(\lambda_{s_1} - \lambda_{s_2})^2}{2\sigma\sqrt{\pi}|v_{s_2} - v_{s_1}|^2}\frac{\sign{(v_{s_2} - v_{s_1})}}{\exp\left({(\lambda_{s_1} -\lambda_{s_2})\Delta t}\right)-1-(\lambda_{s_1} -\lambda_{s_2})\Delta t},$$
and
$$\begin{aligned}
k=k_{s_1,s_2,s_1} :=&\ \frac{-\hat{k} e^{\hat{r}}}{4 c^{3/2}}
\\
=&\ \frac{\sigma^2 e^{\hat{r}}(\lambda_{s_1} - \lambda_{s_2})^2}{\sqrt{\pi}|v_{s_2} - v_{s_1}|^2}\frac{-\sign{(v_{s_2} - v_{s_1})}}{\exp\left({(\lambda_{s_1} -\lambda_{s_2})\Delta t}\right)-1-(\lambda_{s_1} -\lambda_{s_2})\Delta t}.
\end{aligned}$$

\subsection{Comparison of the error of the up-to-one-switch approximation with the error of the up-to-two-switch approximation}

\begin{figure*}[!ht]
    \begin{minipage}{0.06\textwidth}
    \textcolor{white}{-}
    \end{minipage}%
    \begin{minipage}{0.47\textwidth}
    \centering
    \begin{minipage}{0.6\textwidth}
    \small \centering
    \textbf{Two-state model}
    \end{minipage}%
    \begin{minipage}{0.3\textwidth}
    \includegraphics[width=1\textwidth]{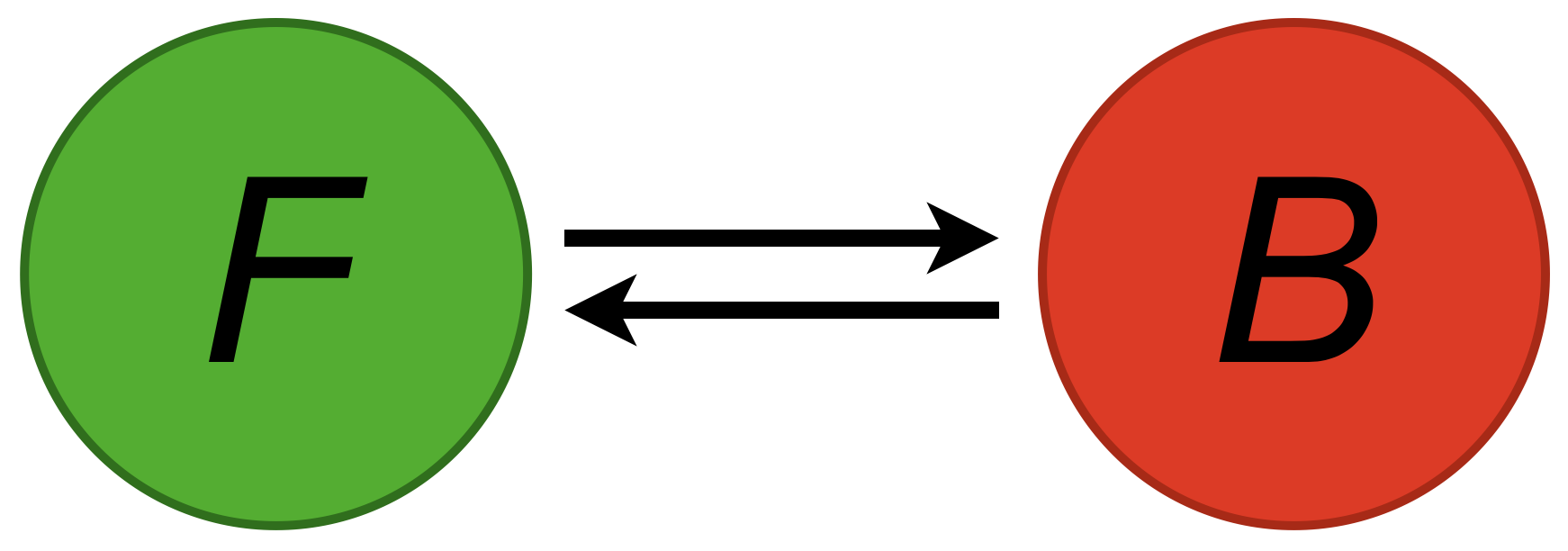}
    \end{minipage}
    \end{minipage}%
    \begin{minipage}{0.47\textwidth}
    \centering
    \begin{minipage}{0.6\textwidth}
    \small \centering 
    \textbf{Three-state model}
    \end{minipage}%
    \begin{minipage}{0.3\textwidth}
    \includegraphics[width=1\textwidth]{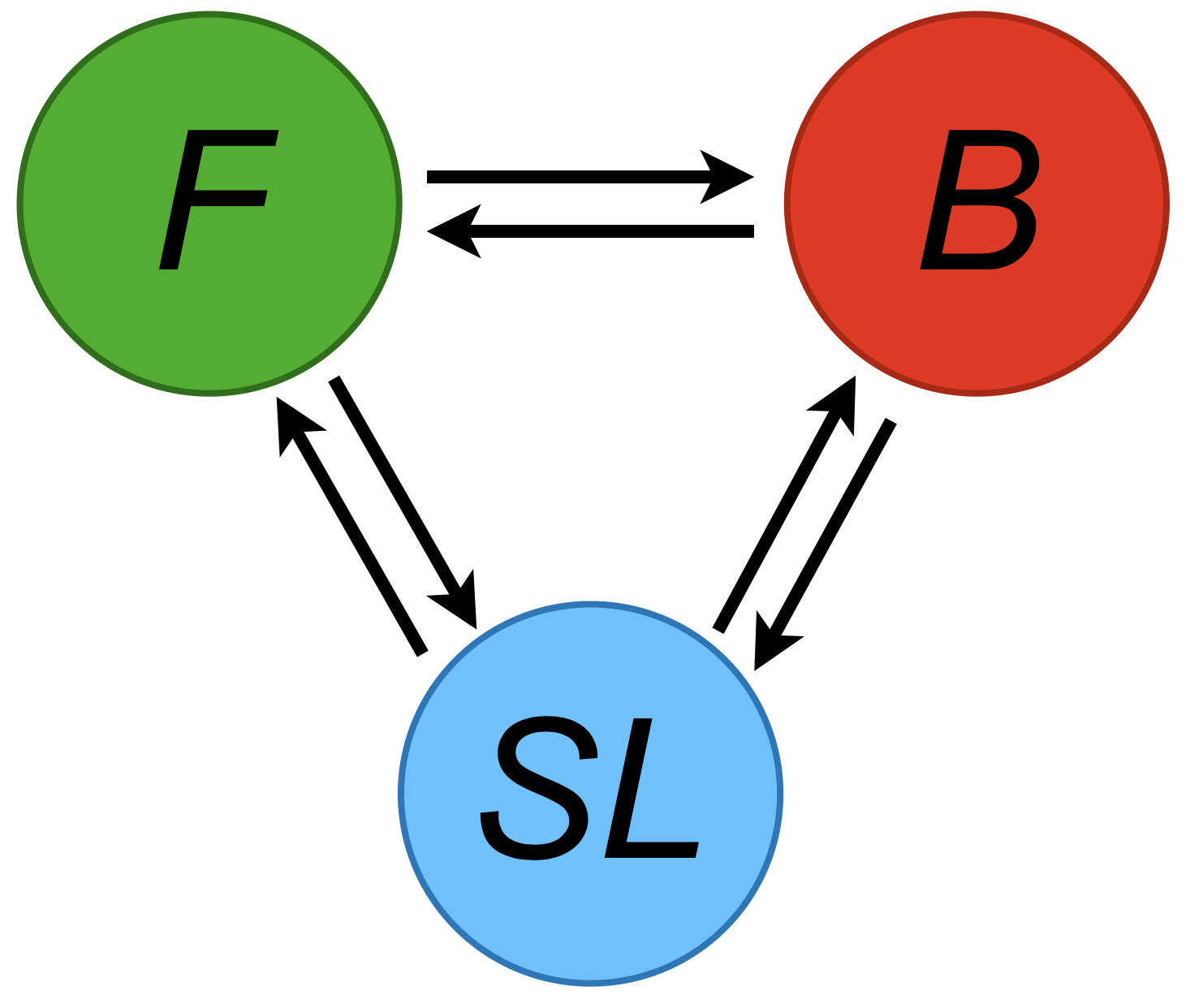}
    \end{minipage}
    \end{minipage}
    \begin{minipage}{0.06\textwidth}
    \centering
    $\boldsymbol{\lambda}$
    \end{minipage}%
    \begin{minipage}{0.47\textwidth}
    \includegraphics[width=1\textwidth]{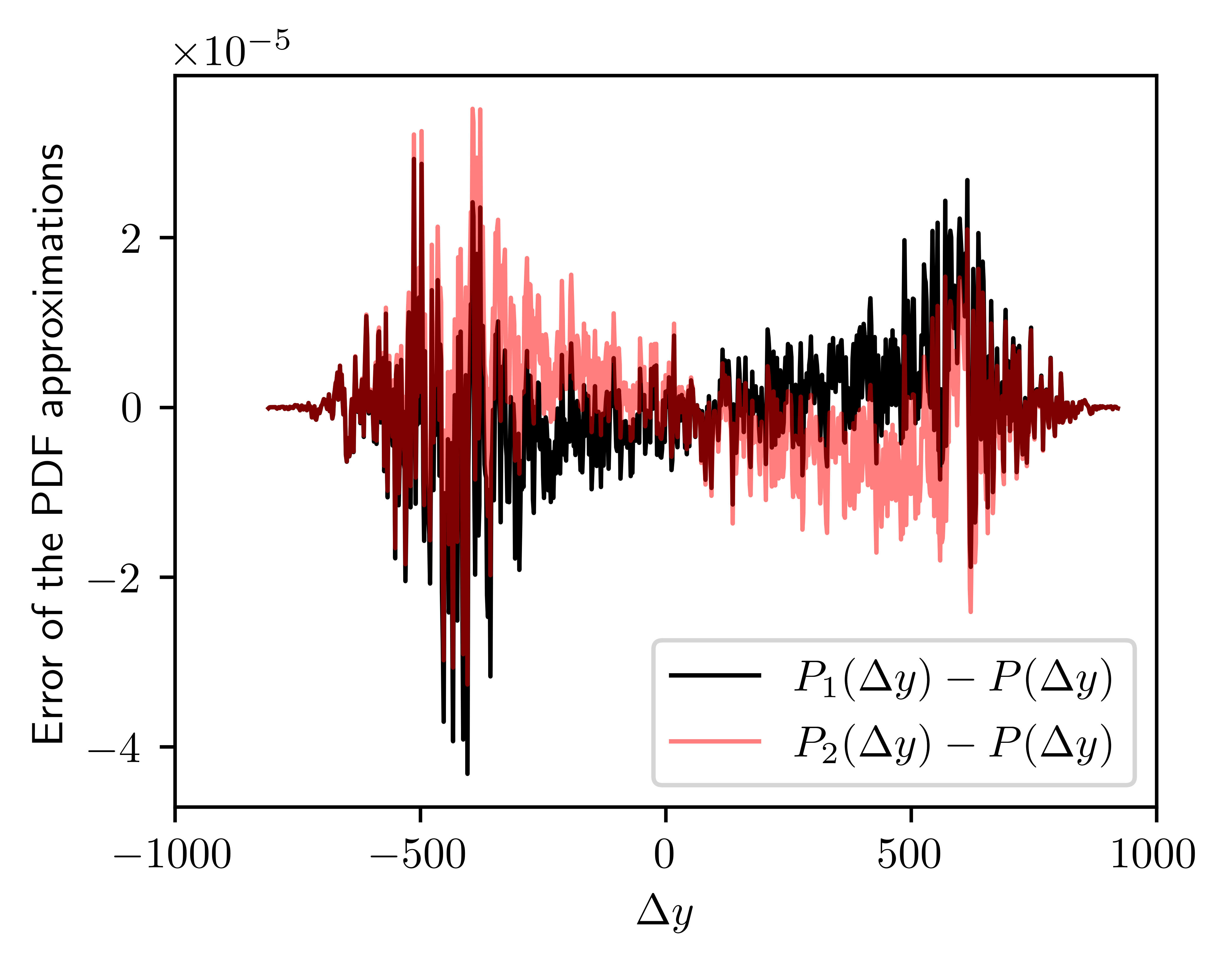}
    \end{minipage}%
    \begin{minipage}{0.47\textwidth}
    \includegraphics[width=1\textwidth]{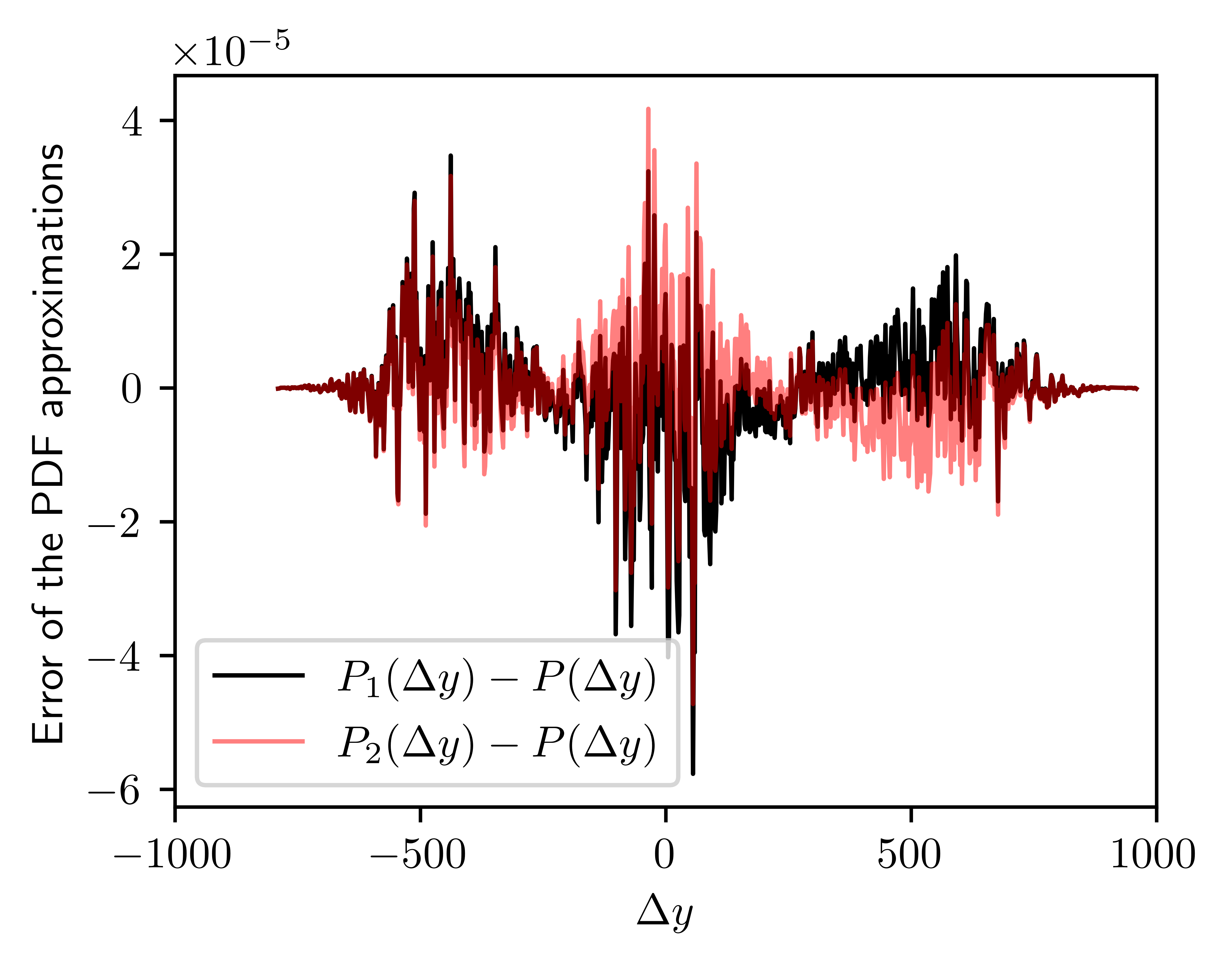}
    \end{minipage}
    \begin{minipage}{0.06\textwidth}
    \centering
    $5\boldsymbol{\lambda}$
    \end{minipage}%
    \begin{minipage}{0.47\textwidth}
    \includegraphics[width=1\textwidth]{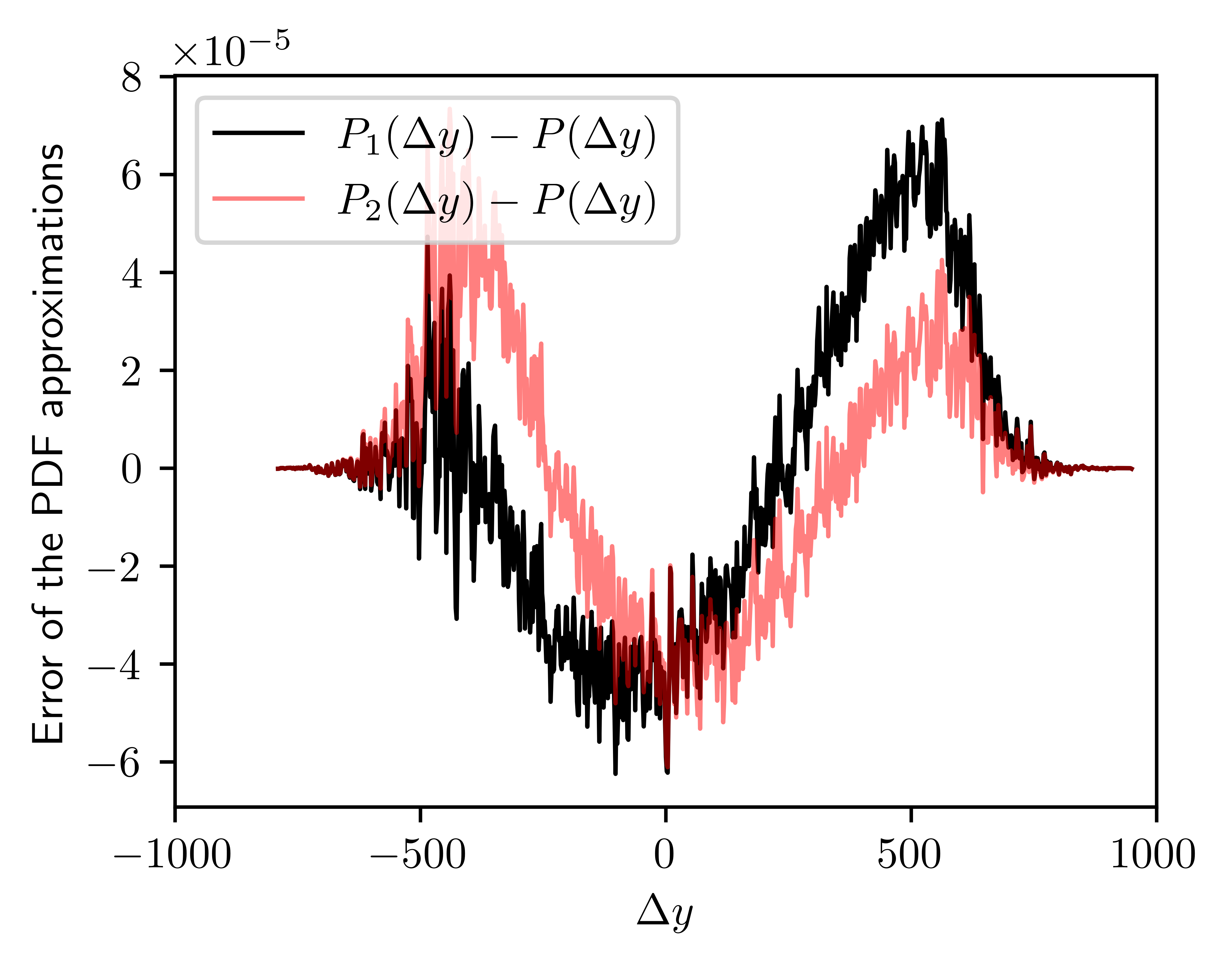}
    \end{minipage}%
    \begin{minipage}{0.47\textwidth}
    \includegraphics[width=1\textwidth]{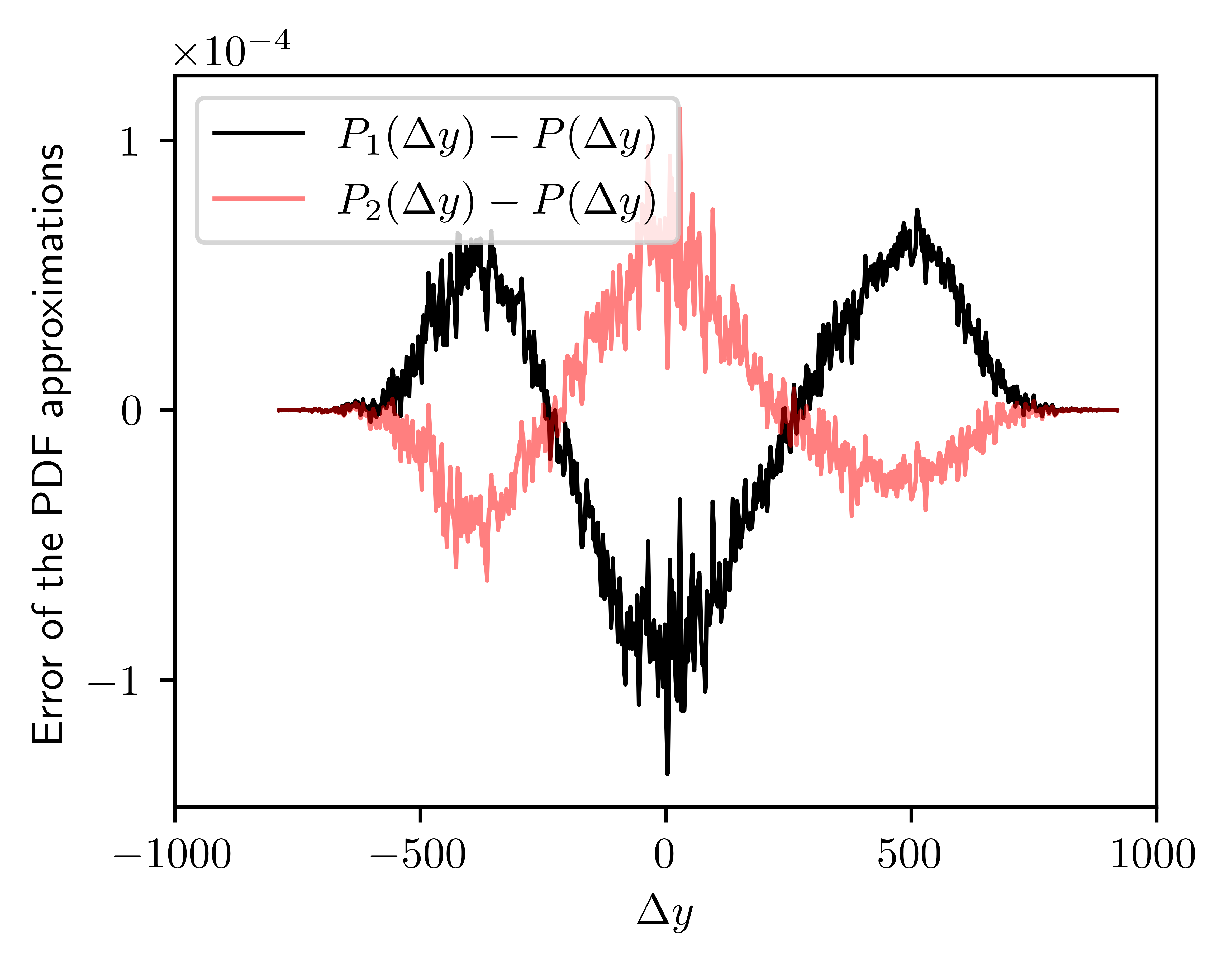}
    \end{minipage}
    \begin{minipage}{0.06\textwidth}
    \centering
    $10\boldsymbol{\lambda}$
    \end{minipage}%
    \begin{minipage}{0.47\textwidth}
    \includegraphics[width=1\textwidth]{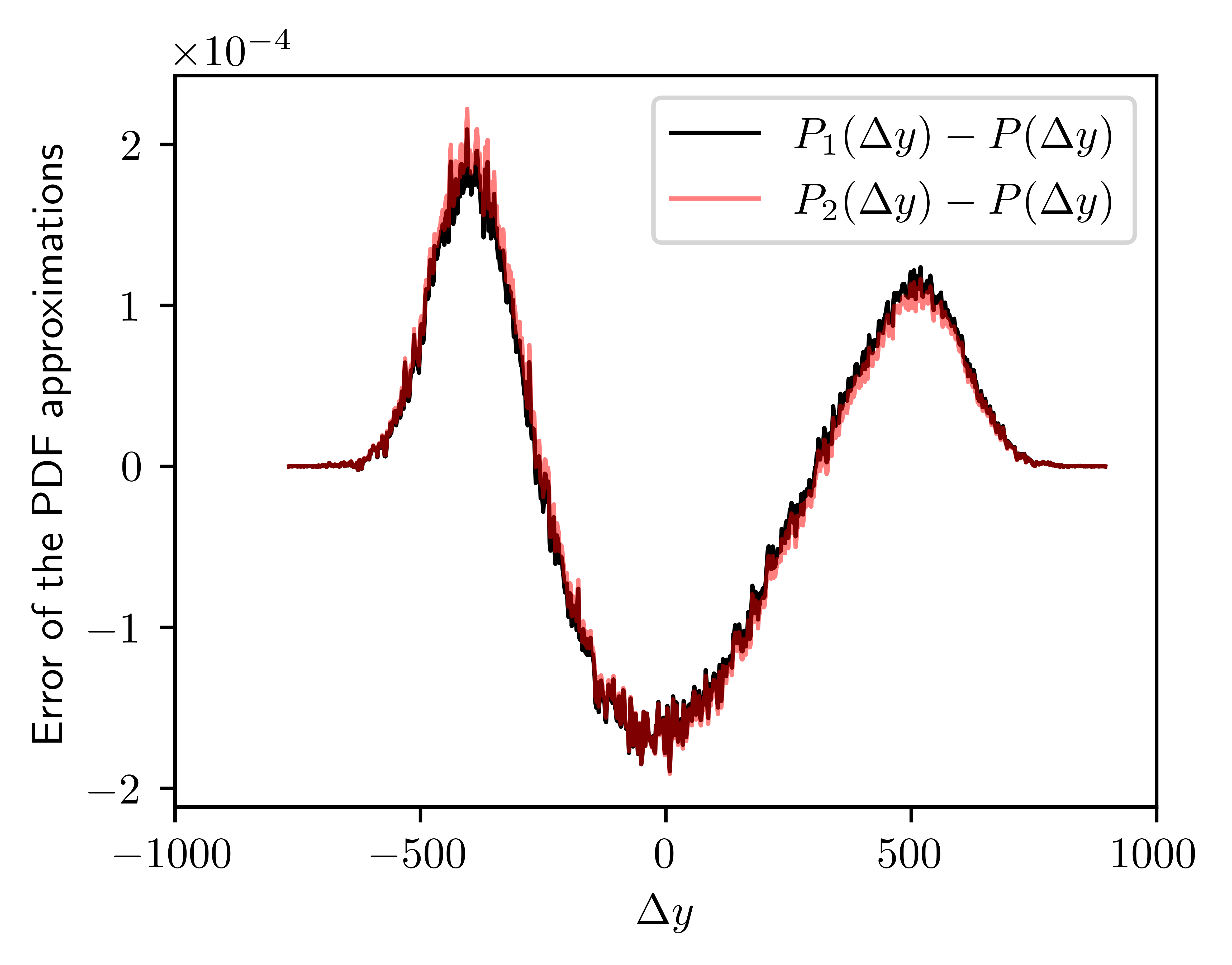}
    \end{minipage}%
    \begin{minipage}{0.47\textwidth}
    \includegraphics[width=1\textwidth]{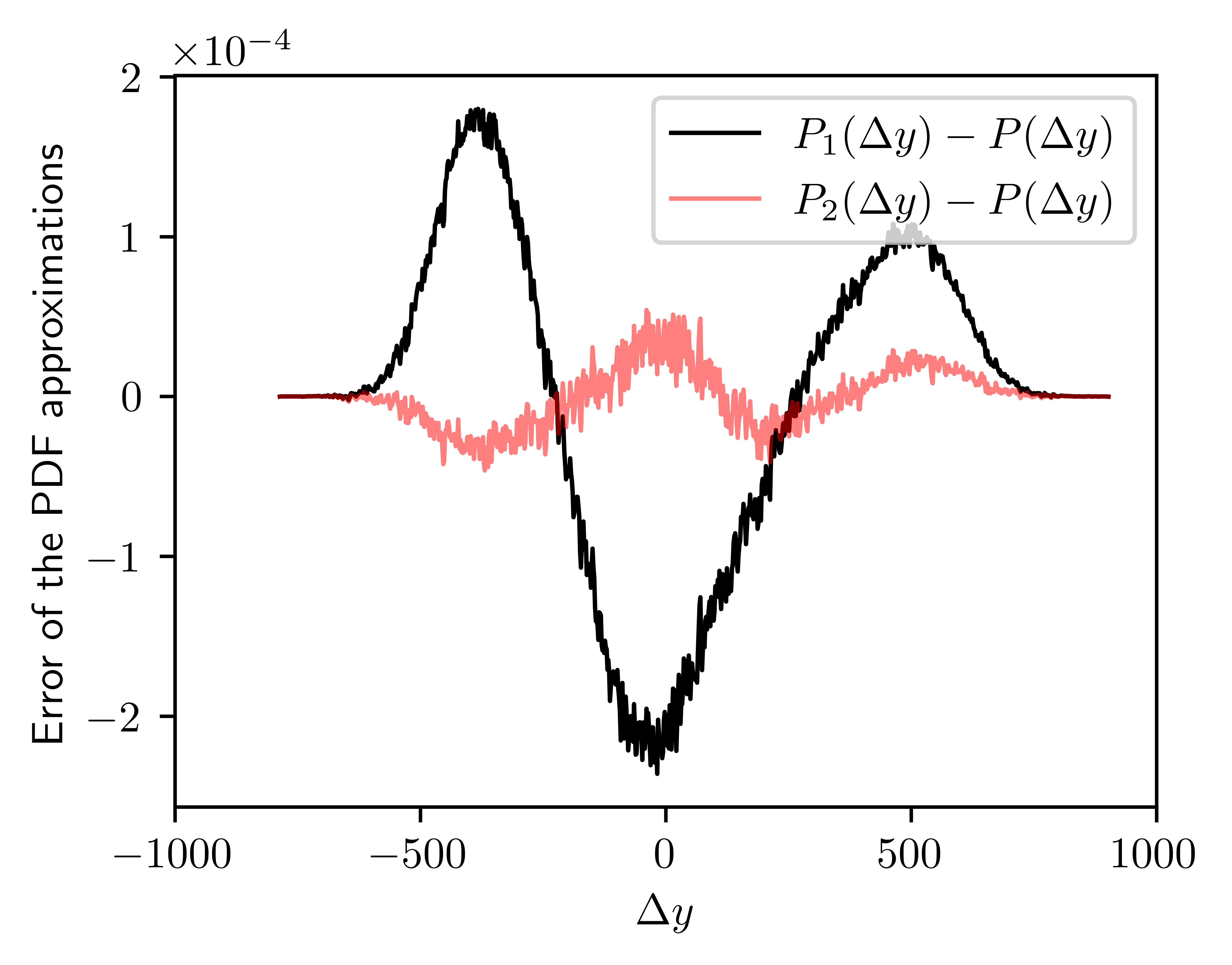}
    \end{minipage}
    \caption{The error of the approximation $P_1(\Delta y)$, defined as $P_1(\Delta y)-P(\Delta y)$ where $P(\Delta y)$ is the
    empirical PDF for $\Delta y$, is compared with the error of the approximation $P_1(\Delta y)$, defined as $P_1(\Delta y)-P(\Delta y)$. The panels at the top are obtained using the parameters as specified in the Supplementary Information Figure~\ref{SI:Fig:networks}, while the panels in the middle and at the bottom are obtained with the same parameters except the rates which are multiplied by 5 and 10, respectively.}
    \label{SI:Fig:Comparison_error_PDFs_2_3}
\end{figure*}

Figure~\ref{SI:Fig:Comparison_error_PDFs_2_3} shows a comparison of the up-to-one-switch approximation error, $P_1(\Delta y)-P(\Delta y)$, with the up-to-two-switch approximation error, $P_1(\Delta y)-P(\Delta y)$, for the two-state model and the three-state model.

\begin{figure*}[!ht]
    \centering
    \begin{minipage}{0.5\textwidth}
    \centering
    \textbf{\small Two-state model}
    \end{minipage}%
    \begin{minipage}{0.5\textwidth}
    \centering
    \textbf{\small Three-state model}
    \end{minipage}
    \begin{minipage}{0.5\textwidth}
    \includegraphics[width=1\textwidth]{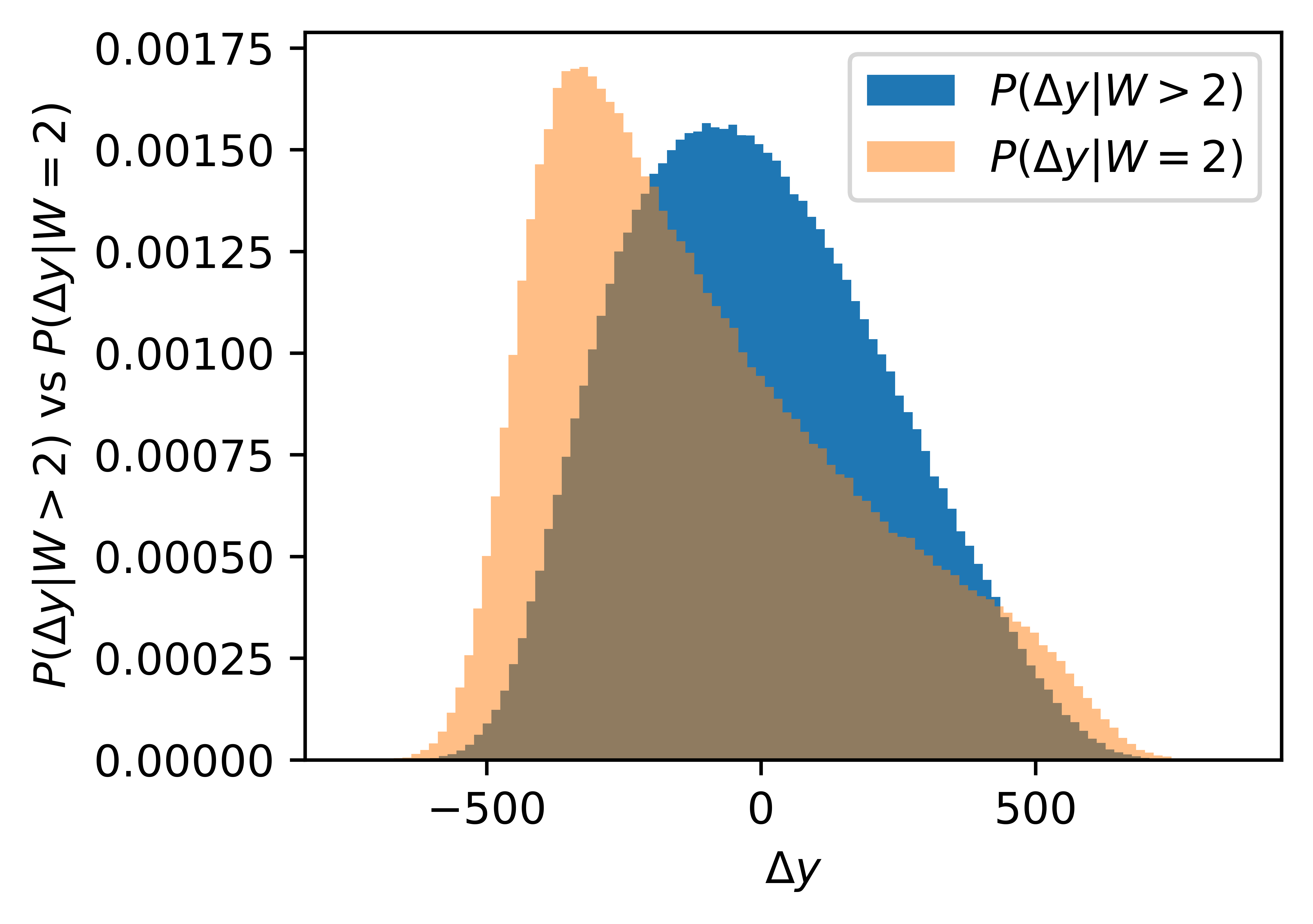}
    \end{minipage}%
    \begin{minipage}{0.5\textwidth}
    \includegraphics[width=1\textwidth]{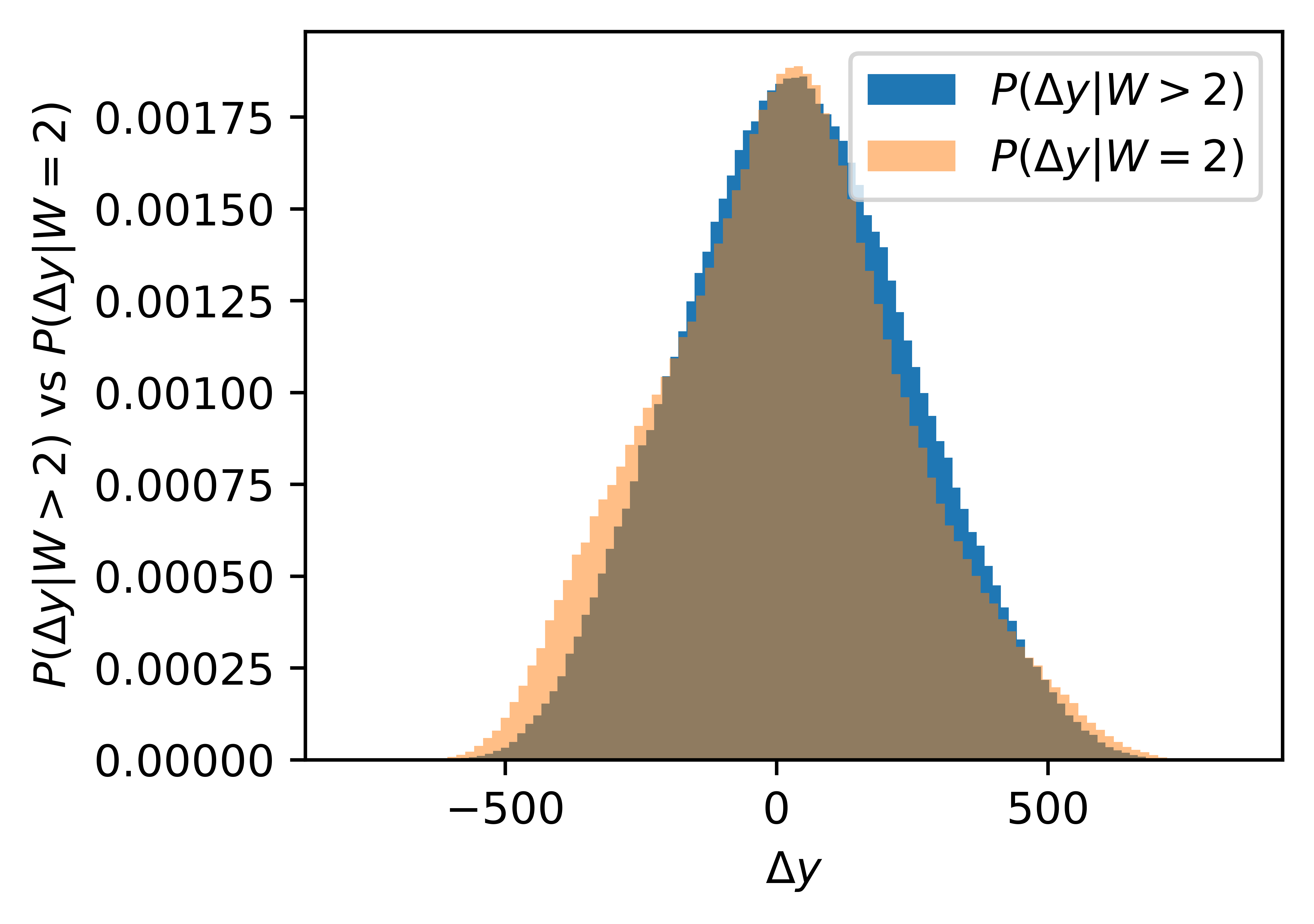}
    \end{minipage}
    \caption{\textcolor{black}{Comparison of the empirical distributions for $\mathbb{P}(\Delta y|W=2)$ and $\mathbb{P}(\Delta y|W>2)$, using the rates $10\boldsymbol{\lambda}$ for $\boldsymbol{\lambda}$ specified in Supplementary Information Figure~\ref{SI:Fig:networks}, for the two-state model (from  Figure~\ref{Fig:in silico_track_examples}\textbf{A}) and three-state model (from Figure~\ref{Fig:in silico_track_examples}\textbf{B}).}}
    \label{SI:Fig:newfig}
\end{figure*}

\textcolor{black}{Figure~\ref{SI:Fig:newfig} shows a comparison between the empirical distributions for $\mathbb{P}(\Delta y|W=2)$ and $\mathbb{P}(\Delta y|W>2)$, using the rates $10\boldsymbol{\lambda}$ for $\boldsymbol{\lambda}$ specified in Supplementary Information Figure~\ref{SI:Fig:networks}, for the two-state model (from  Figure~\ref{Fig:in silico_track_examples}\textbf{A}) and three-state model (from Figure~\ref{Fig:in silico_track_examples}\textbf{B}). Comparing the two panel suggests that the approximation used ($\mathbb{P}(\Delta y\,|\,W>2)\approx \mathbb{P} (\Delta y\,|\,W=2)$) is less accurate for the two-state model in this parameter regime, compared to the three-state model. And this causes the up-to-two-switch approximation to be more accurate for the three-state model (panel \textbf{F} in Figure~\ref{Fig:PDFs_2_3}) than for the two-state model in (panel \textbf{E} in Figure~\ref{Fig:PDFs_2_3}) for frequent switching.}

\section{Up-to-one-switch approximation for the probability distribution function of a set of noisy subsequent location increments }\label{SI:Sec:P(track)}

In this section, we extend the results obtained in Section~\ref{Subsec:up-to-one-switch} to compute an approximation for the PDF a set of $N$ noisy subsequent location increments $\mathbb{P}(\boldsymbol{\Delta y}_N)$. Here, we compute explicitly all the results obtained in Section~\ref{Subsec:up-to-two-switch}. The up-to-two-switch approximation of a single location increment is provided as a Python code in the function \textit{approx\_pdf\_track\_up\_to\_1\_switch} in the file \textit{functions.py}.

The PDF $\mathbb{P}(\boldsymbol{\Delta y}_N)$ is the joint distribution of a set of subsequent location increments, and it can be rewritten in terms of conditional distributions
\begin{equation*}
\begin{aligned}
    \mathbb{P}(\boldsymbol{\Delta y}_N) = \mathbb{P}(\Delta y_1)\mathbb{P}(\Delta y_2\,|\,\boldsymbol{\Delta y}_1)\cdots \mathbb{P}(\Delta y_{N-1}\,|\,\boldsymbol{\Delta y}_{N-2}) \mathbb{P}(\Delta y_N \,|\, \boldsymbol{\Delta y}_{N-1}),
\end{aligned}
\end{equation*}
where we note that, for an increment $\Delta y$, the PDF $\mathbb{P}(\Delta y)$ is its marginal distribution, approximated in Section~\ref{Subsec:up-to-one-switch} and Section~\ref{Subsec:up-to-two-switch}. We obtain that for any $N\ge 2$ we can approximate the joint distribution of a set of subsequent location increments using a recursive method. We denote $S^j_{i}$ the $i$-th state attained and $W^j$ the number of switches the during the $j$-th measured interval of the track (as in Figure~\ref{Fig:states_diagram_track}).

We compute an iterative formula for
$\mathbb{P}(\Delta y_N \,|\, \boldsymbol{\Delta y}_{N-1})$
that involves the approximations for 
$$\mathbb{P}(\Delta y_{N} \,|\,S^{N}_1=s^{N}_1),\ \mathbb{P}(W^{N}=0 \,|\,S^{N}_1=s^{N}_1), \ \mathbb{P}(W^{N}\ge 1 \,|\,S^{N}_1=s^{N}_1),\ \mathbb{P}(S^{N}_1=s^{N}_1 \,|\, \boldsymbol{\Delta y}_{N-1}),$$
using $\mathbb{P}(\Delta y_{N-1} \,|\, \boldsymbol{\Delta y}_{N-2})$ and 
$$\begin{aligned}
        &\mathbb{P}(\Delta y_{N-1} \,|\,S^{N-1}_1=s^{N-1}_1),\ \mathbb{P}(W^{N-1}=0 \,|\,S^{N-1}_1),
        \\& \mathbb{P}(W^{N-1}\ge 1 \,|\,S^{N-1}_1=s^{N-1}_1),\ \mathbb{P}(S^{N-1}_1=s^{N-1}_1 \,|\, \boldsymbol{\Delta y}_{N-2}).
    \end{aligned}$$
We note that, for $N=2$, $\mathbb{P}(\Delta y_{N-1} \,|\, \boldsymbol{\Delta y}_{N-2})$ is the marginal PDF, for which we have computed approximations. For simplicity, we use the up-to-one-switch approximation presented in Section~\ref{Subsec:up-to-one-switch} which we denote by $P_1$, however, the result could be extended to incorporate more switches. 

Using the state at the beginning of the $N$-th interval $S^N_1$, we write 
$$\begin{aligned}
    \mathbb{P}(\Delta y_N \,|\, \boldsymbol{\Delta y}_{N-1})&=\sum_{s^N_1=1}^n\mathbb{P}(\Delta y_N \,|\, S^N_1=s^{N}_1, \boldsymbol{\Delta y}_{N-1})\mathbb{P}(S^N_1=s^{N}_1 \,|\, \boldsymbol{\Delta y}_{N-1}).
\end{aligned}$$
In the exact increments form, we have
$$\mathbb{P}(\Delta x_N \,|\, S^N_1=s^{N}_1, \boldsymbol{\Delta x}_{N-1})= \mathbb{P}(\Delta x_N \,|\, S^N_1=s^{N}_1),$$
which follows from the Markov property of the continuous-time process, and holds since the information on $\Delta x_N$ given by the previous increments $\boldsymbol{\Delta x}_{N-1}$ is fully expressed by the final state at the end of the $(N-1)$-th interval which corresponds to the one at the beginning of the $N$-th interval $S^N_1$. In case of the noisy increments, we use the approximation
\begin{equation}\label{Eq:Approx_no_noise_1}
    \mathbb{P}(\Delta y_N \,|\, S^N_1=s^N_1, \boldsymbol{\Delta y}_{N-1})\approx \mathbb{P}(\Delta y_N \,|\, S^N_1=s^N_1),
\end{equation}
which is not exact since the previous increments contain additional information on the noise $\epsilon_{N-1}$. Hence, we obtain
$$\mathbb{P}(\Delta y_N \,|\, \Delta y_{N-1}, \ldots, \Delta y_2, \Delta y_1)\approx\sum_{s^N_1=1}^n\mathbb{P}(\Delta y_N \,|\, S^N_1=s^{N}_1)\mathbb{P}(S^N_1=s^{N}_1 \,|\, \boldsymbol{\Delta y}_{N-1}).$$

We obtain $\mathbb{P}(S^N_1=s^{N}_1 \,|\, \boldsymbol{\Delta y}_{N-1})$ by using the state at the beginning of the previous interval $S^{N-1}_1$, 
$$\begin{aligned}
    \mathbb{P}(S^N_1=s^{N}_1 \,|\, \boldsymbol{\Delta y}_{N-1}) = \sum_{s^{N-1}_1=1}^n\mathbb{P}(S^N_1=s^{N}_1 \,|\,S^{N-1}_1=s^{N-1}_1, \boldsymbol{\Delta y}_{N-1})\mathbb{P}(S^{N-1}_1=s^{N-1}_1 \,|\, \boldsymbol{\Delta y}_{N-1}).
\end{aligned}$$
Here, we note that the state $S^{N-1}_1$ given $\boldsymbol{\Delta x}_{N-1}$ only depends on the the increment $\Delta x_{N-1}$; thus we obtain the property
$$\mathbb{P}(S^N_1=s^{N}_1 \,|\,S^{N-1}_1=s^{N-1}_1, \boldsymbol{\Delta x}_{N-1})= \mathbb{P}(S^N_1=s^{N}_1 \,|\,S^{N-1}_1=s^{N-1}_1, \Delta x_{N-1}),$$
which again follows from the Markov property of the continuous-time process, and it holds since the state $S^{N-1}_1$ given $\boldsymbol{\Delta x}_{N-1}$ only depends on the the increment $\Delta x_{N-1}$.
For the noisy increment $\Delta y$ we use the approximation (similar to Equation~\eqref{Eq:Approx_no_noise_1})
\begin{equation}\label{Eq:Approx_no_noise_2}
    \mathbb{P}(S^N_1=s^{N}_1 \,|\,S^{N-1}_1=s^{N-1}_1, \boldsymbol{\Delta y}_{N-1})\approx \mathbb{P}(S^N_1=s^{N}_1 \,|\,S^{N-1}_1=s^{N-1}_1, \Delta y_{N-1}),
\end{equation}
to write
$$\begin{aligned}
    \mathbb{P}(S^N_1=s^{N}_1 \,|\, \boldsymbol{\Delta y}_{N-1})
    \approx &\ \sum_{s^{N-1}_1=1}^n\mathbb{P}(S^N_1=s^{N}_1 \,|\,S^{N-1}_1=s^{N-1}_1, \Delta y_{N-1})\mathbb{P}(S^{N-1}_1 \,|\, \boldsymbol{\Delta y}_{N-1})
    \\=&\ \sum_{s^{N-1}_1=1}^n\frac{\mathbb{P}(\Delta y_{N-1}, S^N_1=s^{N}_1 \,|\,S^{N-1}_1=s^{N-1}_1)}{\mathbb{P}(\Delta y_{N-1} \,|\,S^{N-1}_1=s^{N-1}_1)}\mathbb{P}(S^{N-1}_1=s^{N-1}_1 \,|\, \boldsymbol{\Delta y}_{N-1}),
\end{aligned}$$
where the equality is obtained by definition of conditional probability. The denominator ${\mathbb{P}(\Delta y_{N-1} \,|\,S^{N-1}_1=s^{N-1}_1)}$ will be simplified later. Moreover, using Bayes' theorem, we write
$$
\begin{aligned}
    \mathbb{P}(S^{N-1}_1=s^{N-1}_1 \,|\, \boldsymbol{\Delta y}_{N-1}) &= \mathbb{P}( \Delta y_{N-1}\,|\, S^{N-1}_1=s^{N-1}_1, \boldsymbol{\Delta y}_{N-2})\frac{\mathbb{P}(S^{N-1}_1=s^{N-1}_1 \,|\, \boldsymbol{\Delta y}_{N-2})}{\mathbb{P}( \Delta y_{N-1}\,|\, \boldsymbol{\Delta y}_{N-2})}
    \\
    &=\mathbb{P}( \Delta y_{N-1}\,|\, S^{N-1}_1=s^{N-1}_1)\frac{\mathbb{P}(S^{N-1}_1=s^{N-1}_1 \,|\, \boldsymbol{\Delta y}_{N-2})}{\mathbb{P}( \Delta y_{N-1}\,|\, \boldsymbol{\Delta y}_{N-2})},
\end{aligned}$$
where the second equality is obtained using the property in Equation~\eqref{Eq:Approx_no_noise_1} (at the $(N-1)$-th step).
By substitution, and, by simplifying $\mathbb{P}(\Delta y_{N-1} \,|\,S^{N-1}_1=s^{N-1}_1)$, we obtain
$$\begin{aligned}
    \mathbb{P}(S^N_1=s^{N}_1 \,|\, \boldsymbol{\Delta y}_{N-1}) \approx \sum_{s^{N-1}_1=1}^n\mathbb{P}(\Delta y_{N-1}, S^N_1=s^{N}_1 \,|\,S^{N-1}_1=s^{N-1}_1)\frac{\mathbb{P}(S^{N-1}_1=s^{N-1}_1 \,|\, \boldsymbol{\Delta y}_{N-2})}{\mathbb{P}( \Delta y_{N-1}\,|\, \boldsymbol{\Delta y}_{N-2})}.
\end{aligned}$$
Here, the numerator and denominator of the fractions are computed at the $(N-1)$-th induction step.

Finally, we need to compute $\mathbb{P}(\Delta y_{N-1}, S^N_1=s^{N}_1 \,|\,S^{N-1}_1=s^{N-1}_1)$ by conditioning on the number of switches during the $(N-1)$-th interval, $W^{N-1}$, and we use the definition of conditional probability to obtain
\begin{equation}\label{SI:Eq:Track_P_inf}
    \begin{aligned}
    \mathbb{P}(\Delta y_{N-1}, S^N_1=s^{N}_1 \,|\,S^{N-1}_1=s^{N-1}_1)=  \sum_{w=0}^{\infty}&\ \mathbb{P}(\Delta y_{N-1}\,|\, S^N_1=s^{N}_1, S^{N-1}_1=s^{N-1}_1, W^{N-1}=w)\\&
    \times \mathbb{P}(S^N_1=s^{N}_1\,|\,S^{N-1}_1=s^{N-1}_1, W^{N-1}=w)
    \\&
    \times \mathbb{P}(W^{N-1}=w \,|\,S^{N-1}_1=s^{N-1}_1).
\end{aligned}
\end{equation}
We use the up-to-one-switch approximation and get
\begin{equation}\label{SI:Eq:Use_P_1}
\begin{aligned}
    &\mathbb{P}(\Delta y_{N-1}, S^N_1=s^{N}_1 \,|\,S^{N-1}_1=s^{N-1}_1)
    \\
    &\approx P_1(\Delta y_{N-1}, S^N_1=s^{N}_1 \,|\,S^{N-1}_1=s^{N-1}_1)
    \\
    &
    =\begin{dcases}
    \begin{aligned}
        &\mathbb{P}(\Delta y_{N-1}\,|\, S^N_1=s^{N}_1, S^{N-1}_1=s^{N-1}_1, W^{N-1}=0)
        \\
        & \times\mathbb{P}(W^{N-1}=0 \,|\,S^{N-1}_1=s^{N-1}_1),
    \end{aligned}
     & \text{if } s^{N}_1=s^{N-1}_1,
    \\
    \begin{aligned}
        &\mathbb{P}(\Delta y_{N-1}\,|\, S^N_1=s^{N}_1, S^{N-1}_1=s^{N-1}_1, W^{N-1}=1)
        \\
        &\times \mathbb{P}(S^{N-1}_2=s^{N}_1\,|\,S^{N-1}_1=s^{N-1}_1)\mathbb{P}(W^{N-1}\ge 1 \,|\,S^{N-1}_1=s^{N-1}_1),
    \end{aligned}
    & \text{if } s^{N}_1\ne s^{N-1}_1,
\end{dcases}
\end{aligned}
\end{equation}
since if $W^{N-1}=0$, then $S^{N}_1=S^{N-1}_1$ which gives a unitary probability of keeping the same state
$$\mathbb{P}(S^N_1=s^{N}_1\,|\,S^{N-1}_1=s^{N-1}_1, W^{N-1}=0)=1,$$
while if $W^{N-1}=1$, then we obtain $S^{N}_1=S^{N-1}_2\ne S^{N-1}_1$ and in the usual notation we have rewritten
$$\mathbb{P}(S^N_1=s^{N}_1\,|\,S^{N-1}_1=s^{N-1}_1, W^{N-1}=1)=\mathbb{P}(S^{N-1}_2=s^{N}_1\,|\,S^{N-1}_1=s^{N-1}_1).$$

\subsection{Error of the up-to-one-switch approximation for the probability distribution function of two noisy subsequent location increments}

Figure~\ref{SI:Fig:Err_P(Delta y_1, Delta y_2)} compares the error of the up-to-one-switch approximation for the PDF of two noisy subsequent location increments, defined as $|P_1(\Delta y_1,\Delta y_2)-P(\Delta y_1,\Delta y_2)|$, with the error for the approximation consisting of the product of the marginals, defined as $|P_1(\Delta y_1)P_1(\Delta y_2)-P(\Delta y_1,\Delta y_2)|$.

\begin{figure*}
    \centering
    \begin{minipage}{0.9\textwidth}
    \begin{minipage}{0.16\textwidth}
    \small \centering 
    \textbf{Two-state model}
    \end{minipage}%
    \begin{minipage}{0.14\textwidth}
    \includegraphics[width=\textwidth]{IMAGES/TRACKS/2.two-state_model_simplified.png}
    \end{minipage}%
    \begin{minipage}{0.35\textwidth}
    \includegraphics[width=\textwidth]{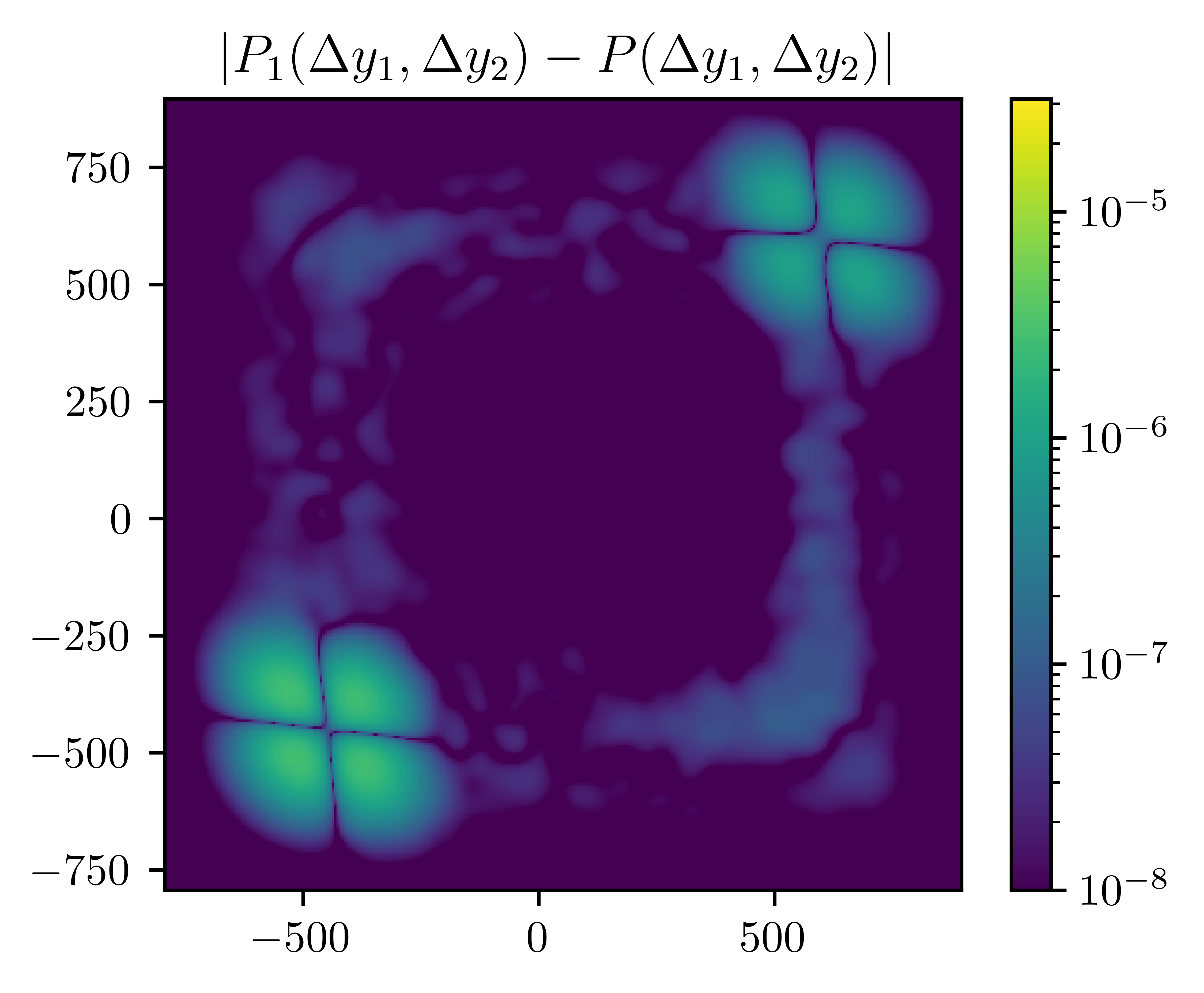}
    \end{minipage}%
    \begin{minipage}{0.35\textwidth}
    \includegraphics[width=\textwidth]{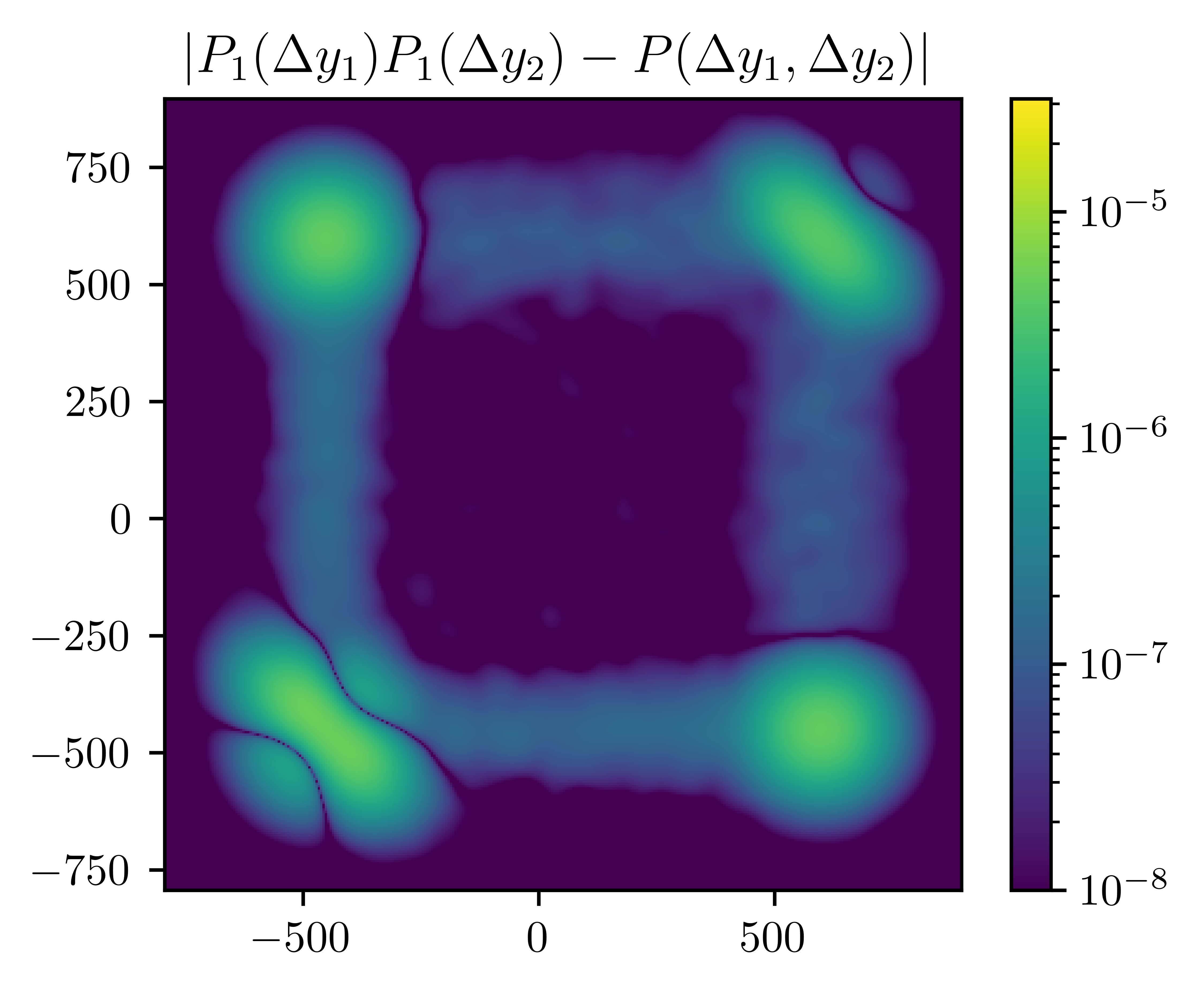}
    \end{minipage}
    \begin{minipage}{0.16\textwidth}
    \small \centering 
    \textbf{Two-state model}
    \end{minipage}%
    \begin{minipage}{0.14\textwidth}
    \includegraphics[width=\textwidth]{IMAGES/TRACKS/2.two-state_model_simplified.png}
    \centering
    $10\boldsymbol{\lambda}$
    \end{minipage}%
    \begin{minipage}{0.35\textwidth}
    \includegraphics[width=\textwidth]{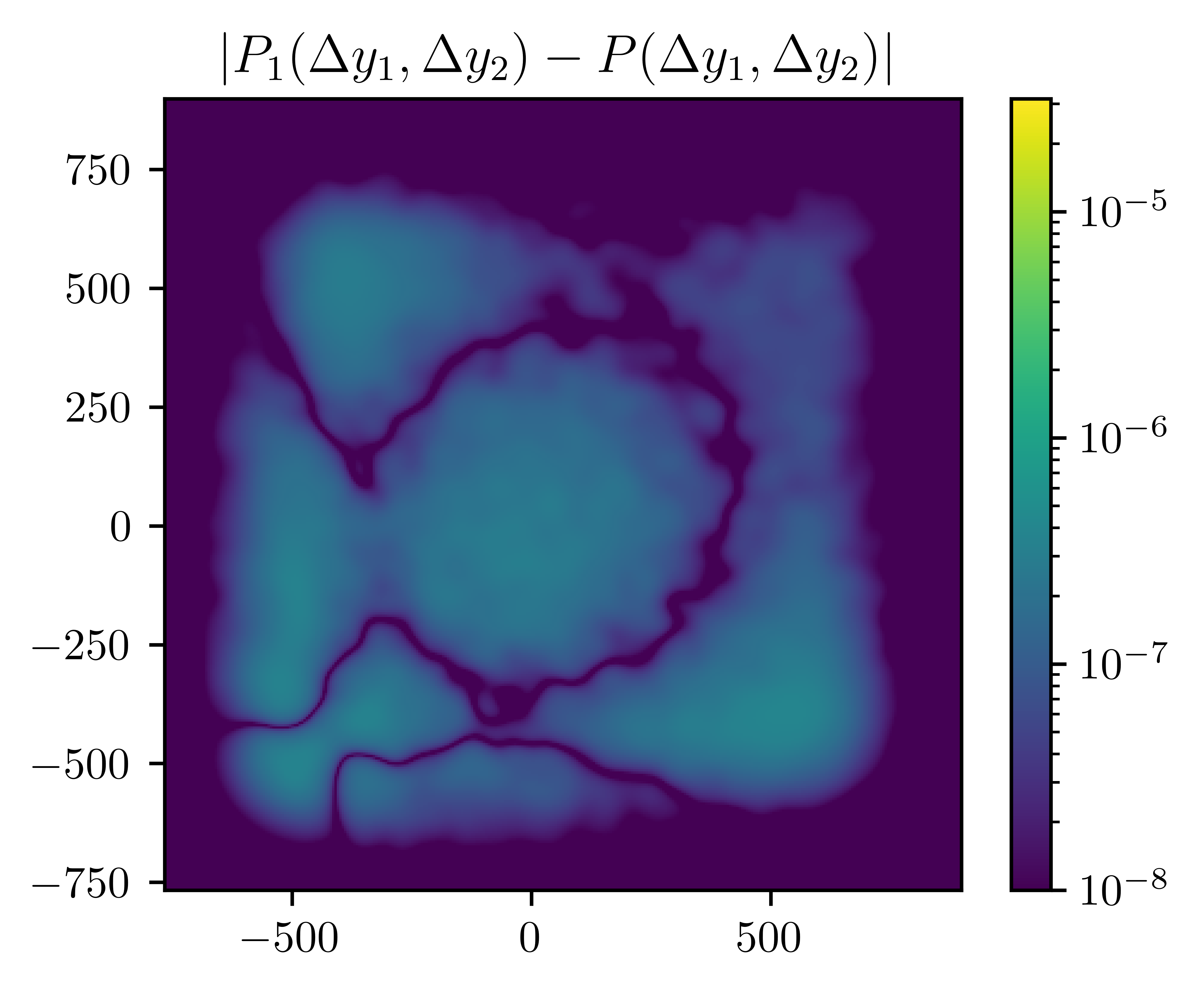}
    \end{minipage}%
    \begin{minipage}{0.35\textwidth}
    \includegraphics[width=\textwidth]{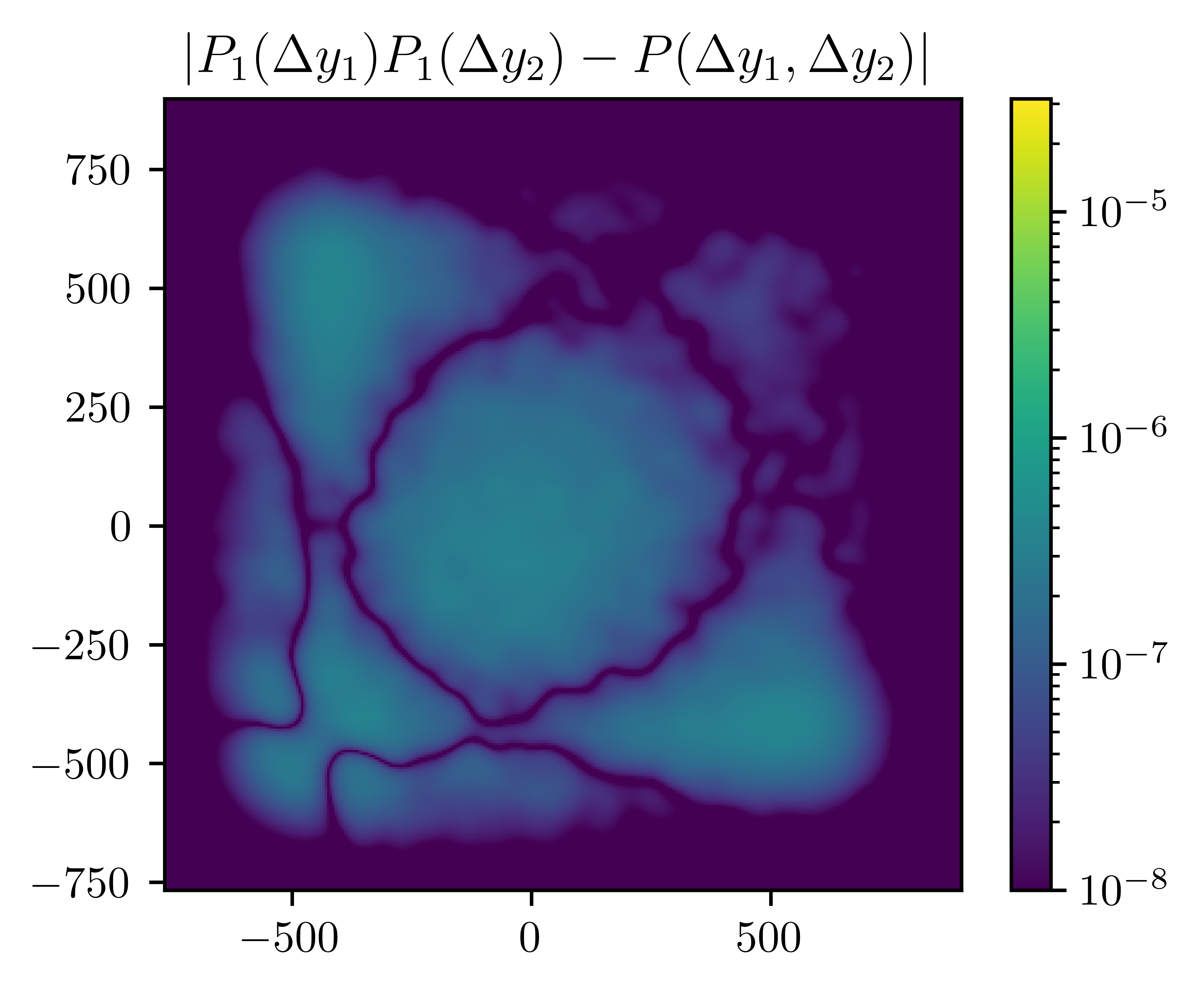}
    \end{minipage}
    \begin{minipage}{0.16\textwidth}
    \small \centering 
    \textbf{Three-state model}
    \end{minipage}%
    \begin{minipage}{0.14\textwidth}
    \includegraphics[width=\textwidth]{IMAGES/TRACKS/3.three-state_model_simplified.png}
    \end{minipage}%
    \begin{minipage}{0.35\textwidth}
    \includegraphics[width=\textwidth]{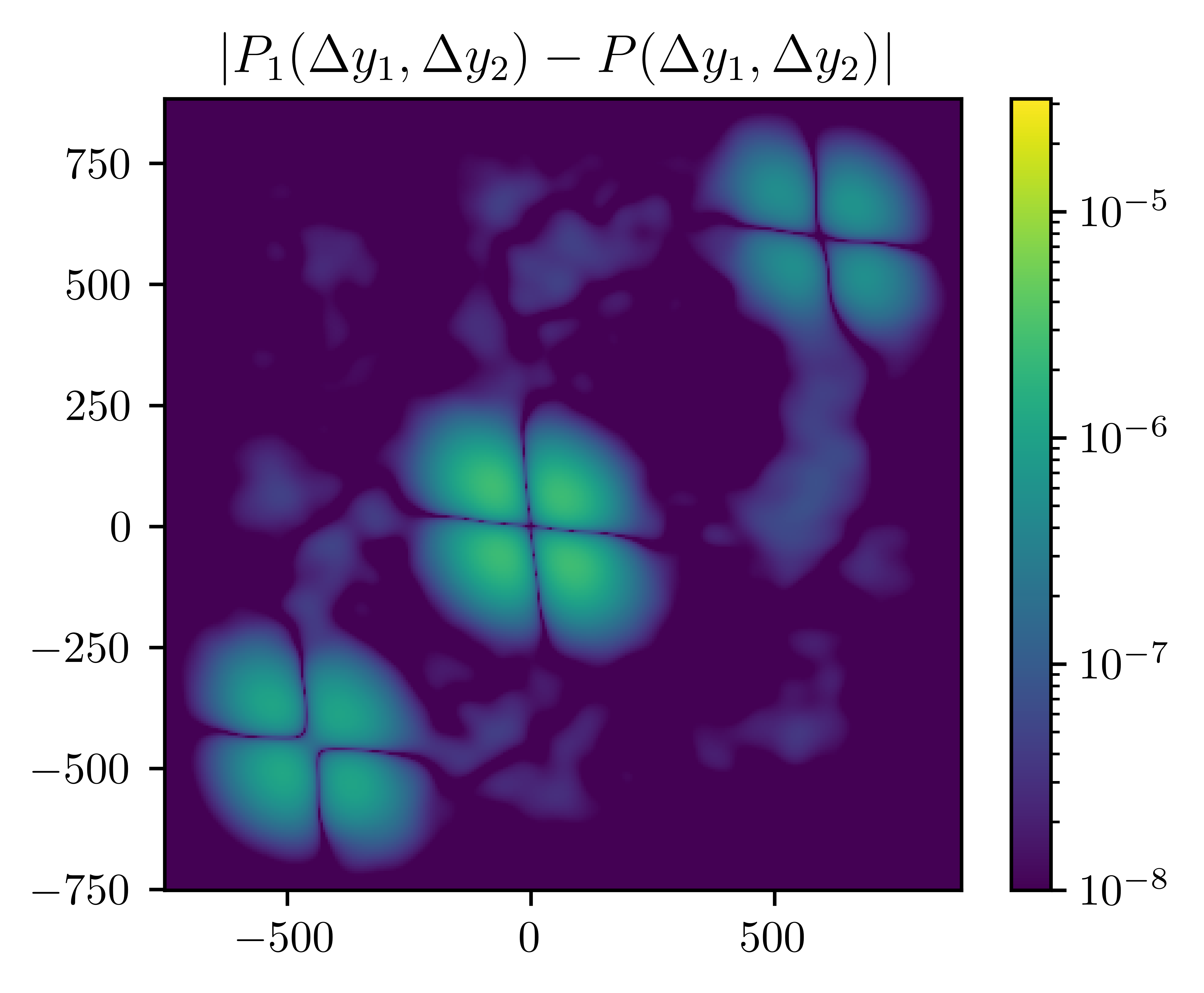}
    \end{minipage}%
    \begin{minipage}{0.35\textwidth}
    \includegraphics[width=\textwidth]{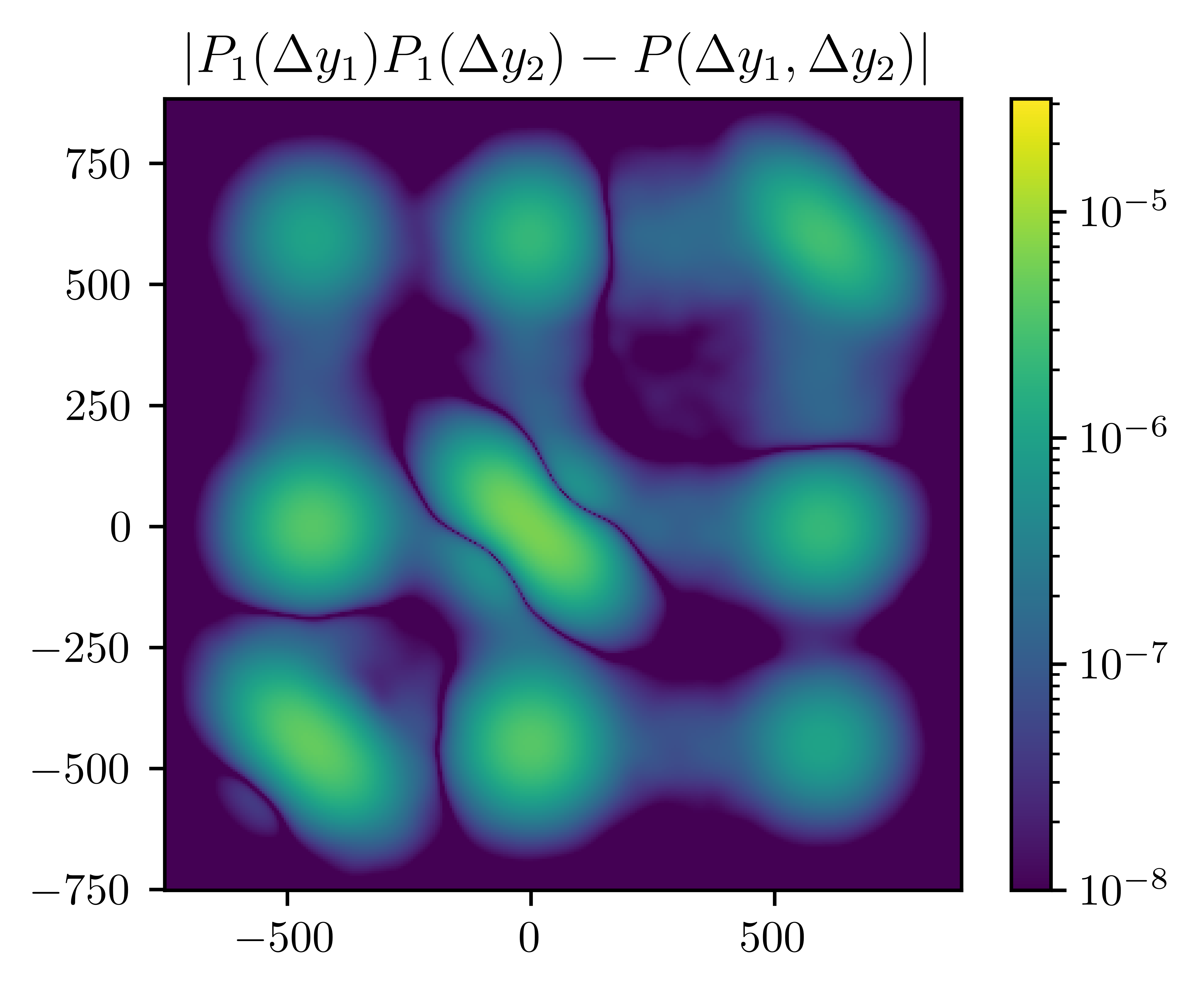}
    \end{minipage}
    \begin{minipage}{0.16\textwidth}
    \small \centering 
    \textbf{Four-state model}
    \end{minipage}%
    \begin{minipage}{0.14\textwidth}
    \includegraphics[width=\textwidth]{IMAGES/TRACKS/4.four-state_model_simplified.png}
    \end{minipage}%
    \begin{minipage}{0.35\textwidth}
    \includegraphics[width=\textwidth]{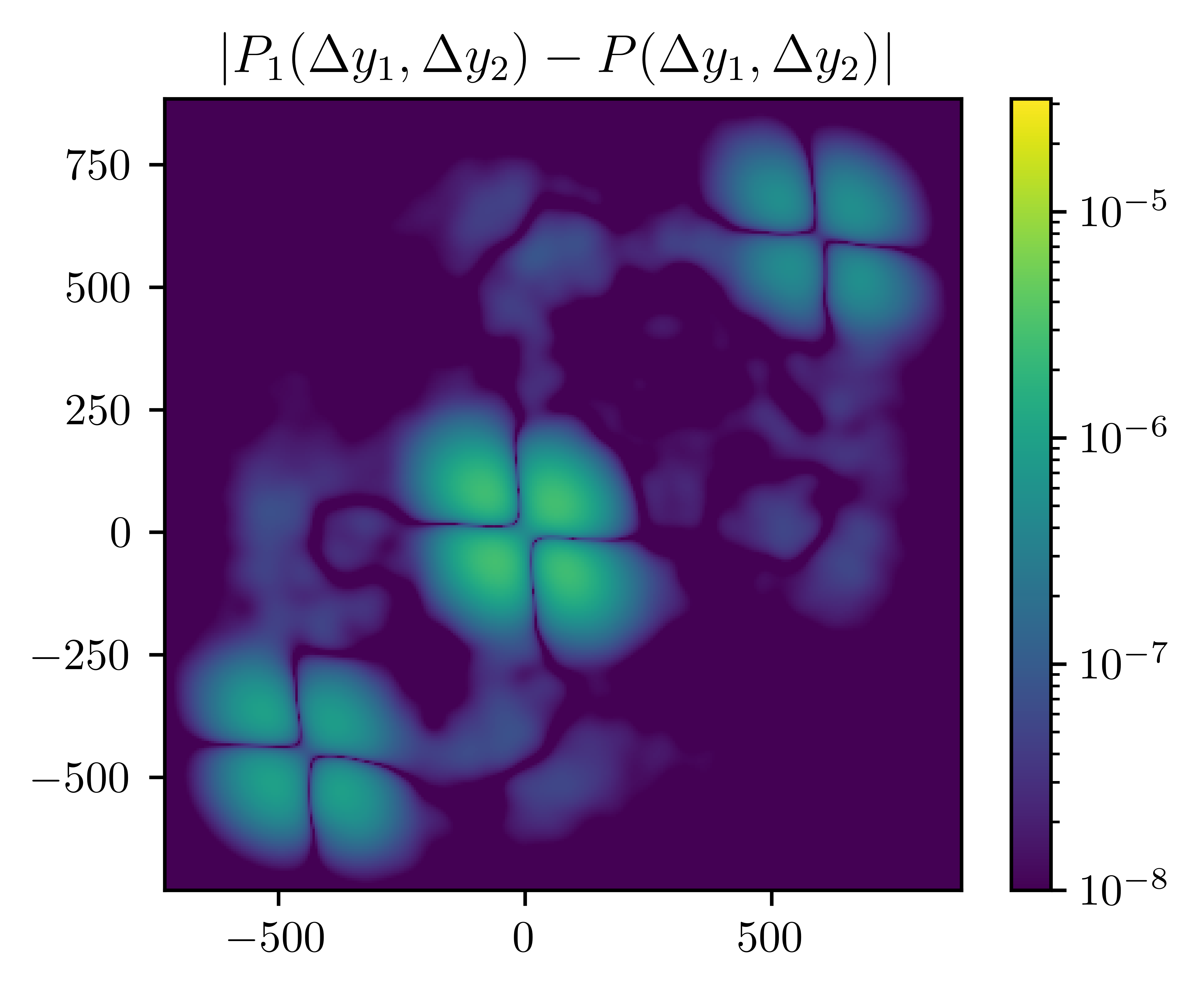}
    \end{minipage}%
    \begin{minipage}{0.35\textwidth}
    \includegraphics[width=\textwidth]{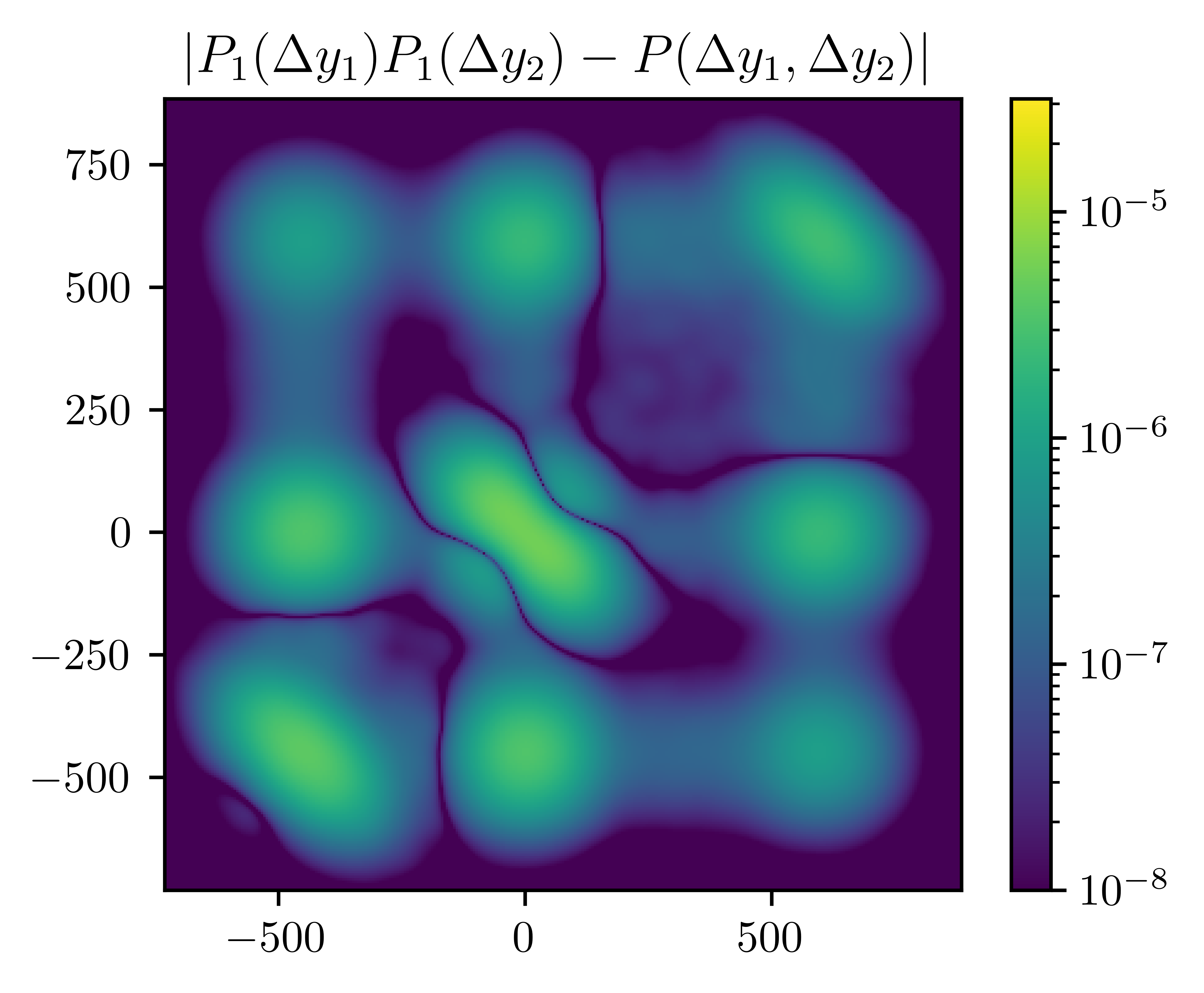}
    \end{minipage}
    \begin{minipage}{0.16\textwidth}
    \small \centering 
    \textbf{Six-state model}
    \end{minipage}%
    \begin{minipage}{0.14\textwidth}
    \includegraphics[width=\textwidth]{IMAGES/TRACKS/6.six-state_model_simplified.png}
    \end{minipage}%
    \begin{minipage}{0.35\textwidth}
    \includegraphics[width=\textwidth]{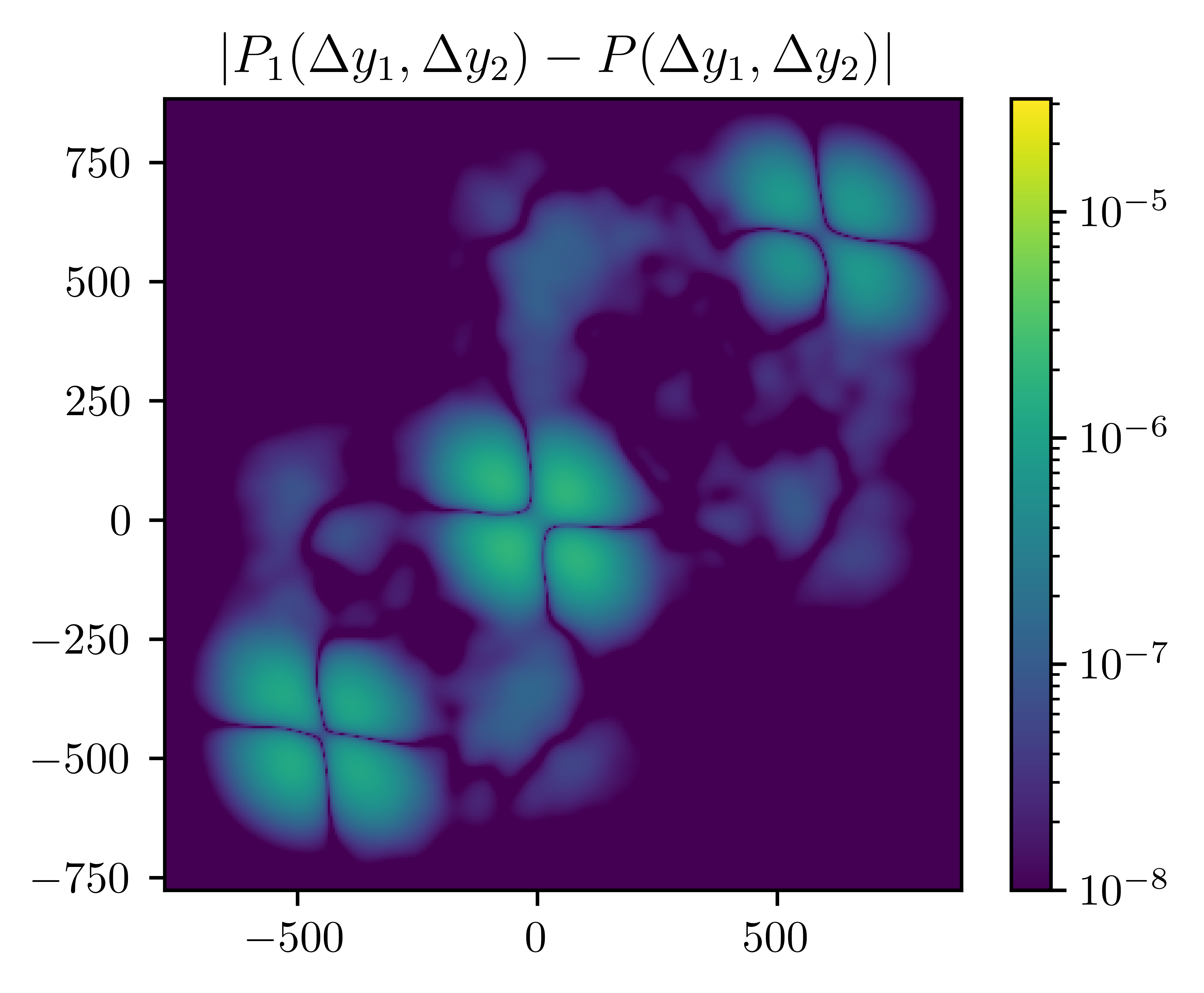}
    \end{minipage}%
    \begin{minipage}{0.35\textwidth}
    \includegraphics[width=\textwidth]{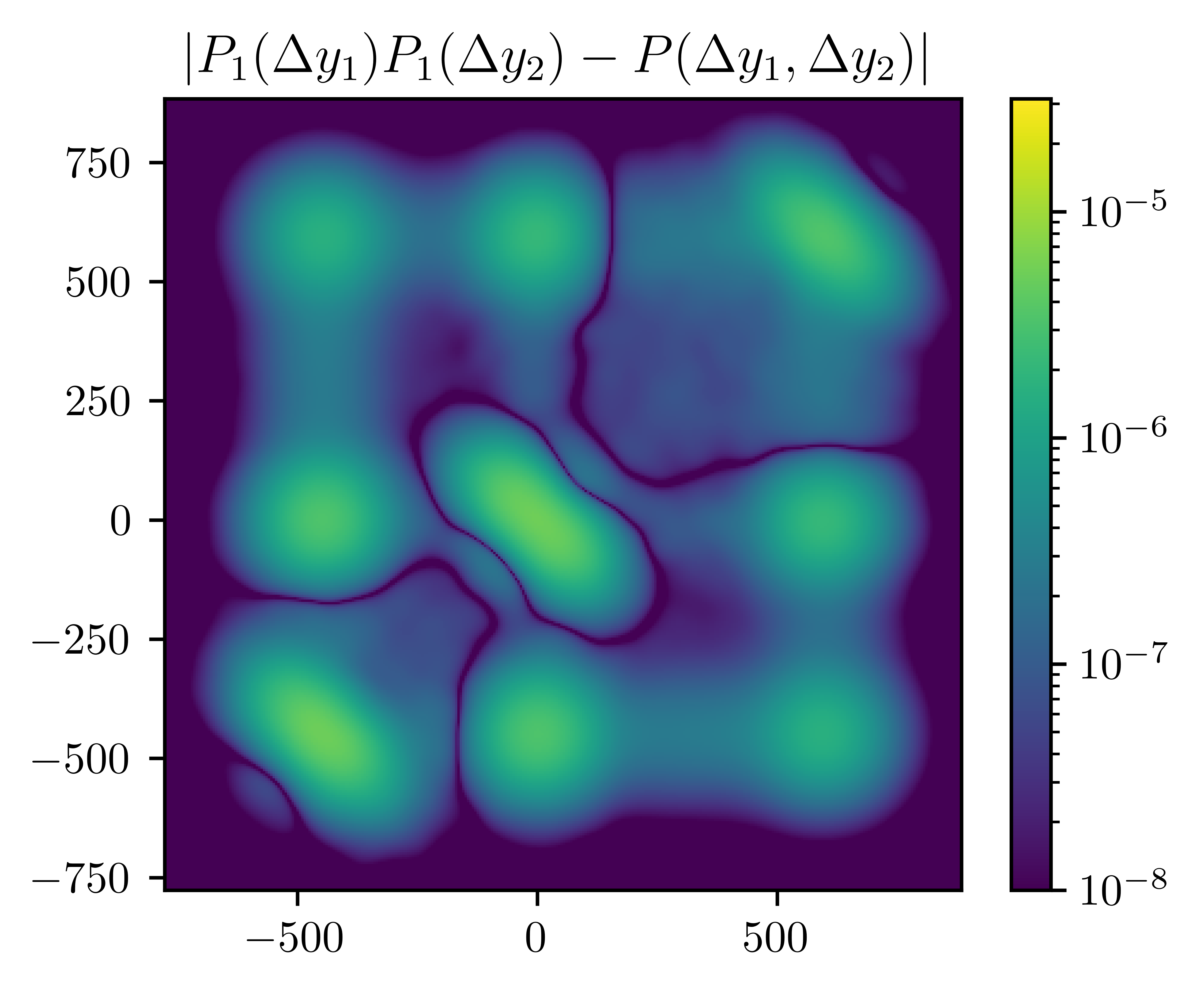}
    \end{minipage}
    \end{minipage}
    \caption{The absolute value of the error of the approximation of the joint PDF $P_1(\Delta y_1, \Delta y_2)$, defined as $P_1(\Delta y_1, \Delta y_2)-P(\Delta y_1, \Delta y_2)$ where $P(\Delta y_1, \Delta y_2)$ is the
    empirical PDF for $[\Delta y_1, \Delta y_2]$, is compared with the absolute value of the error of the approximation obtained as the product of the marginals $P_1(\Delta y_1)P_1(\Delta y_2)$, defined as $P_1(\Delta y_1)P_1(\Delta y_2)-P(\Delta y_1, \Delta y_2)$. The panels at the top are obtained using the parameters as specified in the Supplementary Information Figure~\ref{SI:Fig:networks}.}
    \label{SI:Fig:Err_P(Delta y_1, Delta y_2)}
\end{figure*}

\subsection{Extension to the up-to-two-switch approximation}

The maximum number of switches considered in the approximation for the PDF of a set of noisy subsequent location increments can be increased by modifying Equation~\eqref{SI:Eq:Use_P_1} conditioning on a higher amount of states from Equation~\eqref{SI:Eq:Track_P_inf}. Using an up-to-two-switch approximation we write
$$\begin{aligned}
    \mathbb{P}(\Delta y_{N-1}, S^N_1=s^{N}_1 \,|\,S^{N-1}_1=s^{N-1}_1)
    &=\mathbb{P}(\Delta y_{N-1}\,|\, S^N_1=s^{N}_1, S^{N-1}_1=s^{N-1}_1)\mathbb{P}(S^N_1=s^{N}_1 \,|\,S^{N-1}_1=s^{N-1}_1)
    \\
    &\approx \mathbb{P}(\Delta y_{N-1}\,|\, S^N_1=s^{N}_1, S^{N-1}_1=s^{N-1}_1, W^{N-1}=0) 
    \\
    & \qquad\times \mathbb{P}(S^N_1=s^{N}_1 \,|\,S^{N-1}_1=s^{N-1}_1, W^{N-1}=0) 
    \\
    &\qquad\times \mathbb{P}(W^{N-1}=0\,|\,S^{N}_1=s^{N}_1, S^{N-1}_1=s^{N-1}_1)
    \\
    &\quad + \mathbb{P}(\Delta y_{N-1}\,|\, S^{N-1}_1=s^{N}_1, S^{N-1}_1=s^{N-1}_1, W^{N-1}=1) 
    \\
    &\qquad\times \mathbb{P}(S^N_1=s^{N}_1 \,|\,S^{N-1}_1=s^{N-1}_1, W^{N-1}=1) 
    \\
    &\qquad\times \mathbb{P}(W^{N-1}=1 \,|\,S^{N}_1=s^{N}_1, S^{N-1}_1=s^{N-1}_1)
    \\
    &\quad + \mathbb{P}(\Delta y_{N-1}\,|\, S^N_1=s^{N}_1, S^{N-1}_1=s^{N-1}_1, W^{N-1}=2) 
    \\
    &\qquad\times \mathbb{P}(S^N_1=s^{N}_1 \,|\,S^{N-1}_1=s^{N-1}_1, W^{N-1}=2)
    \\
    &\qquad\times \mathbb{P}(W^{N-1}\ge 2 \,|\,S^{N}_1=s^{N}_1, S^{N-1}_1=s^{N-1}_1).
\end{aligned}$$
Finally, this can be written as
$$\begin{aligned}
&\mathbb{P}(\Delta y_{N-1}, S^N_1=s^{N}_1 \,|\,S^{N-1}_1=s^{N-1}_1)
\\
&\approx \ \mathbb{P}(\Delta y_{N-1}\,|\, S^{N-1}_1=s^{N-1}_1, W^{N-1}=0) 
    \\
    & \qquad\times \mathbb{P}(s^{N}_1=s^{N-1}_1) 
    \\
    &\qquad\times \mathbb{P}(W^{N-1}=0\,|\,S^{N-1}_1=s^{N-1}_1)
    \\
    &\quad+ \mathbb{P}(\Delta y_{N-1}\,|\, S^{N-1}_2=s^{N}_1, S^{N-1}_1=s^{N-1}_1, W^{N-1}=1) 
    \\
    &\qquad\times \mathbb{P}(S^{N-1}_2=s^{N}_1 \,|\,S^{N-1}_1=s^{N-1}_1) 
    \\
    &\qquad\times \mathbb{P}(W^{N-1}=1 \,|\,S^{N-1}_2=s^{N}_1, S^{N-1}_1=s^{N-1}_1)
    \\
    &\quad+ \sum_{\substack{s_2^{N-1}=1\\
    s_2^{N-1}\ne s_1^{N-1}\\s_2^{N-1}\ne s_1^{N}}}^n \mathbb{P}(\Delta y_{N-1}\,|\, S^{N-1}_3=s^{N}_1, S^{N-1}_2=s_2^{N-1}, S^{N-1}_1=s^{N-1}_1, W^{N-1}=2) 
    \\
    &\qquad\qquad\qquad\quad\times \mathbb{P}(S^{N-1}_3=s^{N}_1 \,|\,S^{N-1}_2=s^{N-1}_2)
    \\
    &\qquad\qquad\qquad\quad\times \mathbb{P}(W^{N-1}\ge 2 \,|\,S^{N-1}_2=s^{N-1}_2, S^{N-1}_1=s^{N-1}_1)
    \\
    &\qquad\qquad\qquad\quad\times \mathbb{P}(S^{N-1}_2=s^{N-1}_2 \,|\,S^{N-1}_1=s^{N-1}_1),
\end{aligned}$$
for which we can compute all of the terms.

\section{\textcolor{black}{Comparison to the Fokker-Planck equation}}

\textcolor{black}{The model presented corresponds to the Fokker-Planck equation (or Forward-Kolmogorov equation) in the case with no diffusion \citep{gardiner2009markov}. In particular, the evolution of the probability density function for the exact particle location $p(x,t;s)$ in state $s\in\{1,2,\ldots,n\}$ and at location $x$ at time $t\ge 0$ and is described by the following equation
\begin{equation*}
\frac{\partial p(x,t;s)}{\partial t} + v_s \frac{\partial p(x,t;s)}{\partial x} = \sum_{u=1}^n q_{us}p(x,t;u),
\end{equation*}
where $q_{us}$ is the transition rate from state $u$ to state $s$, and $v_s$ is the velocity in state $s$. We note that the probability density function for the exact particle location over time is obtained by adding the probability densities over all possible states
\begin{equation*}
p(x,t) :=\sum_{s=1}^n p(x,t;s).
\end{equation*}}

\bibliography{sn-bibliography_rev}